\documentclass[12pt,preprint]{aastex}
\def\bm#1{\mbox{\boldmath $#1$}} 
\usepackage{pslatex}
\usepackage{graphicx}

\begin{document}

\title{Convection and AGN feedback in clusters of galaxies}
\author{Benjamin D. G. Chandran}
\email{benjamin.chandran@unh.edu} 
\affil{Space Science Center and
Department of Physics, University of New Hampshire}

\author{Yann Rasera}
\email{yann.rasera@unh.edu} 
\affil{Space Science Center and
Department of Physics, University of New Hampshire}

\begin{abstract}
A number of studies have shown that the convective stability criterion
for the intracluster medium (ICM) is very different from the
Schwarzchild criterion due to the effects of anisotropic thermal
conduction and cosmic rays. Building on these studies, we develop a
model of the ICM in which a central active galactic nucleus (AGN)
accretes hot intracluster plasma at the Bondi rate and produces cosmic
rays that cause the ICM to become convectively unstable. The resulting
convection heats the intracluster plasma and regulates its temperature
and density profiles. By adjusting a single parameter in the model
(the size of the cosmic-ray acceleration region), we are able to
achieve a good match to the observed density and temperature profiles
in a sample of eight clusters. Our results suggest that convection is
an important process in cluster cores. An interesting feature of our
solutions is that the cooling rate is more sharply peaked about the
cluster center than is the convective heating rate. As a result, in
several of the clusters in our sample, a compact cooling flow arises
in the central region with a size~$r_{\rm cf}$ that is typically a
few~kpc. The cooling flow matches onto a Bondi flow at smaller
radii. The mass accretion rate in the Bondi flow is equal to, and
controlled by, the rate at which mass flows in through the cooling
flow. Our solutions suggest that the AGN regulates the mass accretion
rate in these clusters by controlling $r_{\rm cf}$: if the AGN power
rises above the equilibrium level, $r_{\rm cf}$ decreases, the mass
accretion rate drops, and the AGN power drops back down to the
equilibrium level.
\end{abstract}

\maketitle

\section{Introduction}

Active galactic nuclei (AGNs) have enormous mechanical and radiative
luminosities. If an AGN's power can be transferred to the surrounding
interstellar and intergalactic media, the resulting heating can have a
large effect on the ambient plasma. There has recently been
great interest in this process of ``AGN feedback,'' its role in galaxy
formation, and the possibility that AGN feedback solves the
over-cooling problem (Suginohara \& Ostriker 1998, Lewis et al 2000,
Tornatore et al 2003, Nagai \& Kravtsov 2004) and cooling-flow problem
for clusters of galaxies (B\"ohringer et al 20001; David et al 2001;
Tamura et al 2001; Molendi \& Pizzolato 2001; Blanton, Sarazin, \&
McNamara 2003; Peterson et al 2001, 2003).

One of the main unsolved problems for AGN feedback is
to understand how AGN power is transferred to the
diffuse ambient plasma.  A number of mechanisms have
been investigated, including Compton heating (Binney \&
Tabor 1995; Ciotti \& Ostriker 1997, 2001; Ciotti,
Ostriker, \& Pellegrini~2004, Sazonov et al 2005), shocks (Tabor \& Binney
1993, Binney \& Tabor 1995), magnetohydrodynamic (MHD)
wave-mediated plasma heating by cosmic rays
(B\"{o}hringer \& Morfill~1988; Rosner \& Tucker~1989;
Loewenstein, Zweibel, \& Begelman~1991), and cosmic-ray
bubbles produced by the central AGN (Churazov et
al~2001, 2002; Reynolds 2002; Br\"{u}ggen~2003;
Reynolds et~al~2005), which can heat intracluster
plasma by generating turbulence (Loewenstein \& Fabian
1990, Churazov et~al~2004) and sound waves (Fabian
et~al~2003; Ruszkowski, Br\"{u}ggen, \& Begelman
2004a,b) and by doing $pdV$ work (Begelman 2001, 2002;
Ruszkowski \& Begelman~2002; Hoeft \&
Br\"{u}ggen~2004). Despite this substantial progress,
it is still not clear how AGN feedback controls the
density and temperature profiles of the ambient plasma
in a way that is both self-regulating and consistent
with observations.

In this paper, we focus on clusters of galaxies and explore the
hypothesis that central AGNs heat and regulate the intracluster plasma
by causing the intracluster medium to become convectively unstable, a
scenario that was investigated in two earlier studies [Chandran (2004)
  (hereafter Paper~I) and Chandran~(2005) (hereafter Paper~II)].  At
first glance, this hypothesis seems obviously incorrect, since
observations show that the specific entropy~$s$ in intracluster
plasmas increases with radius~$r$.  However, several recent studies
have shown that the Schwarzchild criterion ($ds/dr>0$) does not apply
to low-density, magnetized plasmas such as those found in clusters, in
which the charged-particle gyroradii are much less than the Coulomb
mean free path. In such plasmas, heat and charged particles diffuse
primarily along magnetic field lines, and only weakly across the
magnetic field.  This anisotropy turns out to have a profound effect
on convective stability, as shown analytically by Balbus (2000, 2001)
and numerically by Parrish \& Stone (2005, 2007). These authors
considered a stratified plasma in which the gravitational acceleration
is in the $-z$~direction and the equilibrium magnetic field is in the
$xy$-plane and showed that the convective stability criterion
is~$dT/dz >0$, not $ds/dz>0$, where $T$ is the temperature. When
cosmic rays are present, the convective
stability criterion becomes $n k_B dT/dz + dp_{\rm cr}/dz >0$, as
shown analytically by Chandran \& Dennis (2006) and numerically by
Rasera \& Chandran (2007). Here, $n$ and $p_{\rm cr}$ are the
thermal-plasma number density and cosmic-ray pressure, respectively.
In galaxy clusters, the gravitational acceleration is in the~$-r$
direction, and the convective stability criterion is
\begin{equation}
n k_B \frac{dT}{dr} + \frac{dp_{\rm cr}}{dr} >0.
\label{eq:scintro}
\end{equation}
(Paper~II and appendix~\ref{ap:mlt} provide a more extensive
discussion.) Although~$dT/dr>0$ in cluster cores,
equation~(\ref{eq:scintro}) shows that cosmic rays produced by an AGN
at the center of a cluster can lead to convective
instability, since centrally produced cosmic rays
satisfy~$dp_{\rm cr}/dr<0$.

In this paper, we construct a spherically symmetric,
steady-state model of convective intracluster plasmas
using mixing-length theory, and compare this model to observations.
We assume that a central supermassive black hole accretes hot
intracluster plasma at the Bondi rate, and converts a small fraction
of the accreted rest-mass energy into cosmic rays that are accelerated
by shocks within some distance~$r_{\rm source}$ of the center of the
cluster.  The resulting cosmic-ray pressure gradient leads to
convection, which in turn heats the thermal plasma in the cluster core
by advecting internal energy inwards and allowing the cosmic rays to do~$pdV$
work on the thermal plasma. The model also includes thermal conduction,
cosmic-ray diffusion, and radiative cooling. The model 
involves much less emission from plasma at temperatures below
one-third of the cluster's average temperature than the cooling
flow model (Fabian 1994), and thus
offers a possible solution to the cooling-flow problem.

We compare the density and temperature profiles predicted by the model
to the profiles inferred from X-ray observations of eight clusters. We
adjust a single parameter, the size $r_{\rm source}$ of the cosmic-ray
acceleration region, to optimize the fit. The model solutions match
the observations well, with the exception of the density with the
central $\simeq 50$~kpc of Sersic~159-03, which is underestimated by
the model. We suggest a possible explanation for this discrepancy in
section~\ref{sec:comp}.  We also find that the cosmic-ray luminosities
of the AGN in our sample are strongly correlated with the
observationally inferred mechanical luminosities of these AGN.  Our
results suggest that AGN-driven convection is an important process in
cluster cores.

An attractive feature of this model and other models based on AGN
feedback and Bondi accretion is that they are self-regulating.  One
argument for why Bondi accretion is self-regulating was advanced by
Nulsen~(2004) and B\"ohringer et al~(2004), who noted that the Bondi
accretion rate is a monotonically decreasing function of the specific
entropy near the center of the cluster.  Thus, if the central plasma
becomes too cool, the Bondi accretion rate rises, the AGN feedback
heating increases, and the specific entropy of the central plasma
rises back to its equilibrium value.  In this paper, we offer an
additional explanation for how AGN heating on large scales ($\gtrsim
5$~kpc) can regulate the mass accretion rate onto the central black
hole. In our solutions, we find that the radiative cooling rate is
more sharply peaked about the center of a cluster than is the
convective heating rate. As a result, in several of the clusters
in our sample, the central region
becomes a cooling flow.  The radius of this cooling flow, $r_{\rm
  cf}$, is typically a few~kpc in our solutions. At smaller radii, the
flow makes a transition from a cooling flow to a Bondi flow. However,
as in the work of Quataert \& Narayan (2000), the mass accretion rate
of the inner Bondi flow is controlled by the surrounding cooling flow.
In our model, which has no mass dropout, the mass accretion rate is
approximately the plasma mass interior to~$r_{\rm cf}$ divided by the
cooling time at~$r_{\rm cf}$. The AGN then regulates the mass
accretion rate by controlling $r_{\rm cf}$: if the AGN power rises
above the equilibrium level, the size of the central cooling flow
decreases, the mass accretion rate drops, and the AGN power then drops
back down to the equilibrium level.

This paper extends the previous models of paper~I and paper~II in
several ways. In contrast to paper~I, the present paper takes into
account the role of anisotropic thermal conduction and cosmic-ray
diffusion, which strongly modify the convective stability
criterion. In contrast to paper~II, we take the cosmic-ray acceleration
to occur within a relatively small fraction of the total volume at any
given radius, which allows for localized pockets of excess cosmic-ray
pressure that tend to rise buoyantly. We also take into account the
nonzero average radial velocity, and compare the model to a larger
sample of clusters.

The rest of this paper is organized as follows.  We present the basic
equations of the model in section~\ref{sec:model}.  In
section~\ref{sec:comp}, we compare our model calculations to
observations.  In section~\ref{sec:eb} we consider the radial profiles
of the different heating rates and the factors that determine whether
AGN feedback or thermal conduction is the dominant heat source 
at $r \lesssim 100$~kpc. In section~\ref{sec:CF} we
discuss the central cooling flows that arise in our model solutions
for several of the clusters in our sample.  We also comment in
section~\ref{sec:CF} on the viability of the Bondi
accretion model for the AGN at the centers of clusters.  We summarize
our results in section~\ref{sec:disc}.  We present results on the
radial profiles of the turbulent velocity and cosmic-ray pressure in
appendix~\ref{ap:prof}.  In appendix~\ref{ap:mlt} we present a
systematic derivation of the two-fluid mixing-length theory that we
employ in our model.

\section{Model equations}
\label{sec:model} 

We describe the intracluster medium using a standard set of two-fluid
equations for cosmic rays and thermal plasma (Drury \& Volk 1981,
Jones \& Kang 1990), modified to include thermal conduction, viscous
dissipation, and radiative cooling:
\begin{equation}
\frac{d \rho}{dt} = - \rho \nabla \cdot \bm{v},
\label{eq:cont} 
\end{equation} 
\begin{equation}
\rho\frac{d \bm{v}}{dt} = - \nabla(p+p_{\rm cr}) - \rho \nabla \Phi - \nabla \cdot \Pi_{\rm visc},
\label{eq:momentum} 
\end{equation} 
\begin{equation} 
\frac{dp}{dt} = - \gamma p \nabla \cdot \bm{v} + (\gamma -1)[H_{\rm visc} + \nabla
\cdot ({\bf \mathsf{\kappa}}\cdot \nabla T) - R],
\label{eq:pe} 
\end{equation} 
and 
\begin{equation}
\frac{dp_{\rm cr}}{dt} = - \gamma_{\rm cr} p_{\rm cr} \nabla \cdot \bm{v}  + \nabla \cdot (
{\bf \mathsf{D}} \cdot \nabla p_{\rm cr}) + (\gamma_{\rm cr} -1 )\dot{E}_{\rm source},
\label{eq:cre} 
\end{equation} 
where 
\begin{equation}
\frac{d}{dt} \equiv \frac{\partial }{\partial t} + \bm{v} \cdot \nabla,
\end{equation} 
$\rho$ is the plasma density, $\bm{v}$ is the bulk velocity of the
two-fluid mixture, $p$ and $p_{\rm cr}$ are the plasma and cosmic-ray
pressures, $T$ is the plasma temperature, $\Phi$ is the gravitational
potential, $\Pi_{\rm visc}$ is the viscous stress tensor, $\gamma$ and
$\gamma_{\rm cr}$ are the plasma and cosmic-ray adiabatic indices
(which are treated as constants), $H_{\rm visc}$ is the rate of
viscous heating, $\bf \mathsf{\kappa}$ is the thermal conductivity
tensor, $R$ is the radiative cooling rate, $\dot{E}_{\rm source}$ is the
rate of injection of cosmic-ray energy per unit volume by the central
radio source, and $\bf \mathsf{D}$ is an effective momentum-averaged
cosmic-ray diffusion tensor. For the calculations presented in
section~\ref{sec:comp}, we set~$\gamma=5/3$ and $\gamma_{\rm cr}=4/3$.
We ignore radiative cooling of cosmic rays, which is reasonable if
protons make the dominant contribution to the cosmic-ray pressure.  We
also neglect Coulomb interactions between cosmic rays and thermal
plasma, as well as wave-mediated heating of the thermal plasma by
cosmic rays.  As discussed below, we take into account the effects of
the magnetic field on~$\bf\mathsf{\kappa}$ and~$\bf\mathsf{D}$, but we
neglect the Lorentz force and resistive dissipation.  In the following
subsections, we describe the approximations we use to solve the above
equations.

\subsection{Mixing length theory}

To account for convection, we
write each fluid quantity 
as an average value plus a turbulent fluctuation:
\begin{equation}
\bm{v}  = \langle \bm{v}  \rangle + \delta \bm{v},
\label{eq:avv} 
\end{equation} 
\begin{equation}
p = \langle p \rangle + \delta p, 
\label{eq:avp} 
\end{equation} 
etc, where $\langle \dots \rangle$ denotes an average over the
turbulent fluctuations. We take the averaged quantities to be
spherically symmetric and independent of time, and we treat
the fluctuating quantities as small. 
To obtain equations for the average cluster
properties, we average equations~(\ref{eq:cont}) through~(\ref{eq:cre}).
We evaluate the averages $\langle \rho \bm{v}\rangle$, $\langle
\mbox{\boldmath$v$} p \rangle$, and $\langle
\mbox{\boldmath$v$} p_{\rm cr}\rangle$ in
equations~(\ref{eq:cont}), (\ref{eq:pe}),
and~(\ref{eq:cre}) using a two-fluid
mixing-length theory that we describe in
appendix~\ref{ap:mlt}. The essential idea behind this theory
is that the amplitudes of the turbulent fluctuations increase
as the average plasma and cosmic-ray profiles move past the point of
marginal stability towards increasing degrees of convective
instability. As this happens, the magnitudes of the internal
energy flux $\langle \mbox{\boldmath$v$} p
\rangle/(\gamma-1)$ and the cosmic-ray energy flux $\langle
\mbox{\boldmath$v$} p_{\rm cr}\rangle/(\gamma_{\rm cr} - 1)$
increase, which in turn affects the density, temperature,
and cosmic-ray-pressure profiles. The two-fluid
mixing length theory provides an approximate way of
determining the resulting profiles as well as the
$r$-dependent turbulent velocity in a self-consistent way.
A key parameter of the model is the mixing length~$l$, which
characterizes the length scale of the convective turbulence.
We set
\begin{equation}
l = 0.4 r.
\end{equation}

\subsection{Hydrostatic equilibrium}

We assume that the convection is subsonic and
confine our model to $r\geq 0.2$~kpc, so that the
average radial velocity remains subsonic throughout our
solutions. As a result, we can to a reasonable
approximation drop the inertial terms in the average of
equation~(\ref{eq:momentum}). The viscous term in
equation~(\ref{eq:momentum}) is important primarily for dissipating
small-scale velocity fluctuations and can also be neglected in the
average of equation~(\ref{eq:momentum}).  The average of
equation~(\ref{eq:momentum}) then reduces to
\begin{equation}
\frac{d}{dr}\langle p_{\rm tot}\rangle = -\langle \rho\rangle \frac{d\Phi}{dr},
\label{eq:heq} 
\end{equation} 
where
\begin{equation}
p_{\rm tot} = p + p_{\rm cr}.
\label{eq:defptot} 
\end{equation}

\subsection{Gravitational potential}

We take the gravitational potential to be the sum
of four components,
\begin{equation}
\Phi = \Phi_{\rm c} + \Phi_{\rm s} +  \Phi_{\rm bh} + \Phi_{\rm p},
\end{equation} 
where $\Phi_c$ is the contribution from the 
the cluster's dark matter, $\Phi_{\rm s}$ is 
the contribution from the stars in the
brightest cluster galaxy (BCG), $\Phi_{\rm bh}$ is the
contribution from the black hole at $r=0$,
and $\Phi_{\rm p}$
is the contribution from the intracluster plasma.
We take the cluster dark matter to have an
NFW density profile (Navarro, Frenk, \& White 1997),
\begin{equation}
\rho_{\rm DM} = \frac{\delta_c \rho_{\rm crit}(z)r_s^3}{r ( r + r_s)^2},
\label{eq:nfw} 
\end{equation} 
where
\begin{equation}
\delta_c = \frac{200}{3} \frac{c^3}{[\ln(1+c)\; -\; c/(1+c)]},
\end{equation} 
$r_s$ is the scale radius, $c$ is the concentration parameter, and
$\rho_{\rm crit} = 3H^2/8\pi G$ is the critical density at the
redshift~$z$ of the cluster. The latter
is calculated assuming~$\Omega_0 = 0.3$,
$\Omega_{\Lambda,0}=0.7$, and~$H_0 =70 \mbox{ km}\;\mbox{s}^{-1}
\mbox{Mpc}^{-1}$. The values of $r_s$, $c$, and $z$ for the
eight clusters we consider in section~\ref{sec:comp}  are
taken from the literature and listed in table~\ref{tab:BCG}.

\begin{table}[h]
\caption{Parameters used in determining the gravitational potential
\label{tab:BCG} }
{\footnotesize 
\begin{tabular}{llrrrrrrrr}
&&&&&&&&&  \\ 
\hline \hline 
\vspace{-0.2cm} 
\\
Cluster & BCG
& \multicolumn{1}{c}{$r_s$ } 
& \multicolumn{1}{c}{$c$}
 & \multicolumn{1}{c}{$z$}  
& \multicolumn{1}{c}{$M_B$ } 
& \multicolumn{1}{c}{$B-V$}
& \multicolumn{1}{c}{$L_B$ } 
& \multicolumn{1}{c}{$R_e$}
& \multicolumn{1}{c}{$M_{\rm bh}$}  \\
& &  \multicolumn{1}{c}{(kpc)}  &
& &
& &  
\multicolumn{1}{c}{($10^{11} L_{B,\sun}$)} &
\multicolumn{1}{c}{(kpc)} &
\multicolumn{1}{c}{($10^9 M_{\sun}$)} \\
\vspace{-0.2cm} 
\\
\hline 
\hline 
\vspace{-0.2cm} 
\\
Virgo & NGC 4486 (M87)      & 560 & 2.8  & (see below) & -21.96 & 0.93 & 0.938 & 5.03 & 1.38  \\
Abell 262 & NGC 0708        &  85 & 8.62 & 0.0155      & -21.08 & 1.06 & 0.417 & 25.6 & 0.555 \\
Sersic 159-03 & ESO 291-009 & 159 & 6.56 & 0.0572      & -22.16 & 1.00 & 1.13  & 29.5 & 1.92  \\
Abell 4059 & ESO 349-010    & 744 & 2.7  & 0.0466      & -22.73 & 1.06 & 1.91  & 24.5 & 4.12  \\
Hydra A & PGC 026269        & 77  & 12.3 & 0.0550      & -22.97 & 0.82 & 2.38  & 39.6 & 4.12  \\
Abell 496 &  PGC 015524     & 129 & 7.75 & 0.0322      & -22.48 & 1.12 & 1.51  & 49.9 & 3.27  \\
Abell 1795 & PGC 049005     & 430 & 4.21 & 0.0639      & -22.04 & 1.00 & 1.01  & 40.3 & 1.66  \\
Perseus & NGC 1275          & 481 & 4.09 & 0.0179      & -22.62 & 0.53 & 1.72  & 15.3 & 1.89  \\
\vspace{-0.2cm} 
\\
\hline
\end{tabular}
\vspace{0.5cm} 

\noindent The NFW parameters $r_s$ and $c$ describe the clusters' dark
matter density profiles.  For Virgo $r_s$ and $c$ are taken from
McLaughlin (1999). For Hydra~A, $r_s$ and $c$ are taken from David
et~al~(2001).  For all other clusters, $r_s$ and $c$ are taken from
table~1 of Piffaretti et~al~(2005).  Redshifts $z$ are taken from
Kaastra et al~(2004), except for Virgo --- Kaastra et~al~(2004) take
the distance to Virgo to be 16~Mpc, and we use the same value.
Absolute B-band magnitudes $M_B$ and $B-V$ color indices for the
brightest cluster galaxies (BCGs) are taken from the ``Hyperleda''
database of Paturel et al (2003). The BCG effective radii $R_e$ are
taken from Schombert~(1987) for Perseus and Abell~1795, from Graham et
al~(1996) for Hydra~A, Abell~262, and Abell~496, and from
``Hyperleda'' for Virgo, Sersic~159-03 and Abell~4059. $L_B$ is the
BCG B-band luminosity.  The black hole masses are determined using the
mass-luminosity relation given in equation~(6) of Lauer et al~(2007).}
\end{table}

We take the stellar mass density to have a Hernquist profile
in which the stellar mass interior to radius~$r$ is
\begin{equation}
M_{\rm stars}(r) = \frac{M_0 r^2}{(r+a)^2},
\label{eq:Mstars} 
\end{equation} 
where $M_0$ is the total stellar mass and $a$ is a scale length equal
to $R_e/1.8153$, where $R_e$ is the radius of the isophote
enclosing half the galaxy's light. (Hernquist 1990) As in Graham et al
(2006), we set $M_0 = \Upsilon_B L_B$, where $L_B$ is the BCG B-band
luminosity, and $\Upsilon_B = 5.3 M_{\sun}/L_{B, \sun}$ is the B-band
stellar mass-to-light ratio for a 12-Gyr-old single stellar population
(Worthey 1994).  We set $L_B/L_{B,\sun} = \displaystyle 10^{0.4
  (M_{B,\sun} - M_B)}$, where $M_{B,\sun}$ and $M_B$ are,
respectively, the solar and BCG absolute B-band magnitudes, and $
M_{B,\sun} = 5.47$ (Cox 2000). The values of $R_e$ and $M_B$ for each
cluster are taken from the literature (see table~\ref{tab:BCG}).

We determine the black-hole mass using the mass-luminosity relation
given in equation~(6) of Lauer et al~(2007):
\begin{equation}
\log\left(\frac{M_{\rm bh}}{M_{\sun}}\right) =
 8.67 - 0.528(M_V + 22),
\label{eq:Mbh} 
\end{equation} 
where $M_V$ is the BCG absolute V-band magnitude. We set $M_V = M_B -
(B-V)$, where $M_B$ and the $B-V$ color index for each cluster are
taken from the ``Hyperleda'' database (Paturel et al 2003) and listed
in table~\ref{tab:BCG}. The resulting values of $M_{\rm bh}$ for each
cluster are also listed in table~\ref{tab:BCG}.

The contribution to the gravitational potential from the intracluster
plasma $\Phi_{\rm p}$ is not determined ahead of time, but is instead
obtained by solving $\nabla^2 \Phi_p = -4\pi G\langle \rho\rangle$,
where $\langle \rho\rangle$ is the average plasma density that results
from solving the model equations.

\subsection{Radiative cooling and chemical composition}

We use the analytic fit of Tozzi \& Norman (2001) to
approximate the full cooling function
for free-free and line emission:
\begin{equation}
R = n_{\rm i} n_{\rm e} \left[0.0086\left(\frac{k_{\rm B} T}{\mbox{1 keV}}\right)^{-1.7}
+ 0.058 \left(\frac{k_{\rm B} T}{\mbox{1 keV}}\right)^{0.5} + 0.063\right]
\cdot 10^{-22} \mbox{ ergs}\mbox{ cm}^3 \mbox{ s}^{-1},
\label{eq:R1}
\end{equation} 
where $n_{\rm i}$ is the ion density, $n_{\rm e}$
is the electron density, $k_{\rm B}$ is the Boltzmann constant,
and the numerical constants
correspond to 30\% solar metallicity.   Because we treat the
turbulent fluctuations as small, we can replace $n_{\rm e}$, $n_{\rm
  i}$, and~$T$ in equation~(\ref{eq:R1}) by their average values
when calculating~$\langle R\rangle$.
We take the intracluster plasma to be fully ionized and
to have a uniform chemical composition, 
with a hydrogen mass
fraction of $X=0.7$ and a helium mass fraction $Y=0.29$.
We take the metals to have a mean charge to mass ratio
equal to that of helium. The mean molecular weight is then
\begin{equation}
\mu \equiv \frac{\rho}{(n_{\rm e}+ n_{\rm i})m_H} = 0.62.
\end{equation} 
The mean molecular weight per electron is then
\begin{equation}
\mu_{\rm e}\equiv  \frac{\rho}{n_{\rm e}m_H} = 1.18.
\end{equation} 
In addition,
\begin{equation}
\frac{n_{\rm i}}{n_{\rm e}} = 0.91,
\end{equation} 
and
\begin{equation}
\frac{n_{\rm e}}{n_{\rm H}} = 1.21,
\label{eq:nenh} 
\end{equation} 
where~$n_{\rm H}$ is the hydrogen number density.

\subsection{Transport}

Cluster magnetic fields are easily strong enough to cause cosmic rays and
heat to diffuse primarily along magnetic field lines, so that
\begin{equation}
{\bf \mathsf{\kappa}} \simeq  \kappa_\parallel\hat{b}\hat{b} ,
\label{eq:kappapar} 
\end{equation} 
and
\begin{equation}
{\bf \mathsf{D}} \simeq D_\parallel \hat{b}\hat{b},
\label{eq:Dpar} 
\end{equation} 
where $\hat{b}$ is the magnetic field unit vector, and $\kappa_\parallel$
and $D_\parallel$ are the parallel conductivity and diffusivity.
We take the parallel conductivity to be the classical
Spitzer thermal conductivity (Spitzer \& Harm 1953, Braginskii 1965),
\begin{equation}
\kappa_\parallel =\kappa_{\rm S} =9.2 \times 10^{30}  n_{\rm e} k_{\rm B} \left(
\frac{k_{\rm B} T}{5 \mbox{ keV}}\right)^{5/2}
\left(\frac{10^{-2} \mbox{ cm}^{-3}}{n_{\rm e}}\right)
\left(\frac{37}{\ln \Lambda_{\rm c}}\right) \frac{\mbox{ cm}^{2}}{\mbox{ s}},
\label{eq:kappas} 
\end{equation} 
where $\ln \Lambda_{\rm c}$ is the Coulomb logarithm.  The local
anisotropy of ${\bf \mathsf{\kappa}}$ and ${\bf \mathsf{D}} $ turns
out to be critical for convective stability, as discussed by Balbus
(2000,2001), Parrish \& Stone (2005,2007), Chandran \& Dennis (2006),
and Rasera \& Chandran (2007), and we take this anisotropy into
account in our mixing length theory for intracluster convection.
[See, e.g., the discussion preceding equation~(\ref{eq:defHtc}).]
However, when we average equations~(\ref{eq:cont})
through~(\ref{eq:cre}) and solve for the structure of the ICM, we are
interested in the transport of heat and cosmic rays over distances
much greater than the correlation length of the magnetic field, $l_B$,
which is $ \sim 1 - 10$~kpc (Kronberg 1994; Taylor et al~2001,~2002;
Vogt \& Ensslin 2003, 2005 - see
Schekochihin et al 2006
and Schekochihin \& Cowley 2006 for
a recent discussion of intracluster
magnetic fields and turbulence).  For transport over such large scales,
averaging over the turbulent magnetic field leads to an effectively
isotropic conductivity, which we denote~$\kappa_T$, that is reduced
relative to~$\kappa_\parallel$ (Rechester \& Rosenbluth 1978, Chandran
\& Cowley~1998). Theoretical studies find that the reduction is by a
factor of~$\sim 5-10$ (Narayan \& Medvedev 2001, Chandran \& Maron
2004, Maron, Chandran, \& Blackman 2004). In this paper, we assume
that
\begin{equation}
\kappa_T = \frac{\kappa_\parallel}{8}.
\label{eq:kappaT} 
\end{equation} 
We take the average of the conductive heating term to be given by
\begin{equation}
\langle \nabla \cdot(\mathsf{\bf \kappa}\cdot \nabla T)\rangle
 = \frac{1}{r^2}\frac{d}{dr} \left[r^2
\kappa_T  \frac{d}{dr}\langle T \rangle\right],
\label{eq:htc} 
\end{equation} 
with $T$ set equal to~$\langle T \rangle$ in equation~(\ref{eq:kappas}).
Similarly,  we assume that
\begin{equation}
\langle \nabla \cdot ({\bf \mathsf{D}} \cdot \nabla p_{\rm cr}) \rangle
= \frac{1}{r^2} \frac{d}{dr} \left[
r^2 D_{\rm cr} \frac{d}{dr}\langle p_{\rm cr}\rangle\right].
\end{equation} 
We take the value of~$D_{\rm cr}$ to be
\begin{equation}
D_{\rm cr} = \sqrt{D_0^2 + v_d^2 r^2},
\label{eq:defDcr} 
\end{equation} 
where $D_0= 10^{28} \mbox{ cm}^2/\mbox{s}$ and~$v_d = 10$~km/s.
The $v_d$~term is loosely motivated by a simplified picture
of cosmic-ray ``self-confinement,'' in which cosmic rays are
scattered by waves generated by the streaming of cosmic rays
along field lines.  If, contrary to fact, the field lines
were purely radial, efficient self-confinement would limit
the average radial velocity of the cosmic rays to the
Alfv\'en speed~$v_{\rm A}$, allowing the cosmic rays to
travel a distance~$r$ in a time~$\sim r/v_{\rm A}$. For
constant $v_{\rm A}$, this scaling can be approximately
recovered by taking the cosmic rays to diffuse isotropically
with~$D_{\rm cr} \propto r$, the scaling that arises from
equation~(\ref{eq:defDcr}) when $v_d r \gg D_0$.  This
self-confinement scenario is too simplistic, since in
clusters field lines are tangled, $v_{\rm A}$ varies in
space, and it is not known whether cosmic rays are primarily
scattered by cosmic-ray-generated waves or by
magnetohydrodynamic (MHD) turbulence excited by large-scale
stirring of the intracluster plasma.  It is not clear,
however, how to improve upon
equation~(\ref{eq:defDcr}). Self-confinement in the presence
of tangled field lines is not well understood, and the
standard theoretical treatment of scattering by MHD
turbulence, which takes the fluctuations to have wave
vectors directed along the background magnetic field, is
known to be inaccurate (Bieber et~al~1994, Chandran 2000,
Yan \& Lazarian 2004). A more definitive treatment must thus await
further progress in our understanding of MHD turbulence and
cosmic-ray transport.  The value of~$D_\parallel$ is needed
in the mixing length theory developed below. We assume
that $D_{\rm cr}/D_\parallel = \kappa_T/\kappa_\parallel$, and thus set
\begin{equation}
D_\parallel = 8 D_{\rm cr}.
\label{eq:dcr} 
\end{equation}

\subsection{The mass accretion rate and cosmic-ray luminosity of the central AGN}

We assume that the black hole at $r=0$ in our model, with a mass~$M_{\rm BH}$ given
by equation~(\ref{eq:Mbh}), accretes intracluster plasma at the Bondi
(1952) rate,
\begin{equation}
\dot{M} = \frac{\pi G^2 M_{\rm BH}^2 \rho}{c_{\rm s} ^3},
\label{eq:mdot} 
\end{equation} 
where $c_{\rm s}$ is the adiabatic sound speed, and $\rho$ and $c_{\rm
  s}$ are evaluated using the average plasma parameters at the
radius~$r_1=0.2$~kpc, which defines the inner boundary of our model
solutions. We assume that this accretion powers a jet that leads to
shocks, which in turn accelerate cosmic rays.  We take the cosmic-ray
luminosity to be
\begin{equation}
L_{\rm cr} = \eta \dot{M} c^2,
\label{eq:lcr} 
\end{equation} 
where
\begin{equation}
\eta = 5\times 10^{-3}.
\label{eq:defeta} 
\end{equation} 

An argument against Bondi accretion in clusters is that the radiative
luminosities of AGNs in elliptical galaxies are typically several
orders of magnitude smaller than the nominal Bondi accretion power,
given by $P_{\rm Bondi}=0.1 \dot{M}_{\rm Bondi} c^2$,
where $\dot{M}_{\rm Bondi}$ is the Bondi accretion rate given in
equation~(\ref{eq:mdot}). (Allen et al 2006) However, the mechanical
luminosities $L_{\rm mech}$ of these AGN are often much larger than
their radiative luminosities.  Moreover,  in a recent study of nine
AGNs in nearby x-ray luminous elliptical galaxies, Allen et~al~(2006)
found a strong correlation between $P_{\rm Bondi}$ (as calculated from
the observed plasma temperature and density profiles) and $L_{\rm
  mech}$ (as inferred from the energies and time scales required to
inflate the observed x-ray cavities). Allen et al (2006) found that $L_{\rm
  mech}$ can be related to $P_{\rm Bondi}$ by a power-law fit of the
form $\log(P_{\rm Bondi}/10^{43} \mbox{ erg}\mbox{ s}^{-1}) = c_1 +
c_2 \log(L_{\rm mech}/10^{43} \mbox{ erg}\mbox{ s}^{-1})$, with $c_1=
0.65\pm 0.16$ and $c_2= 0.77\pm 0.20$, and that the fraction of
$\dot{M}c^2$ that is converted into mechanical luminosity ranges from
1.3\% for a jet power of $10^{42}$~erg/s to 3.7\% for a jet power
of~$10^{44}$~erg/s.  Results consistent with these were also found by
Tan \& Blackman~(2005). These authors reviewed studies of 
M87 and estimated that $L_{\rm mech}$ is about an order of
magnitude larger than the radiative luminosity, and that $L_{\rm mech}
\sim 0.01 \dot{M}_{\rm Bondi} c^2$.  Our choice of $\eta = 0.005$ is
smaller than the accretion efficiencies found in these studies, in
part to provide a more conservative estimate, and in part because only
part of the mechanical energy is converted into cosmic rays.

We note that Bondi accretion in clusters has been considered
previously by a number of authors (e.g., Quataert \& Narayan 2000, Di
Matteo et al 2002, Nulsen~2004, B\"{o}hringer et al~2004, Springel et
al 2005, Cattaneo \& Teyssier 2007). Also, in Tan \& Blackman's (2005)
analysis, part of the reason for the small value of~$\eta$ is that
part of the mass flowing in through the Bondi radius never reaches the
central black hole because it forms stars in a gravitationally
unstable disk. Thus, the Bondi accretion rate in our model may be
significantly higher than the time derivative of the mass of the
central black hole.

Pizzolato \& Soker (2005) and Soker (2006) considered a different
``cold feedback'' scenario for mass accretion, in which cold gas fuels
the central AGN. In section~\ref{sec:CF} we address several issues
related to the question of whether one expects Bondi accretion or some
form of cold feedback in clusters.

\subsection{Cosmic-ray acceleration by the central radio source}
\label{sec:cra} 

The spatial distribution of cosmic-ray injection into the ICM is not
precisely known. Some clues are provided by radio observations, which
show that cluster-center radio sources (CCRS) differ morphologically
from radio sources in other environments. As discussed by Eilek
(2004), roughly half of the CCRS in a sample of 250 sources studied by
Owen \& Ledlow (1997) are ``amorphous,'' or quasi-isotropic,
presumably due to jet disruption by the comparatively high-pressure,
high-density cluster-core plasma. With the exception of Hydra~A, the
CCRS in the Owen-Ledlow (1997) study are smaller than
non-cluster-center sources, with most extending less than 50~kpc from
the center of the host cluster (Eilek 2004). Given these findings, we
take the cosmic-ray acceleration to be concentrated within the cluster
core.

In paper~II, it was assumed that the cosmic rays are accelerated in an
approximately volume-filling manner.  In contrast, in this paper, it
is assumed that cosmic-ray energy is injected into the intracluster
medium in only a fraction of the volume at any given radius.  We then
take
\begin{equation}
\dot{E}_{\rm source} = \langle \dot{E}_{\rm source} \rangle
+ \delta \dot{E}_{\rm source},
\end{equation} 
where 
\begin{equation}
\langle \dot{E}_{\rm source}\rangle = S_0 e^{-r^2/r_{\rm source}^2}
\label{eq:defS} 
\end{equation} 
can be thought of as an average of $\dot{E}_{\rm source}$ over
spherical polar angles.  The constant~$r_{\rm source}$ is a free
parameter that characterizes the size of the cosmic-ray acceleration
region. The constant~$S_0$ is determined on energy grounds from the
equation $L_{\rm cr} = 4\pi \int_0^\infty dr \, r^2 \langle
\dot{E}_{\rm source}(r) \rangle$ and equation~(\ref{eq:lcr}).  After
determining $\langle \dot{E}_{\rm source}(r) \rangle$, we set
\begin{equation}
\delta \dot{E}_{\rm rms}= \eta_2 \langle \dot{E}_{\rm source}\rangle,
\label{eq:eta2} 
\end{equation} 
where $\delta \dot{E}_{\rm rms}$ is the rms value of $\delta
\dot{E}_{\rm source}$, and
$\eta_2$ is a constant that is related to the volume
filling factor of the cosmic-ray acceleration region.  For
example, suppose that $\dot{E}_{\rm source} = C = \mbox{
  constant}$ in a fraction~$f_{\rm cr}$ of the volume between
radius~$r$ and $r+dr$, and that $\dot{E}_{\rm source} = 0$ in
the remainder of the volume between~$r$ and $r+dr$.  In this
case, $\langle \dot{E}_{\rm source}(r) \rangle = f_{\rm cr}C$,
$\langle [\dot{E}_{\rm source}]^2 \rangle = f_{\rm cr}C^2$, and
$\delta \dot{E}_{\rm rms} = \sqrt{\langle
  [\dot{E}_{\rm source}(r) - \langle \dot{E}_{\rm
      source}(r)\rangle]^2\rangle} = \langle \dot{E}_{\rm
  source}(r) \rangle\sqrt{f_{\rm cr}^{-1} - 1}$. For the calculations
presented below, we set $\eta_2=2.5$, which corresponds to $f_{\rm cr} = 0.138$.
These fluctuations in the cosmic-ray
source term drive fluctuations in the fluid quantities and
contribute to convection.  This effect is incorporated into
the two-fluid mixing length theory presented in
appendix~\ref{ap:mlt}.  The fluctuations in $\dot{E}_{\rm
  source}$ result in larger fluctuations (spatial variations)
in $p_{\rm cr}$ and~$\rho$ than in the model of
paper~II, which in some sense represent the
``cosmic-ray bubbles'' or X-ray cavities seen in about
one-fourth of the clusters in the Chandra archive (Birzan et
al 2004).

\subsection{Summary and numerical method}

The approximations described above lead to a set of coupled ordinary
differential equations for the average density, temperature, and
cosmic-ray pressure and the rms turbulent velocity.  These equations
are presented in appendix~\ref{ap:mlt}.  We solve this set of
equations using a shooting method, in which we guess the electron
density, temperature, and cosmic-ray pressure at the inner radius of
our model ($r_1=0.2$~kpc) and then update these guesses until the
model solution satisfies the three boundary conditions at the outer
radius~$r_{\rm outer}$. These outer boundary conditions are the
observed electron density $n_{\rm e, outer}$ and temperature $T_{\rm
  outer}$ at $r_{\rm outer}$, and a condition on~$\langle dp_{\rm
  cr}/dr\rangle$ at $r_{\rm outer}$, which amounts to requiring that
$\langle p_{\rm cr}\rangle \rightarrow 0$ as $r\rightarrow
\infty$. The value of $r_{\rm outer}$ for a cluster is taken to be the
radius of the first observational data point outside the cluster's
cooling radius, $r_{\rm cool}$, given in table~\ref{tab:t0} (except
for Virgo, for which we take $r_{\rm outer}$ to be the outermost data
point, which lies inside of $r_{\rm cool}$.)  The values of $r_{\rm
  outer}$, $n_{\rm outer}$, and $T_{\rm outer}$ are listed in
table~\ref{tab:t1}.  After finding the values of $n_e$, $T$,
and~$p_{\rm cr}$ at $r_1$ needed to match the boundary conditions
at~$r_{\rm outer}$, we integrate the equations out to radii
greater than~$r_{\rm outer}$ as needed to compare to the data. A more
extensive discussion of our numerical method is given in
appendix~\ref{ap:mlt}.

\section{Comparison to observations}
\label{sec:comp} 

We compare our model solutions with observations of the central
regions ($r<0.25 r_{\rm vir}$, where $r_{\rm vir} = c r_s$ is the
virial radius) of eight clusters: Virgo, Abell 262, Sersic 159-03,
Abell 4059, Hydra~A, Abell 496, Abell 1795, and Perseus. Temperature
and hydrogen-number-density ($n_H$) profiles for these clusters are
taken from table~5 of Kaastra et al~(2004). Redshifts ($z$) and
angular-diameter distances~$d_{\rm scdm}$ are given in table~1 of
Kaastra et al~(2004). The data of Kaastra et al~(2004) are obtained
assuming a standard cold dark matter (SCDM) cosmology with $\Omega =
1$ and $H_0=50 \mbox{ km} \mbox{ s}^{-1} \mbox{ Mpc}^{-1}$. We convert
to a $\Lambda$CDM cosmology with $\Omega_0=0.3$, $\Omega_{\Lambda,0} =
0.7$, and $H_0 = 70\mbox{ km} \mbox{ s}^{-1} \mbox{ Mpc}^{-1}$ by
calculating the ratio of angular-diameter distance in the two
cosmologies, $\zeta(z) \equiv d_{\rm scdm}/d_{\rm \Lambda cdm}$, for
each cluster in the sample. We then multiply Kaastra et al's (2004)
values for~$n_H$ by~$\sqrt{\zeta}$ (since the observed X-ray flux and
angular size are fixed) and multiply linear distances
by~$\zeta^{-1}$.  Values of $\zeta$, as well as the cooling radius,
are given in table~\ref{tab:t0}.
[The conversion from $n_H$ to $n_e$ is given by equation~(\ref{eq:nenh}).]

\begin{table}[h]
\caption{Cluster parameters
\label{tab:t0} }
{\footnotesize 
\begin{center}
\begin{tabular}{lrr}
&&  \\ 
\hline \hline 
\vspace{-0.2cm} 
\\
Cluster & 
 \multicolumn{1}{c}{$\displaystyle \frac{d_{\rm scdm}}{d_{\rm \Lambda cdm}}$}
 & \multicolumn{1}{c}{$r_{\rm cool}$}  \\
& & \multicolumn{1}{c}{(kpc)} 
\vspace{0.2cm} 
\\
\hline 
\hline 
\vspace{-0.2cm} 
\\
Virgo & 1.0 & 73 \\
Abell  & 1.39 & 61 \\
Sersic 159-03 & 1.36 & 128 \\
Abell 4059  & 1.37 & 86 \\
Hydra A   & 1.36 & 130 \\
Abell 496  & 1.38 & 89 \\
Abell 1795 & 1.36 & 130 \\
Perseus & 1.39 & 128 \\
\vspace{-0.2cm} 
\\
\hline
\end{tabular}
\end{center}
\noindent  The quantity
$d_{\rm scdm}$ is the angular-diameter distance to each cluster in
the SCDM cosmology employed by Kaastra et
al (2004) in which $\Omega=1$ and $H_0 =50 \mbox{ km}\;\mbox{s}^{-1}
\mbox{Mpc}^{-1}$.  $d_{\rm \Lambda cdm}$ is angular-diameter distance to each cluster
in the $\Lambda$CDM cosmology assumed in this paper, in which
$\Omega=0.7$, $\Omega_\Lambda=0.3$ and $H_0 =70 \mbox{
  km}\;\mbox{s}^{-1} \mbox{Mpc}^{-1}$. For Virgo, $d_{\rm scdm}/d_{\Lambda \rm cdm}$
is set equal to 1.0, since we use the same distance (16~Mpc) to
Virgo as employed by Kaastra et al~(2004).
The cooling radii~$r_{\rm cool}$ are taken
from Kaastra et~al~(2004) but rescaled
to $\Lambda$CDM. Kaastra et al's values of $r_{\rm cool}$
are the radii at which the radiative cooling time $t_{\rm cool}$
is 15~Gyr in SCDM. At the rescaled values of $r_{\rm cool}$,
$t_{\rm cool}$ is 15~Gyr$\cdot (d_{\Lambda\rm cdm}/d_{\rm scdm})^{1/2}$
in $\Lambda$CDM
(because the cooling time scales like $n_e^{-1}$ and
$n_e\propto d^{-1/2}$ for a fixed observed X-ray flux and angular size,
where $d$ is the angular-diameter distance).} 
\end{table}
\normalsize

\begin{figure}[h]
\hspace{-1cm}
\includegraphics[width=1.7in]{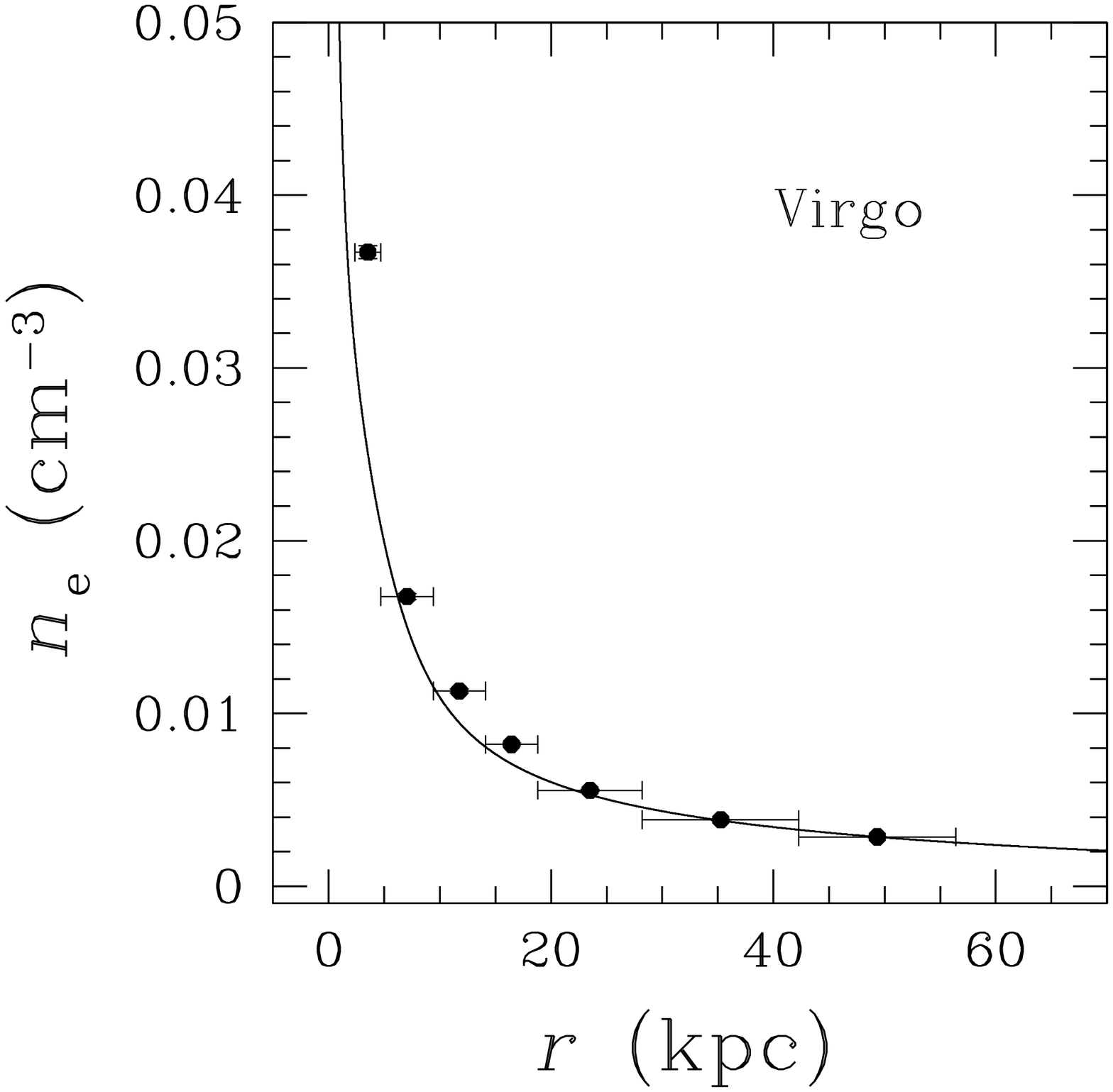}\hspace{0.05cm}\includegraphics[width=1.7in]{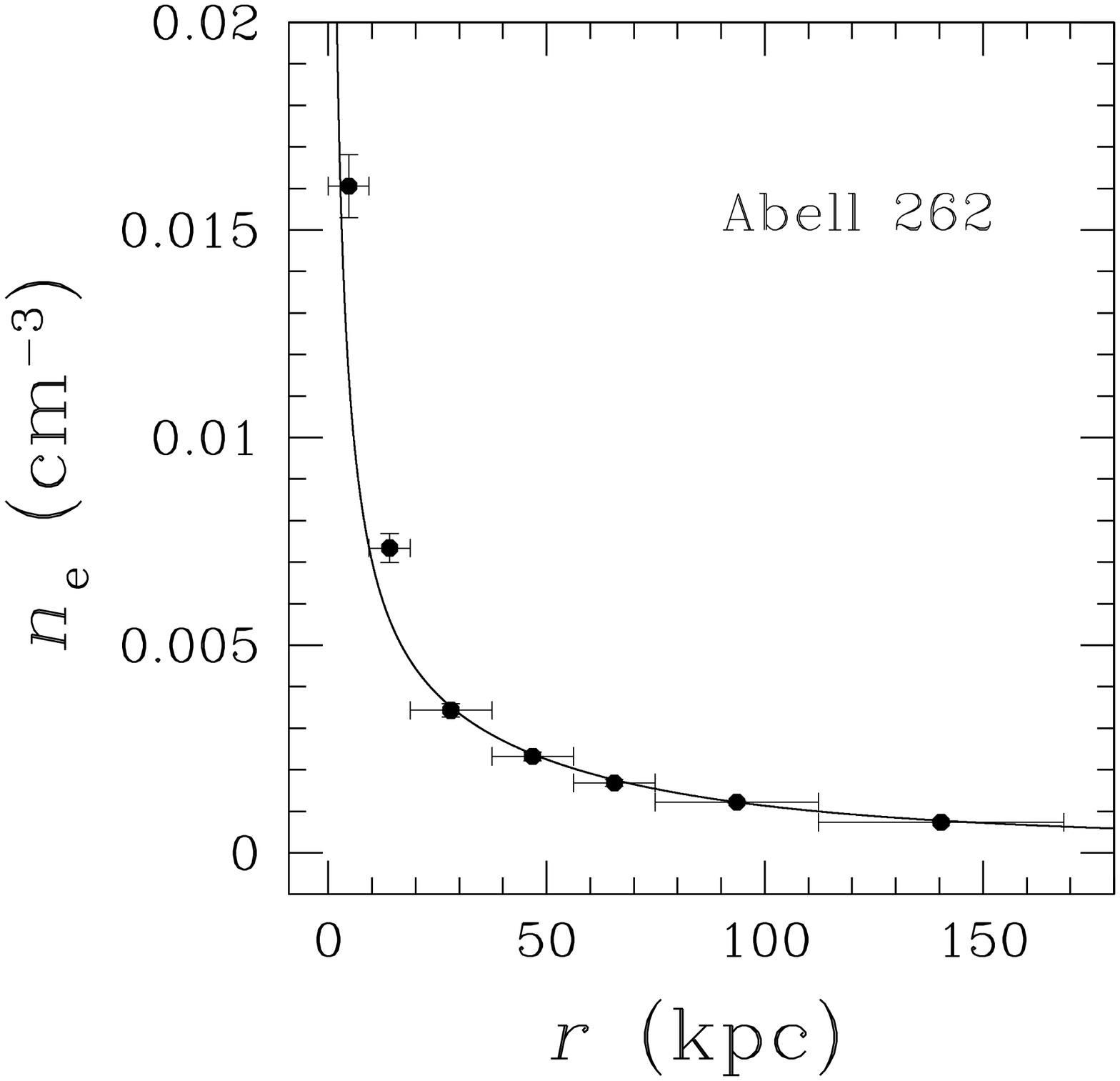}\hspace{0.05cm}\includegraphics[width=1.7in]{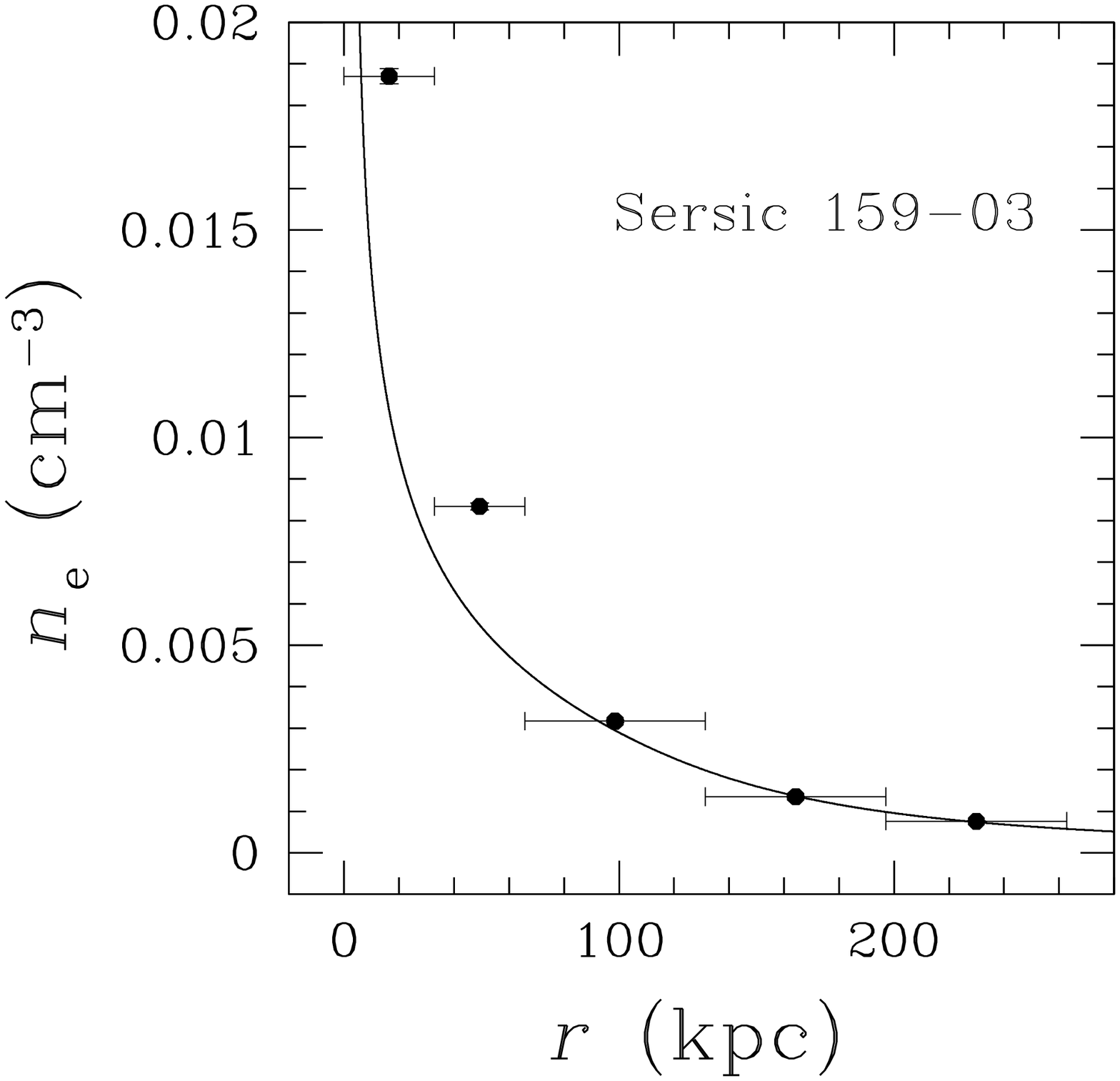}\hspace{0.05cm}\includegraphics[width=1.7in]{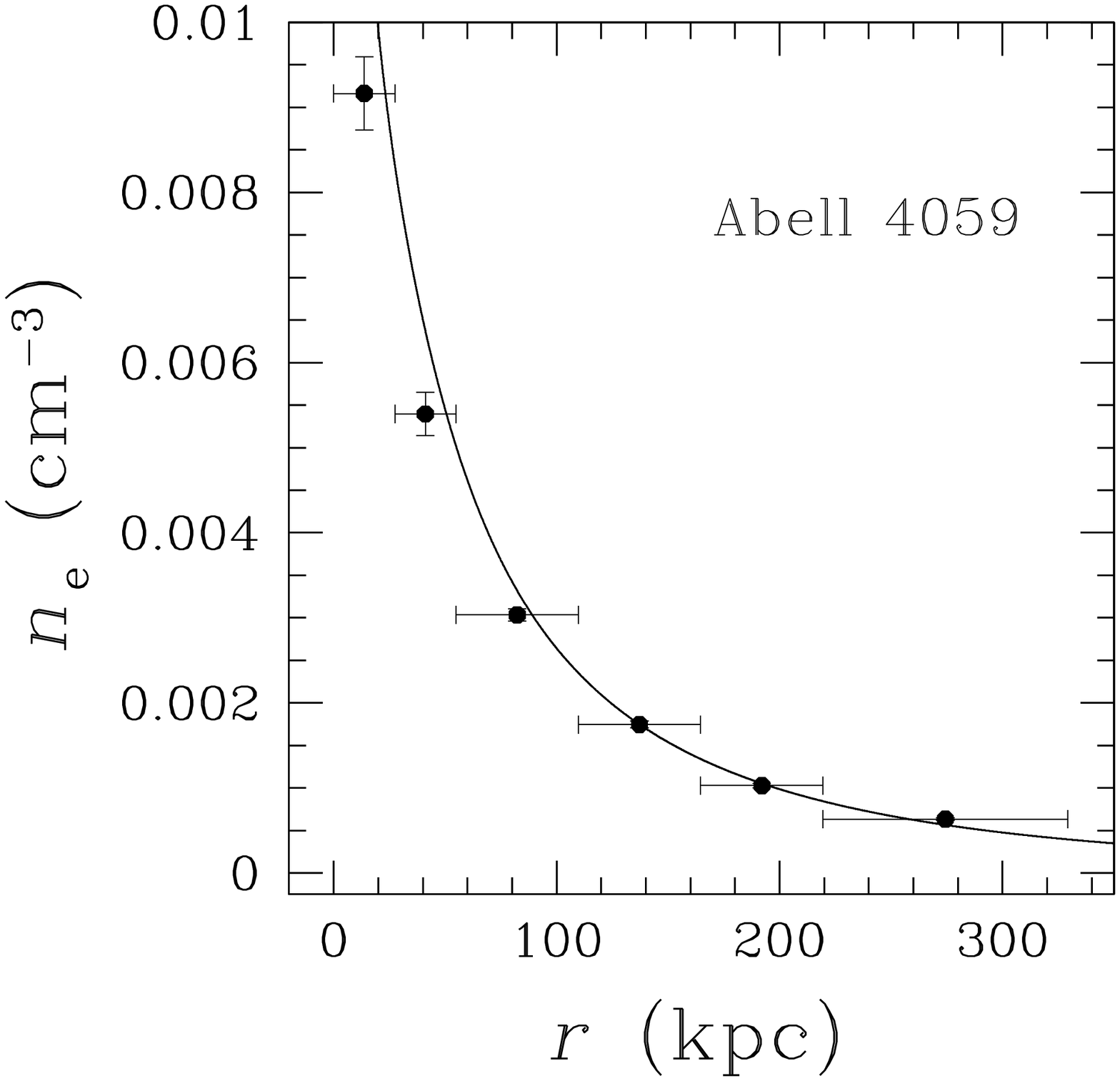}

\hspace{-1cm}
\includegraphics[width=1.7in]{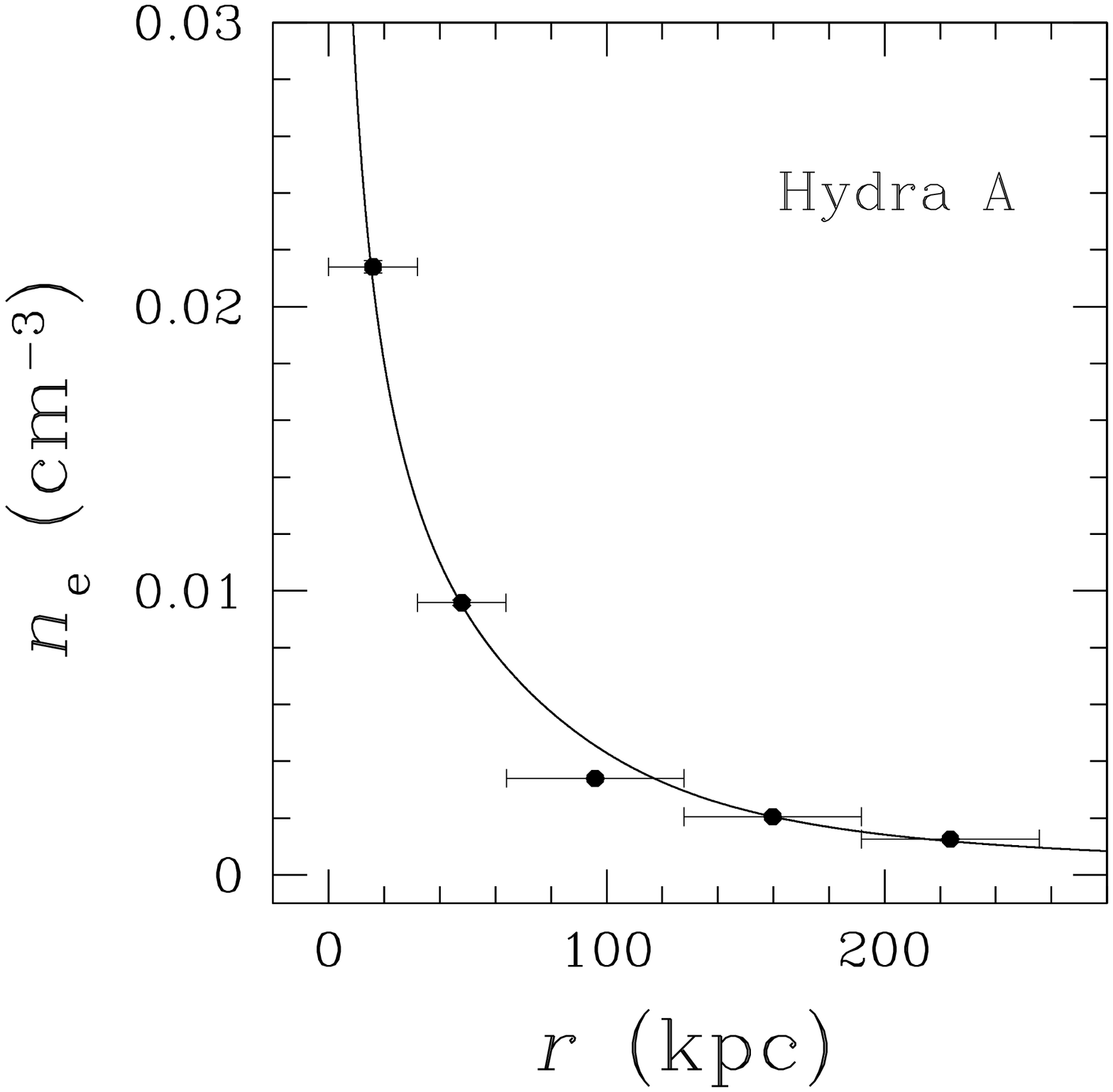}\hspace{0.05cm}\includegraphics[width=1.7in]{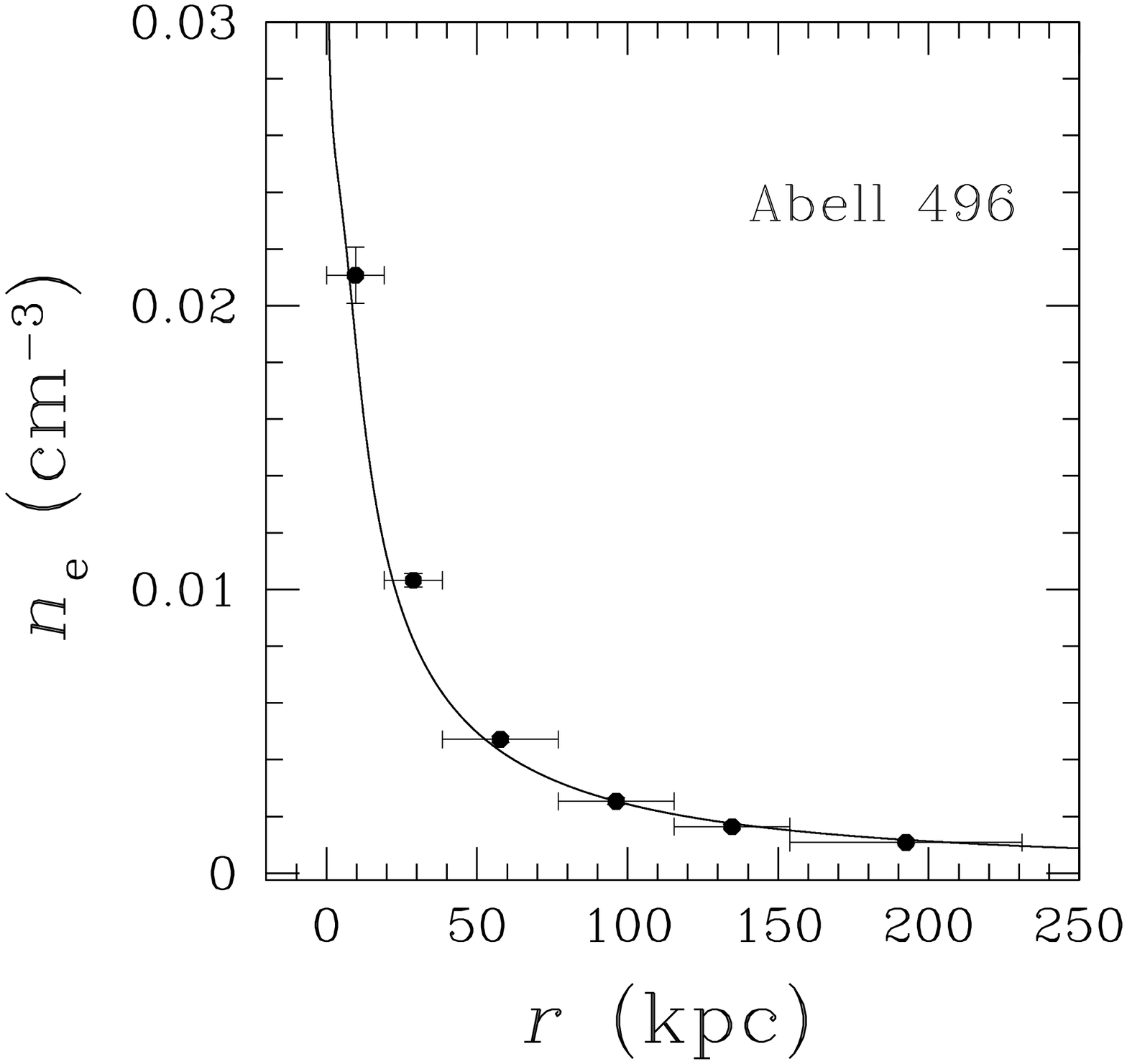}\hspace{0.05cm}\includegraphics[width=1.7in]{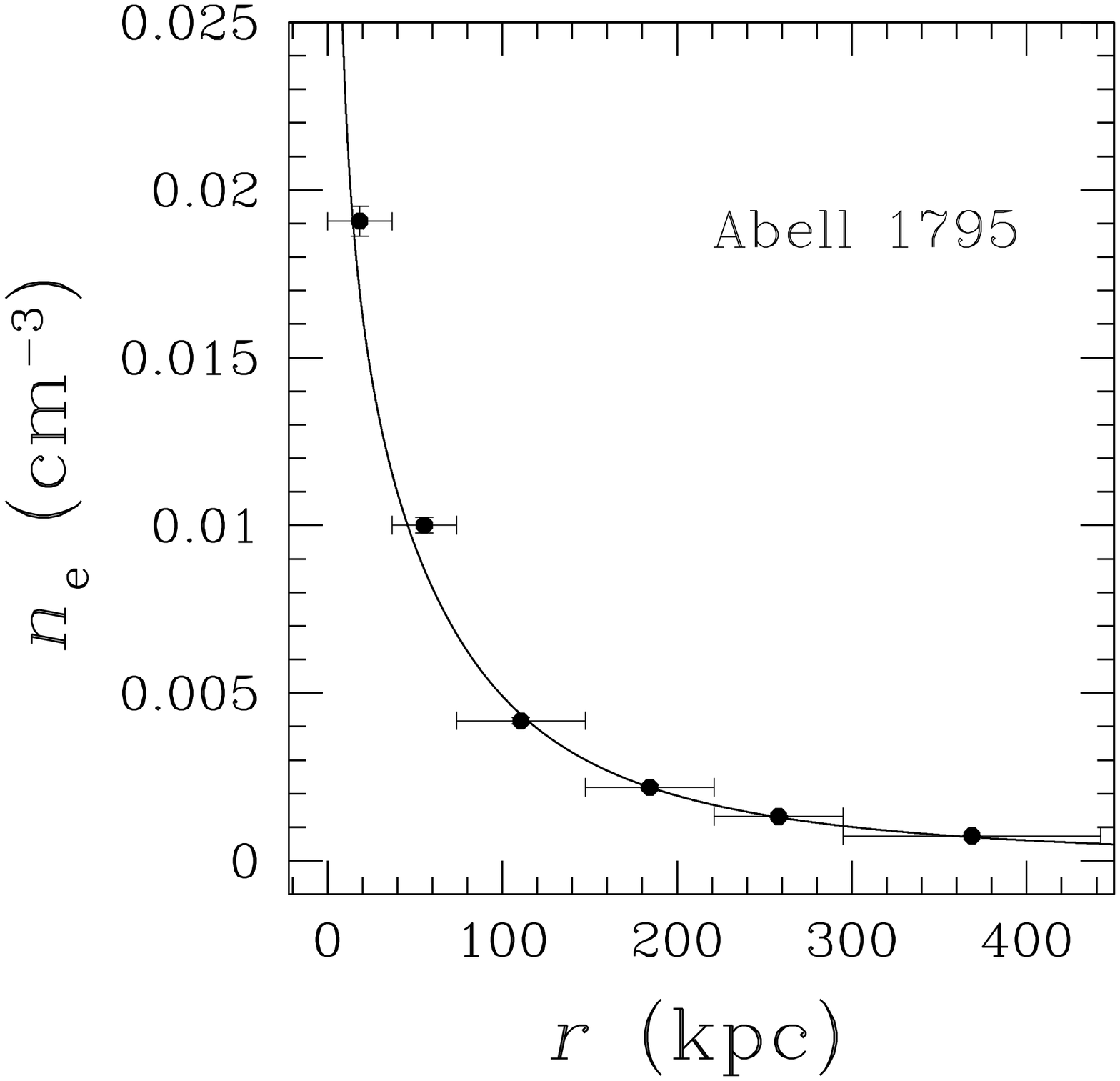}\hspace{0.05cm}\includegraphics[width=1.7in]{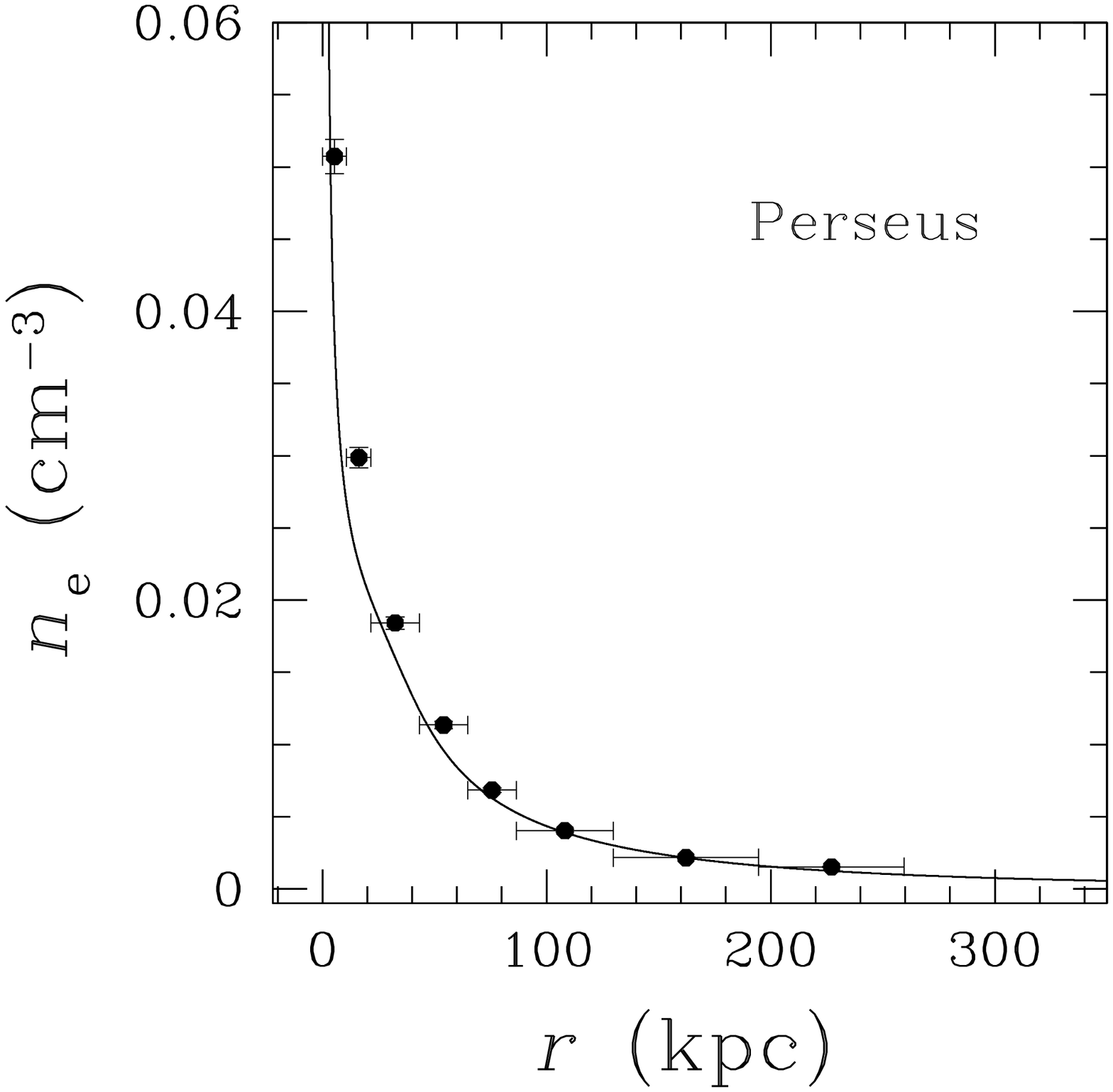}

\caption{\footnotesize The solid lines give the electron density as a
  function of radius in our model solutions for the eight clusters in
  our sample.  The data points are from the observations of
  Kaastra~et~al~(2004).
 \label{fig:f1} }
\end{figure}

\begin{figure}[h]
\hspace{-1cm}
\includegraphics[width=1.7in]{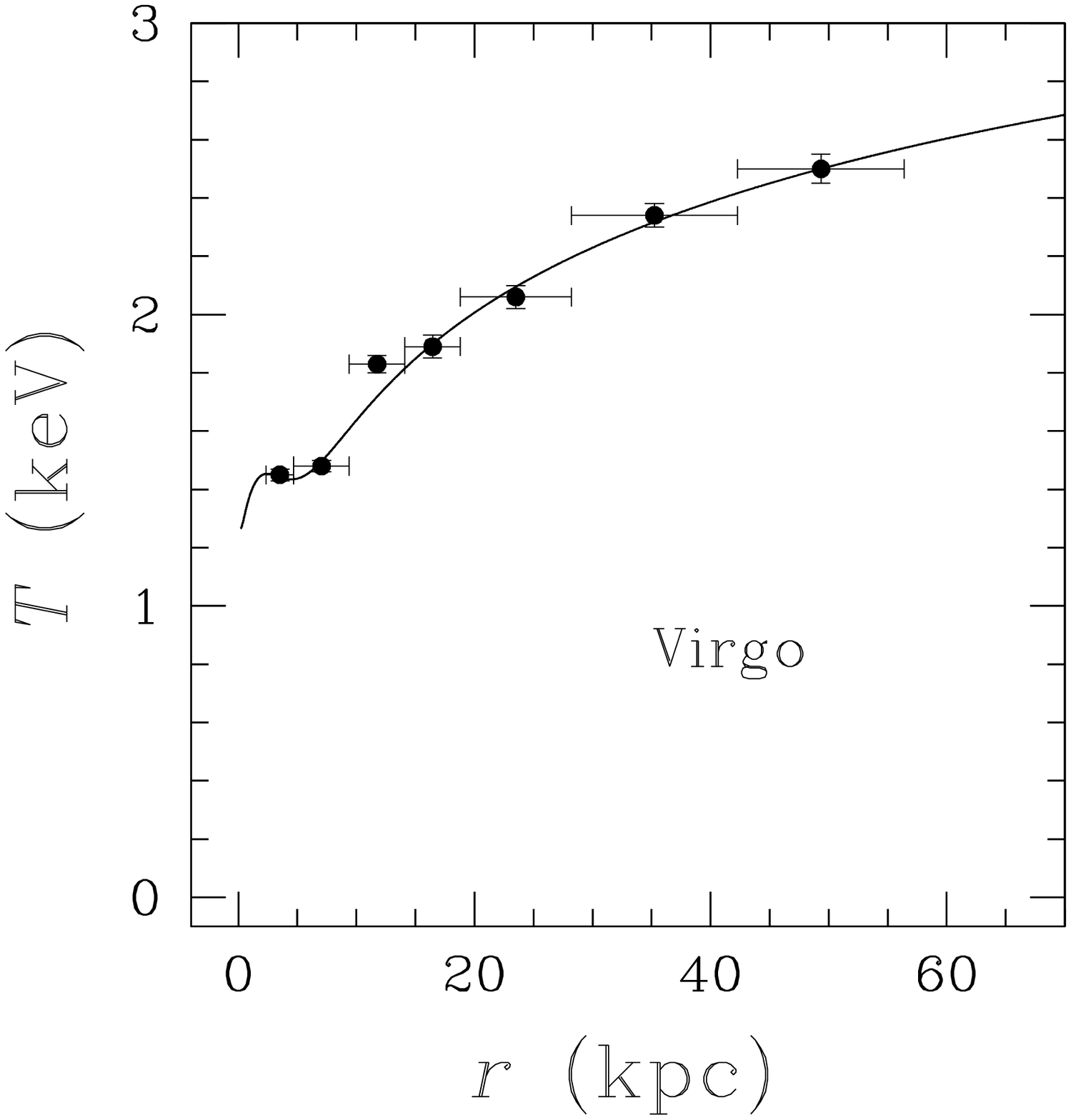}\hspace{0.05cm}\includegraphics[width=1.7in]{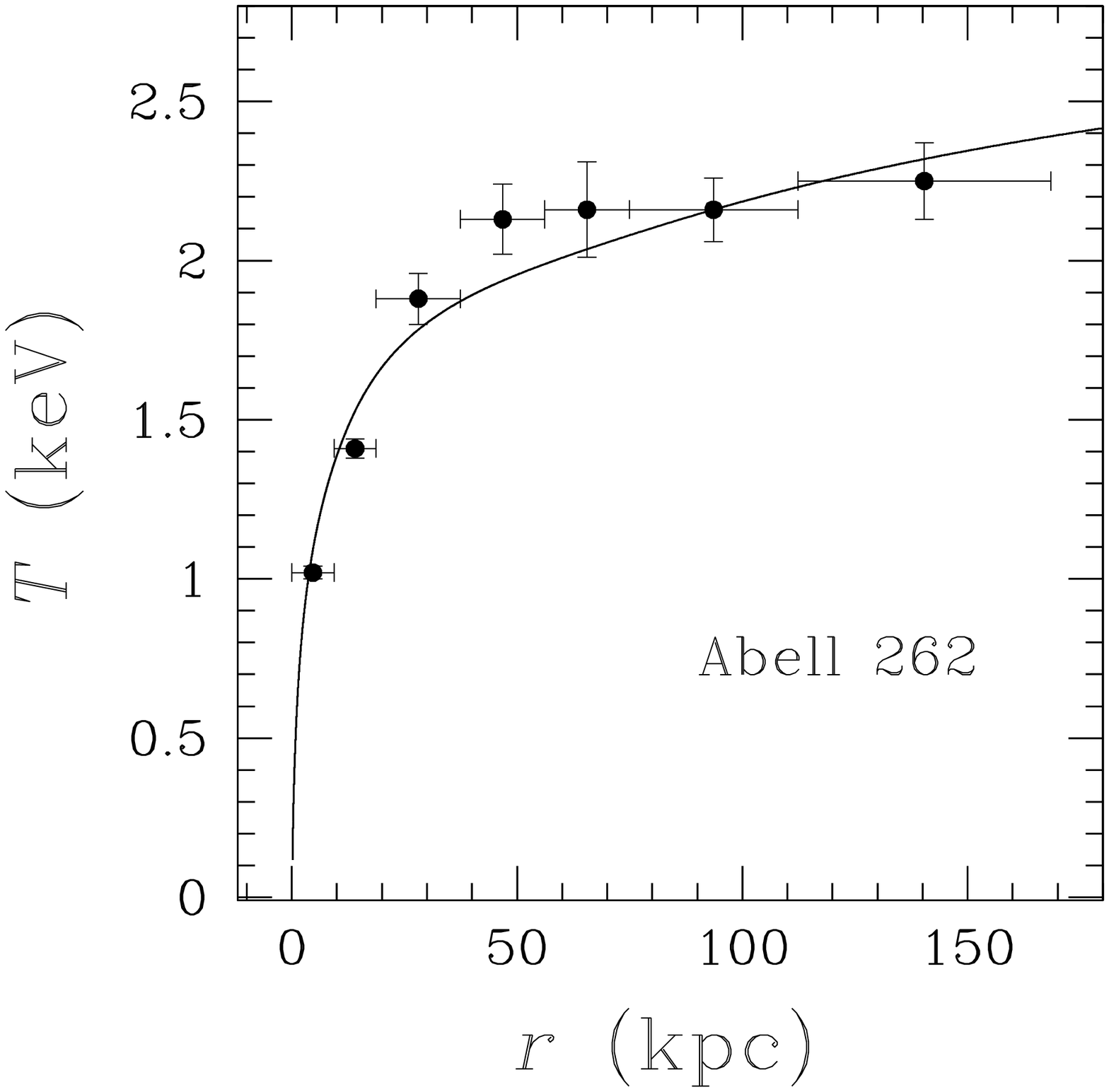}\hspace{0.05cm}\includegraphics[width=1.7in]{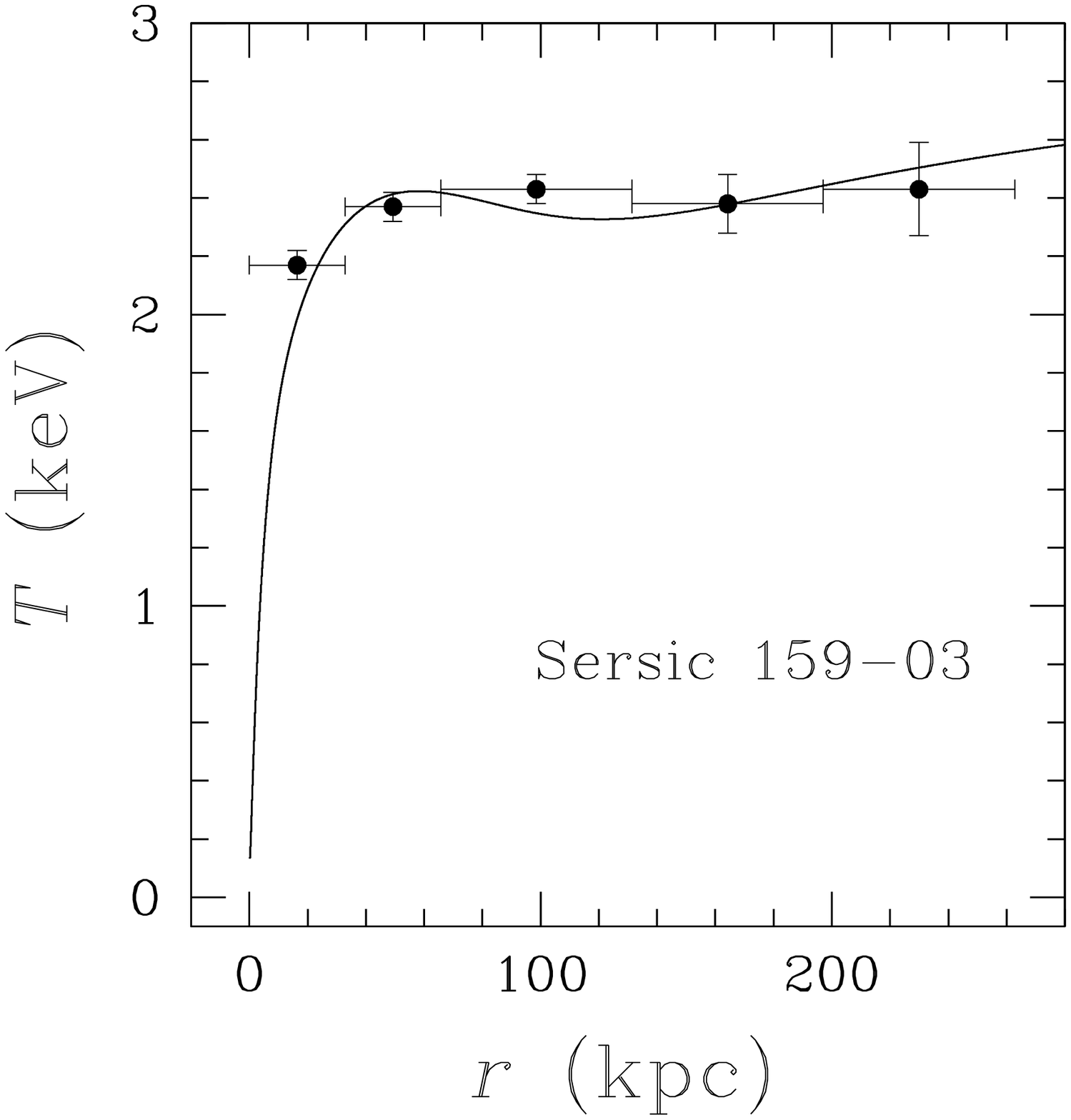}\hspace{0.05cm}\includegraphics[width=1.7in]{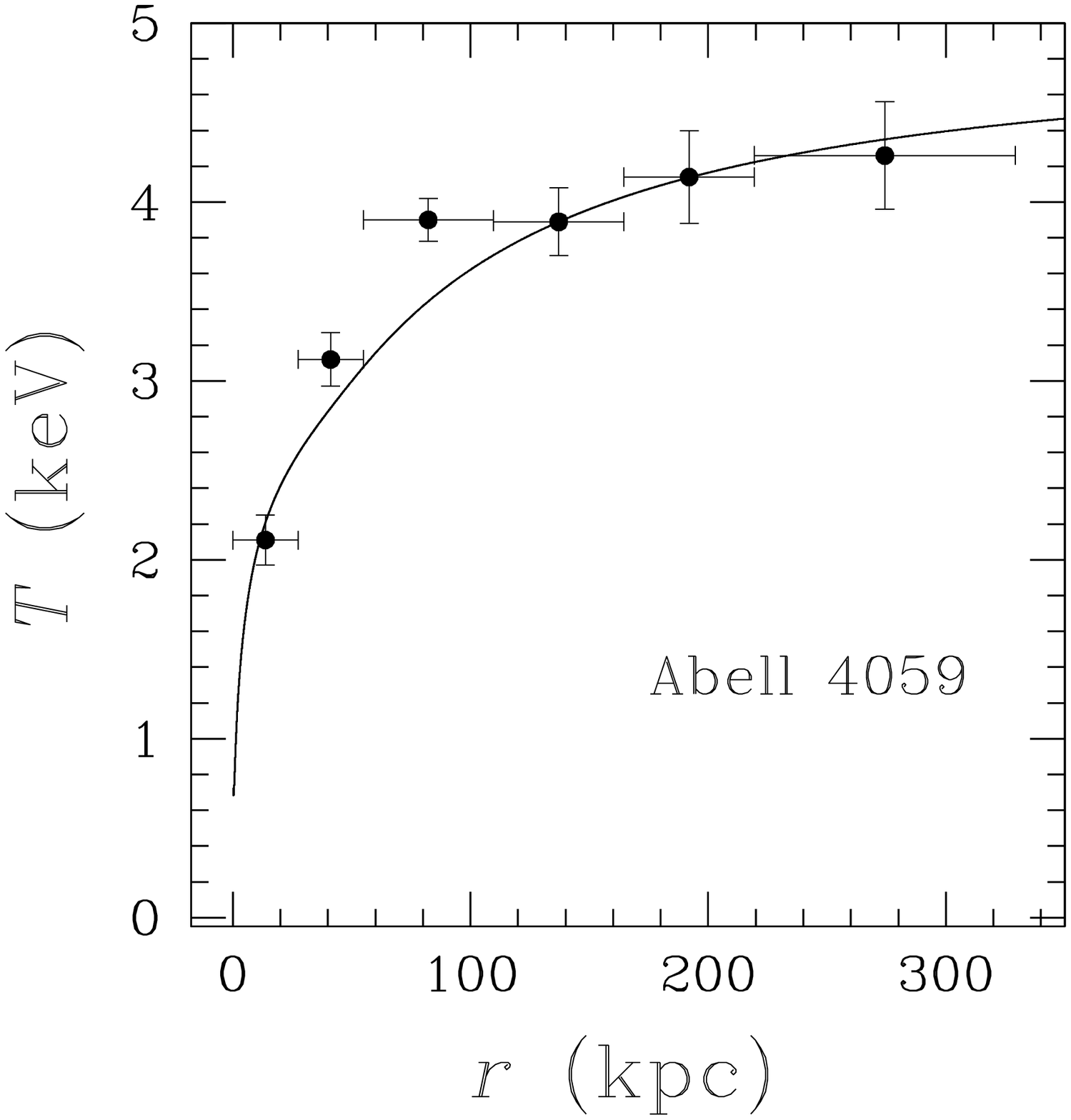}

\hspace{-1cm}
\includegraphics[width=1.7in]{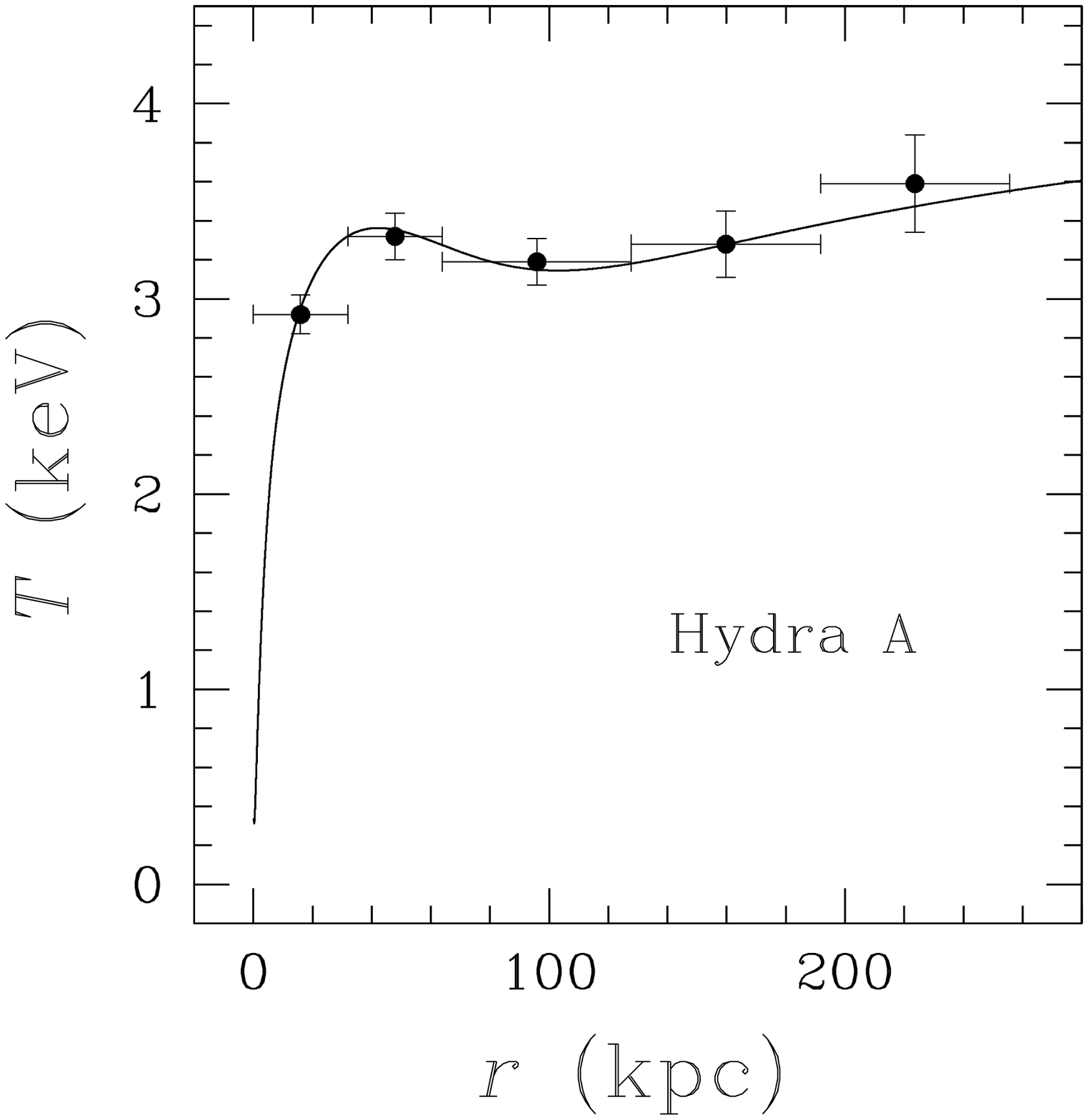}\hspace{0.05cm}\includegraphics[width=1.7in]{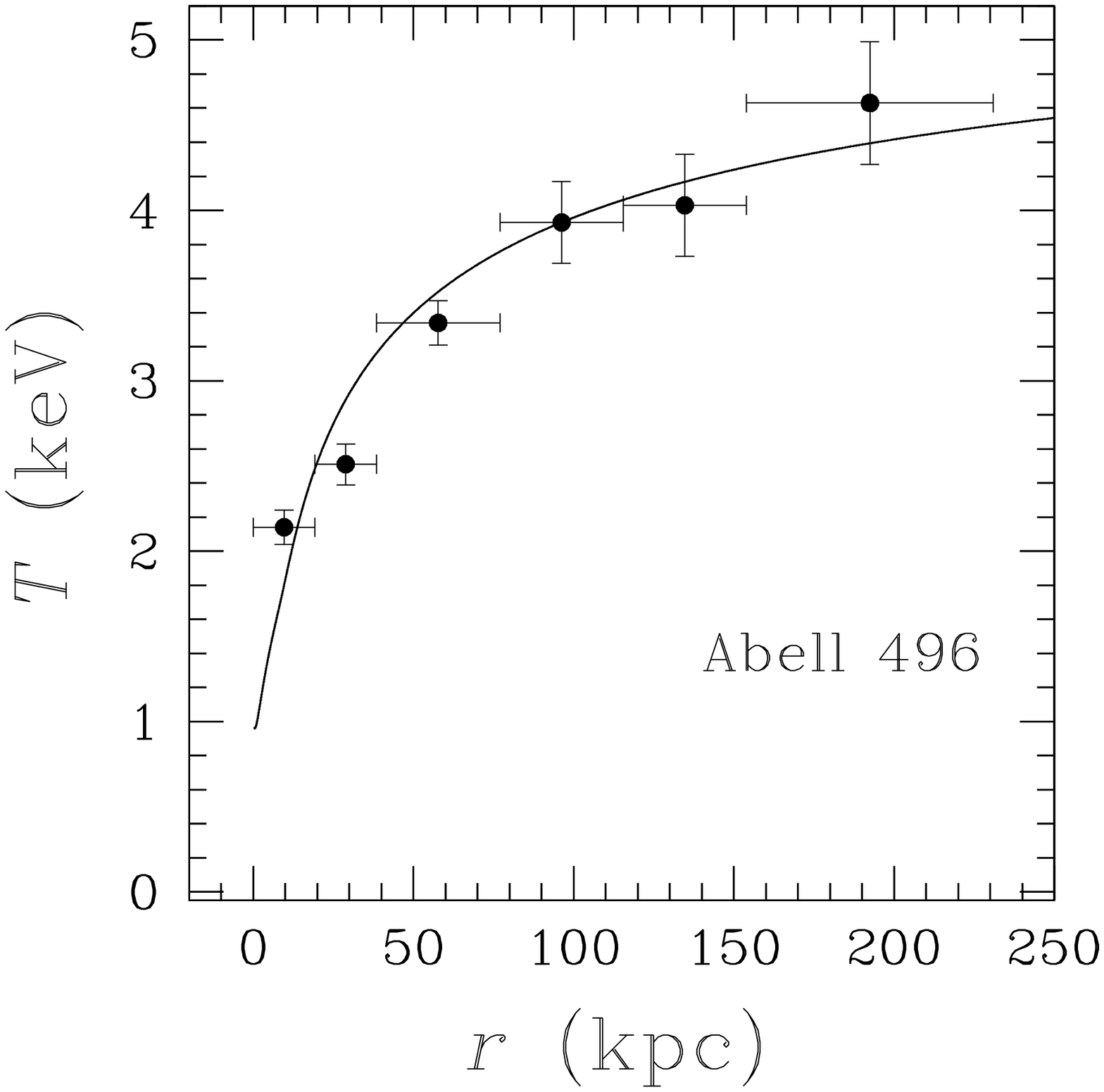}\hspace{0.05cm}\includegraphics[width=1.7in]{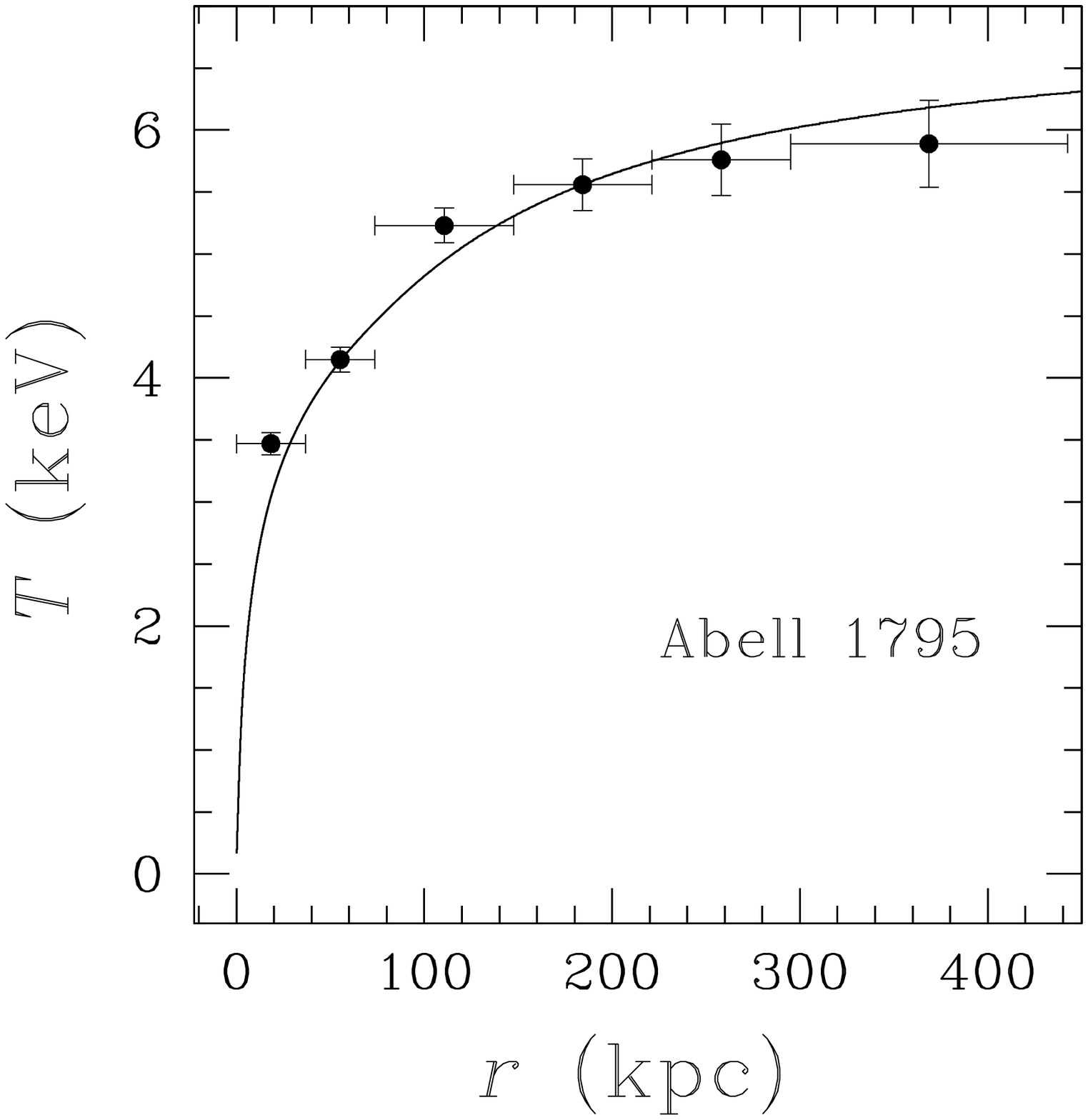}\hspace{0.05cm}\includegraphics[width=1.7in]{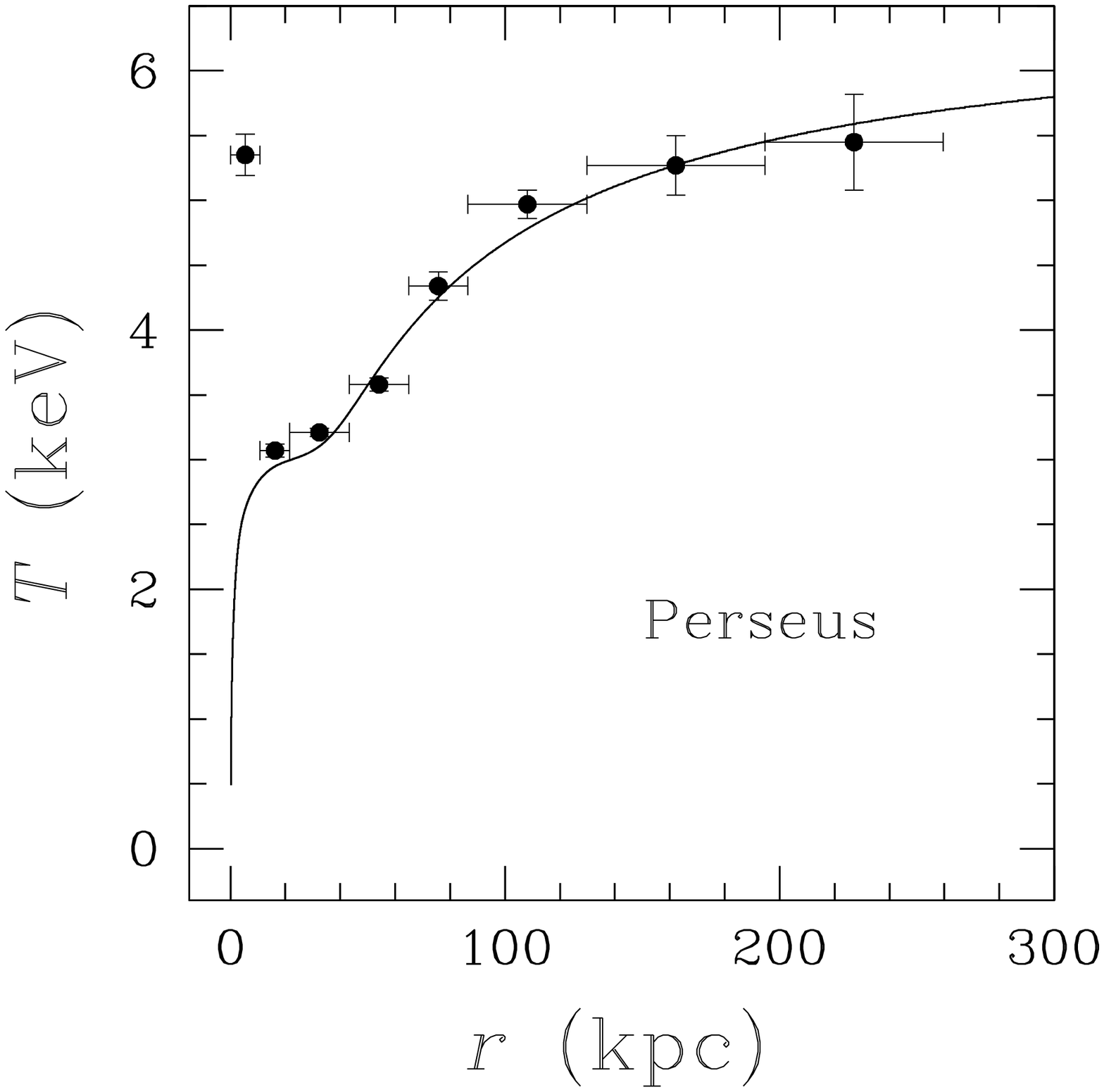}
\caption{\footnotesize The solid lines give the temperature as a function
of radius in our model solutions for the eight clusters in our sample.
 The data points are from the
observations of Kaastra~et~al~(2004).
 \label{fig:f2} }
\end{figure}

We adjust a single parameter, $r_{\rm source}$, to fit to the
observations.  The optimal values for $r_{\rm source}$ are given in
table~\ref{tab:t1}. The temperature and density profiles in the model
solutions are plotted in figures~\ref{fig:f1} and~\ref{fig:f2}.  
We note that the central
peak in the Perseus temperature data is due to the hard power-law
spectrum of the central active galaxy NGC~1275 (E. Churazov, private
communication).  The values of~$\dot{M}$ and $L_{\rm cr}$, as well as
fluid quantities at $r_1=0.2$~kpc and $r_{\rm outer}$ are given in
table~\ref{tab:t1}.  The radial profiles of the rms turbulent velocity
and cosmic-ray pressure are presented in appendix~\ref{ap:prof}.

Overall, the model profiles match the observations quite well, which
suggests that convection is an important process in cluster cores.
The model, however, substantially underestimates the observed density
in Sersic~159-03 at $r\lesssim 50$~kpc.  This discrepancy may be
explained by a recent study by Werner et al (2007). These authors
found that Sersic~159-03 has the largest soft x-ray excess of all
clusters observed by XMM-Newton and argued that the observed excess is
best explained by the presence of a substantial population of
non-thermal electrons that is concentrated in the cluster core. When
they modeled the observed emission as coming from a combination of
thermal plasma and nonthermal electrons, they found that the x-ray
emission between 0.3 and 10~keV from non-thermal electrons is a
substantial fraction of the emission from the thermal plasma at large
radii ($\sim 35-55$\% at $r\simeq 375$~kpc) but only a small fraction
of the emission at small radii ($\sim 1- 7$\% at $r\lesssim
50$~kpc). Thus, the non-thermal contribution to the emission measure
does not lead directly to a large change in the observationally
inferred electron density at $r\lesssim 50$~kpc. On the other hand,
Werner et al (2007) note that if there is a non-thermal proton
population with significantly more pressure than the non-thermal
electrons, the total cluster mass may be significantly underestimated.
Moreover, since the non-thermal pressure inferred by Werner et al
(2007) peaks strongly towards the cluster's center, the actual
gravitational acceleration in the central 50-100~kpc may be much
larger than in an NFW profile calculated neglecting non-thermal
pressure. This is an issue for all the clusters that we consider, but
especially for Sersic~159-03, since its especially large soft excess
indicates a large non-thermal pressure fraction. We note that although
the non-thermal emission is less peaked than the thermal emission in
the results of Werner et al (2007) (i.e., $p_{\rm non-thermal}/n_e^2$
decreases towards the center), the non-thermal pressure is more peaked
than the thermal pressure ($p_{\rm non-thermal}/p$ increases inwards).
If we were to re-calculate our model solutions using a larger
gravitational acceleration in the cluster core, the thermal plasma
density would peak more sharply near the cluster center than in
Figure~\ref{fig:f1}.  Thus, the deviation between the model and
observations of Sersic 159-03 may be due to a significant
underestimate of the gravitational acceleration in this cluster
resulting from its unusually large non-thermal pressure.

\begin{table}[h]
\caption{Physical quantities in the model solutions
\label{tab:t1} }
{\footnotesize 
\hspace{-0.8cm}
\begin{tabular}{lrrrrrrrrrr}
&&&&&&&&&&  \\ 
\hline \hline 
\vspace{-0.2cm} 
\\
Cluster & 
\multicolumn{1}{c}{$r_{\rm source}$} 
& \multicolumn{1}{c}{$ n_{\rm e}(r_1)$}
 &  \multicolumn{1}{c}{$k_{\rm B} T(r_1)$}
 &  \multicolumn{1}{c}{$\displaystyle \frac{p_{\rm cr}(r_1)}{p(r_1)}$}
 &  \multicolumn{1}{c}{$\displaystyle \frac{|\langle v_r(r_1)\rangle|}
{c_s(r_1)}$}
  &\multicolumn{1}{c}{$\dot{M}_{\rm Bondi}$} 
& \multicolumn{1}{c}{$L_{\rm cr}$} 
 & \multicolumn{1}{c}{$r_{\rm outer}$} 
 & \multicolumn{1}{c}{$k_BT_{\rm outer}$} 
 & \multicolumn{1}{c}{$n_{\rm e, outer}$}  \\
& \multicolumn{1}{c}{(kpc)} 
& \multicolumn{1}{c}{($\mbox{cm}^{-3}$)} 
& \multicolumn{1}{c}{(keV)} 
&
&
& \multicolumn{1}{c}{($M_{\sun} \mbox{yr}^{-1} $)} 
& \multicolumn{1}{c}{(erg/s)} 
& \multicolumn{1}{c}{(kpc)} 
 & \multicolumn{1}{c}{(keV)}
 & \multicolumn{1}{c}{($\mbox{cm}^{-3}$)}\\
\\
\hline 
\hline 
\vspace{-0.2cm} 
\\
Virgo  & 1.70 & 0.115 & 1.27 & 0.268 & 0.00207 & 0.00203 & $5.77\times 10^{41}$
 & 49.3 & 2.50 & $2.85\times 10^{-3}$ \\
Abell 262  & 34.0 & 0.231 & 0.118 & 0.153 & 0.0388 & 0.0232 & $6.59\times 10^{42}$ & 93.6 & 2.16
 & $1.22\times 10^{-3}$ \\
Sersic 159-03 & 46.0 & 0.397 & 0.133 & 1.24 &  0.365 & 0.399 & $1.13\times 10^{44}$ & 165
 & 2.38 & $1.35\times 10^{-3}$   \\
Abell 4059  & 24.0 & 0.168 & 0.727 & 0.156 & 0.0562 & 0.0607 & $1.72\times 10^{43}$ & 137 & 3.89 
& $1.75\times 10^{-3}$    \\
Hydra A  & 35.0 & 0.742 & 0.340 & 0.740 & 0.257 & 0.839 & $2.38\times 10^{44}$ & 160
 & 3.28 & $2.05\times 10^{-3}$ \\
Abell 496  & 5.00 & 0.0437 & 0.965 & 0.701 & 0.0201 & 0.00650 & $1.85\times 10^{42}$ & 96.2 & 3.93 & $2.53\times 10^{-3}$ \\
Abell 1795  & 40.0 & 0.671 & 0.165 & 0.700 & 0.177 & 0.363 & $1.03\times 10^{44}$
 & 184 & 5.56 & $2.19\times 10^{-3}$  \\
Perseus   & 16.0 & 2.47 & 0.487 & 0.121 & 0.0264 & 0.343 & $9.75\times 10^{43}$ & 162 & 5.27
 & $2.16\times 10^{-3}$   \\
\vspace{-0.2cm} 
\\
\hline
\end{tabular}
\vspace{0.5cm} 

\noindent $r_{\rm source}$ is the size of the cosmic-ray acceleration
region that leads to the best fit between the mixing-length model and
the observations of Kaastra et~al~(2004).  $n_{\rm e}(r_1)$, $T
(r_1)$, $p(r_1)$, and $p_{\rm cr}(r_1)$ are the electron density,
temperature, thermal pressure, and cosmic-ray pressure at the inner
radius $r_1 = 0.2$~kpc. $v_r$ and $c_{\rm s}$ are the radial velocity
and adiabatic sound speed.  $\dot{M}_{\rm Bondi}$ is the Bondi
accretion rate based on the plasma density and plasma temperature
at~$r=r_1$, and $L_{\rm cr} = 0.005 \dot{M}_{\rm Bondi}c^2$ is the
cosmic-ray luminosity of the central radio source.  $T_{\rm outer}$
and $n_{\rm e, outer}$ are the observed temperature and electron
density at the radius~$r_{\rm outer}$ of the outer boundary used in
our shooting method.}
\end{table}
\normalsize

\begin{figure}[h]
\includegraphics[width=3in]{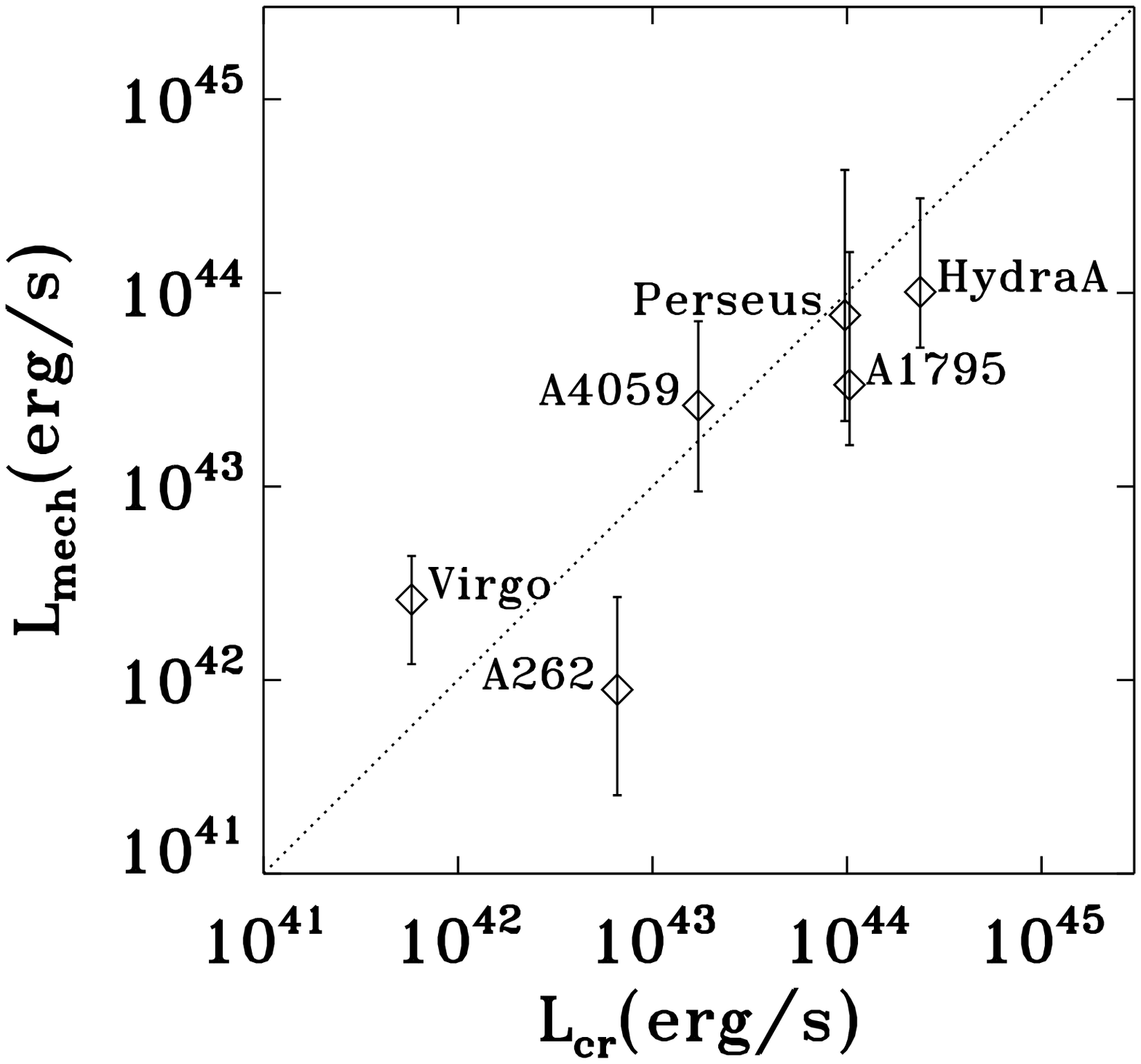}
\caption{\footnotesize  Comparison between the cosmic-ray
luminosities in our model calculations $L_{\rm cr}$ against observationally
inferred values of the mechanical luminosities of the
central AGN in six of the clusters in our sample (B\^irzan et al~2004). The dotted line represents equality between these two quantities.
\label{fig:f3}}
\end{figure}

One of the quantities that is calculated as part of our
solutions is the cosmic-ray
luminosity~$L_{\rm cr}$ of the central AGN.  To further test
the plausibility of our model, we compare the theoretically predicted
values of~$L_{\rm cr}$ from table~\ref{tab:t1} and the observationally
inferred mechanical luminosity~$L_{\rm mech}$ of the central AGN in
the six clusters in our sample for which we were able to find
published values. The mechanical luminosities are taken from B\^irzan
et al~(2004), and are calculated from observations of X-ray cavities,
by assuming that an energy $pV$ (where $p$ is the surrounding pressure
and $V$ is the cavity volume) per cavity is released during a time
equal to the buoyancy time scale. We plot $L_{\rm mech}$
versus~$L_{\rm cr}$ in figure~\ref{fig:f3}.  The error bars in this
figure take into account projection effects on the estimate of the
cavity volume as well as uncertainties in the ages of the
cavities. These two luminosities are strongly correlated over a range
of~$\sim 100-1000$ in luminosity.

\section{The energy budget of the intracluster medium}
\label{sec:eb}

In our model, radiative cooling is balanced by a
combination of thermal conduction, convective heating, 
and radial inflow due to the accretion onto
the central AGN. To distinguish between these last
two mechanisms, we separate the average radial velocity
into two components,
\begin{equation}
\langle v_r \rangle = v_{\rm inflow} + v_{r, \rm turb},
\label{eq:defVr} 
\end{equation} 
where
\begin{equation}
v_{\rm inflow} = - \frac{\dot{M}}{4\pi r^2 \langle \rho\rangle}
\end{equation} 
is the inflow rate that
arises in a laminar radial flow with constant mass accretion rate~$\dot{M}$.
The term $v_{r, \rm turb}$ is an additional average radial velocity that
is induced by the convection. [Its value is given by $- \langle
\delta \rho \delta v_r \rangle/\langle \rho \rangle$, as
in equation~(\ref{eq:defQ}).] With this definition in hand,
we write the average of equation~(\ref{eq:pe}), divided by~$(\gamma-1)$,
as 
\begin{equation}
0 = \langle   H_{\rm inflow} + H_{\rm conv} + H_{\rm visc} + H_{\rm tc} - R \rangle.
\label{eq:avpe} 
\end{equation} 
Here,
\begin{equation}
\left\langle H_{\rm inflow}\right\rangle = 
- \frac{1}{(\gamma-1)r^2} \frac{d}{dr}\left(
r^2 v_{\rm inflow} \langle p \rangle  \right)
-  \frac{\langle p \rangle}{r^2} \frac{d}{dr} \left(r^2 v_{\rm inflow}\right)
\label{eq:defHinflow} 
\end{equation} 
is the source term associated with~$v_{\rm inflow}$. The term
\begin{equation}
\left \langle H_{\rm conv}\right\rangle = 
\left\langle - \frac{\nabla\cdot(\bm{v} p)}{(\gamma-1)}   -  p \nabla \cdot \bm{v} - H_{\rm inflow}
\right\rangle
\label{eq:defHconv}
\end{equation}
is the convective heating rate of the thermal plasma,
excluding viscous dissipation. It 
includes the turbulent diffusion of heat as well as 
the turbulent $pdV$ work done on the thermal plasma by cosmic rays. 
The average of the viscous dissipation term is set equal to
\begin{equation}
\langle H_{\rm visc}\rangle  = \frac{0.42  \rho u_{\rm rms}^3}{l},
\label{eq:hdiss2} 
\end{equation} 
where $l= 0.4 r$ is the mixing length, $u_{\rm rms}$ 
is the rms turbulent velocity defined in
equation~(\ref{eq:urms2}), and the constant 0.42 is
taken from direct numerical simulations of compressible
magnetohydrodynamic turbulence (Haugen, Brandenburg, \&
Dobler 2004).\footnote{The constant 0.42 is obtained by
  taking the mixing length~$l$ to correspond to
  $\pi/k_{\rm p}$ in the simulations of Haugen
  et~al~(2004), where $k_{\rm p}$ is the wave number at
  which $kE(k)$ peaks, and $E(k)$ is the power spectrum
  of the turbulent velocity.}
The average of $H_{\rm tc}$ is given by
equation~(\ref{eq:htc}), and the average of $R$ is given by
equation~(\ref{eq:R1}).

\begin{figure}[t]
\hspace{-1cm}
\includegraphics[width=1.7in]{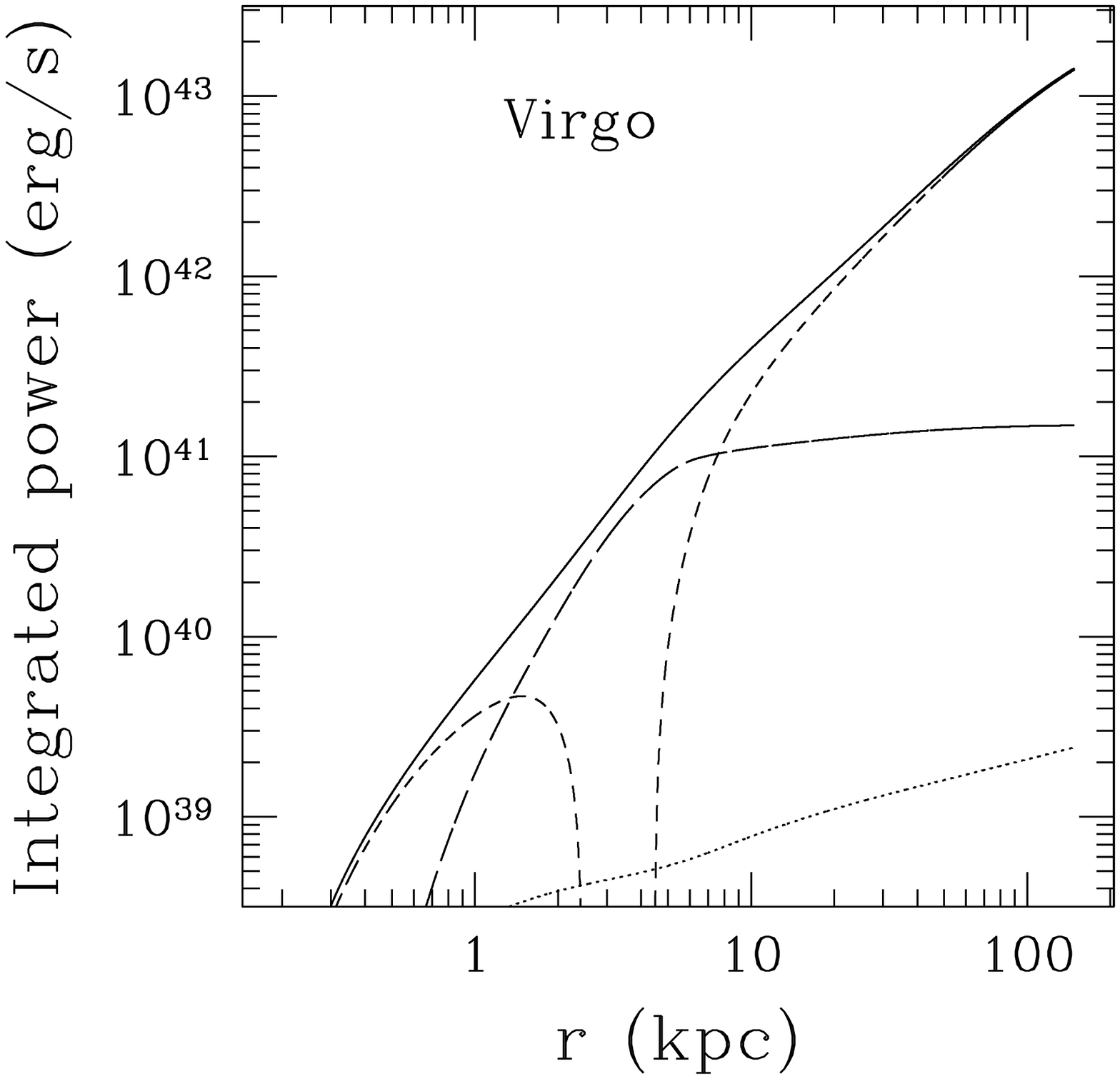}\hspace{0.05cm}\includegraphics[width=1.7in]{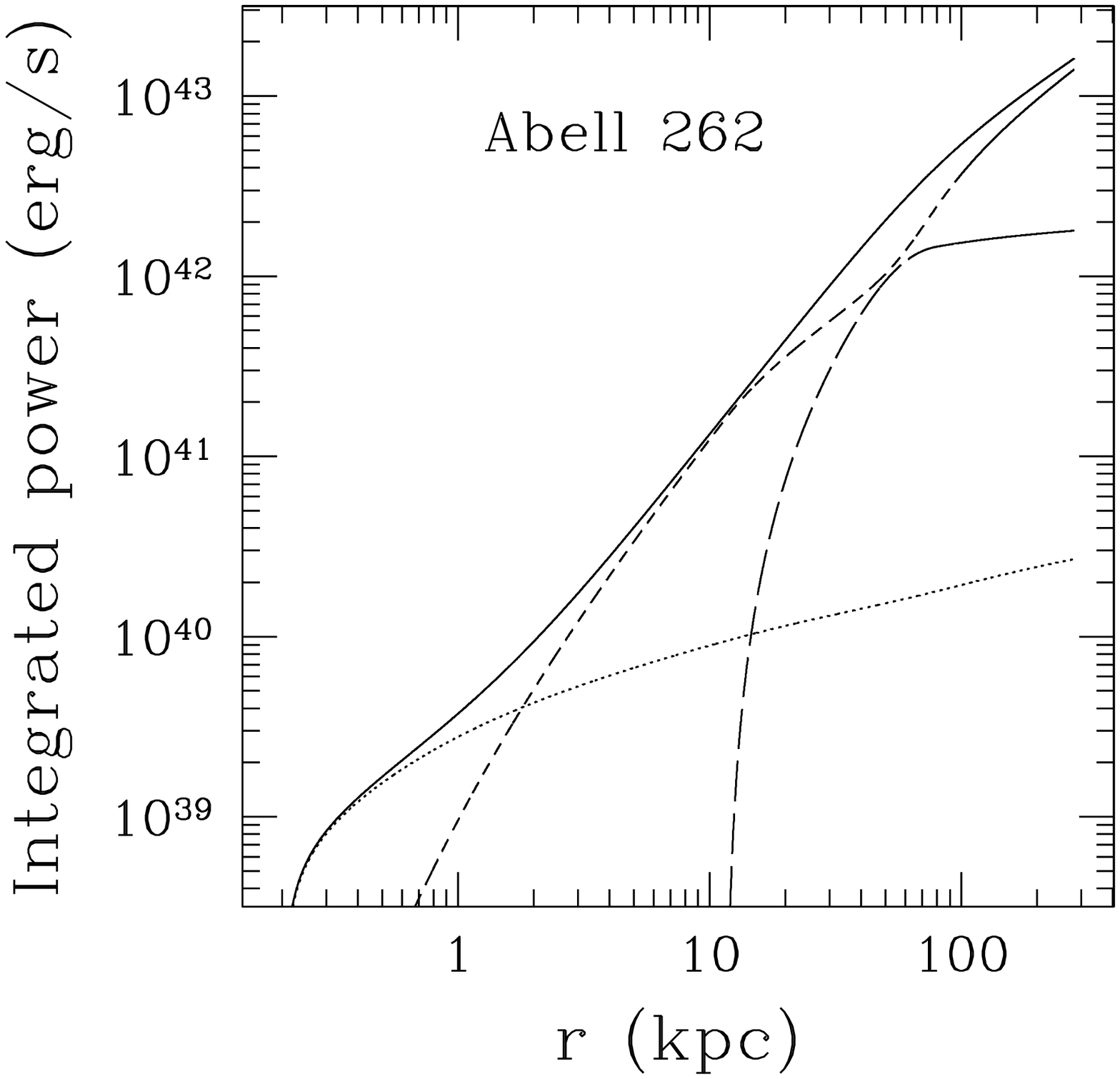}\hspace{0.05cm}\includegraphics[width=1.7in]{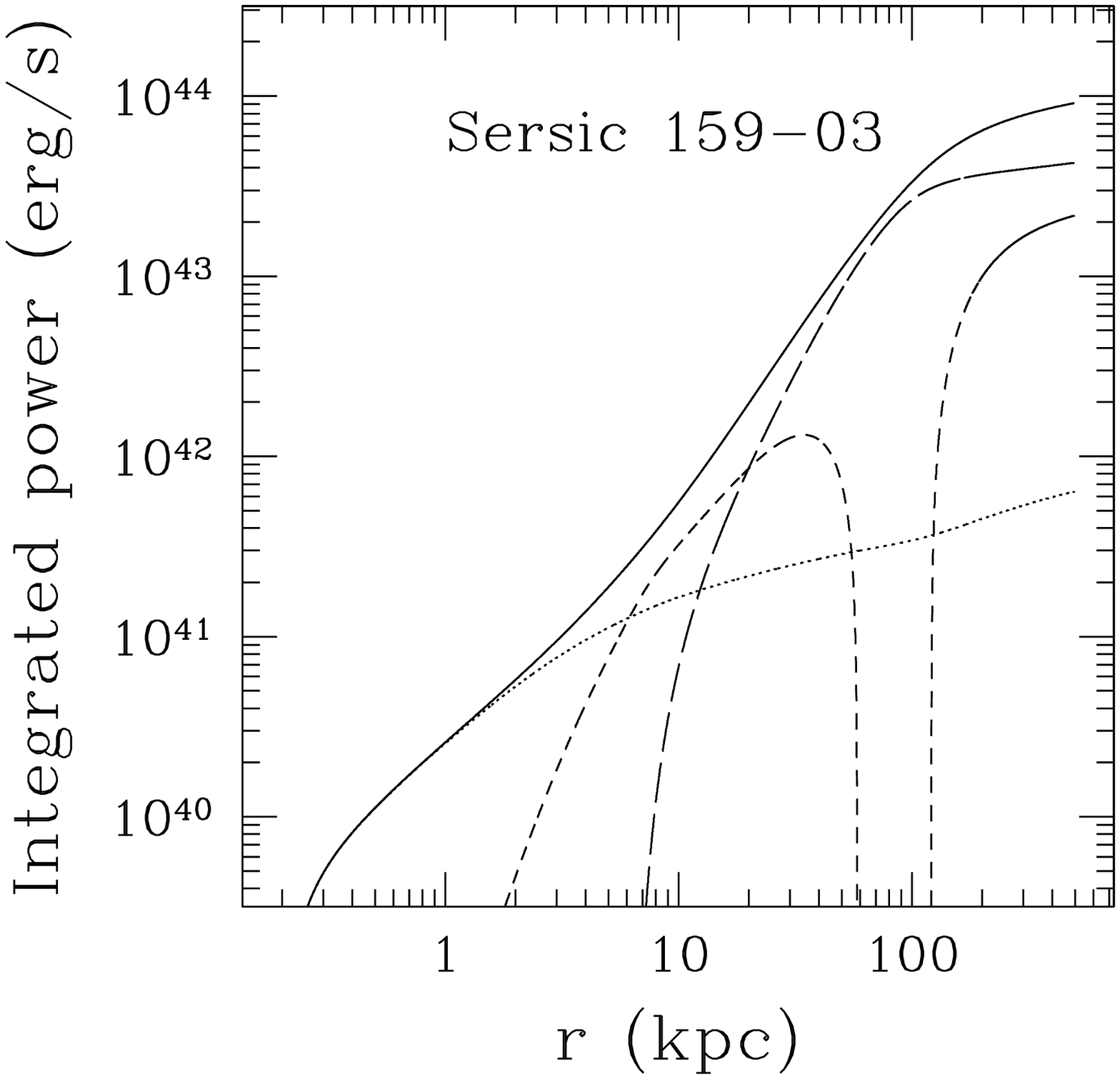}\hspace{0.05cm}\includegraphics[width=1.7in]{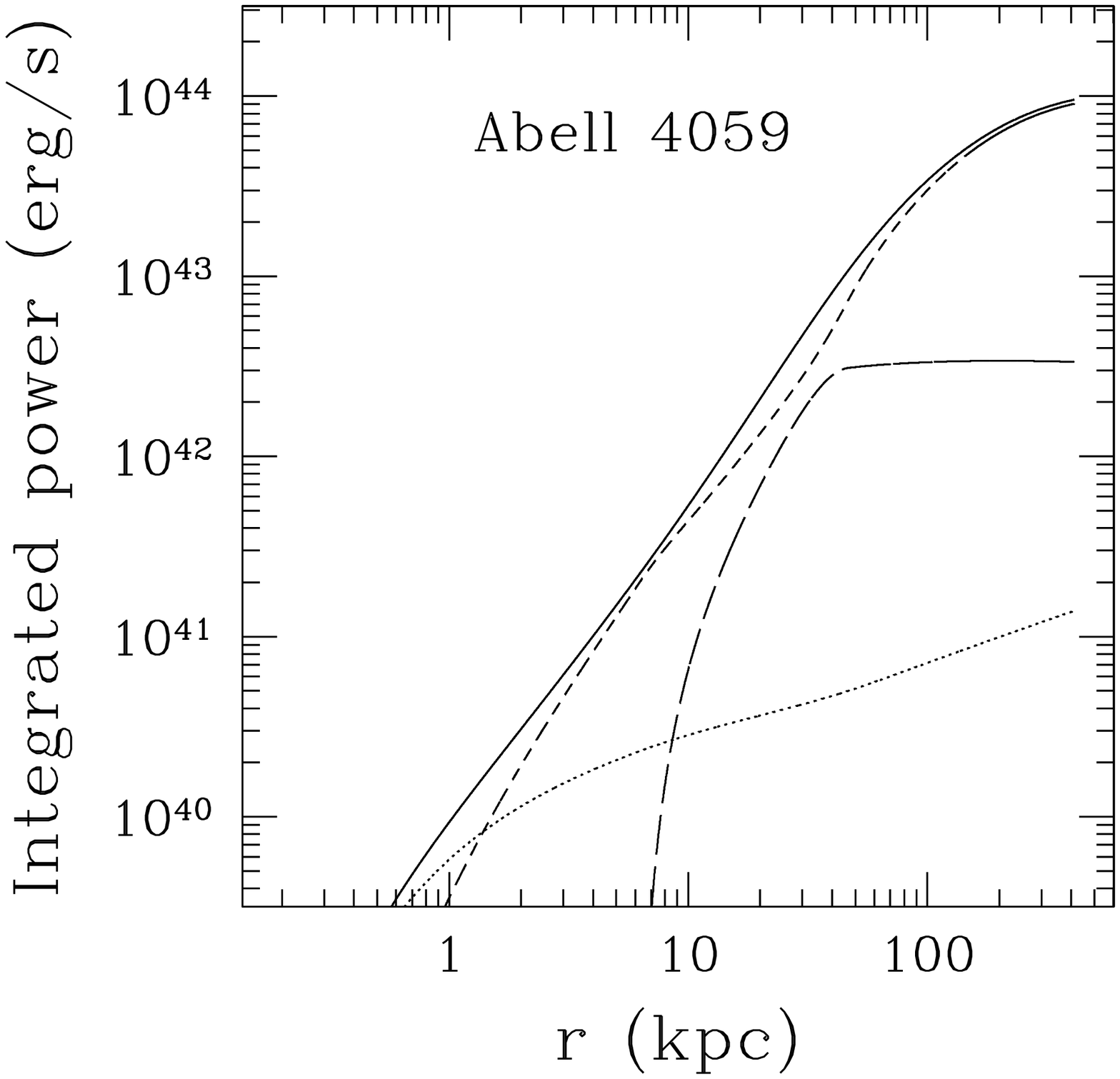}

\hspace{-1cm}
\includegraphics[width=1.7in]{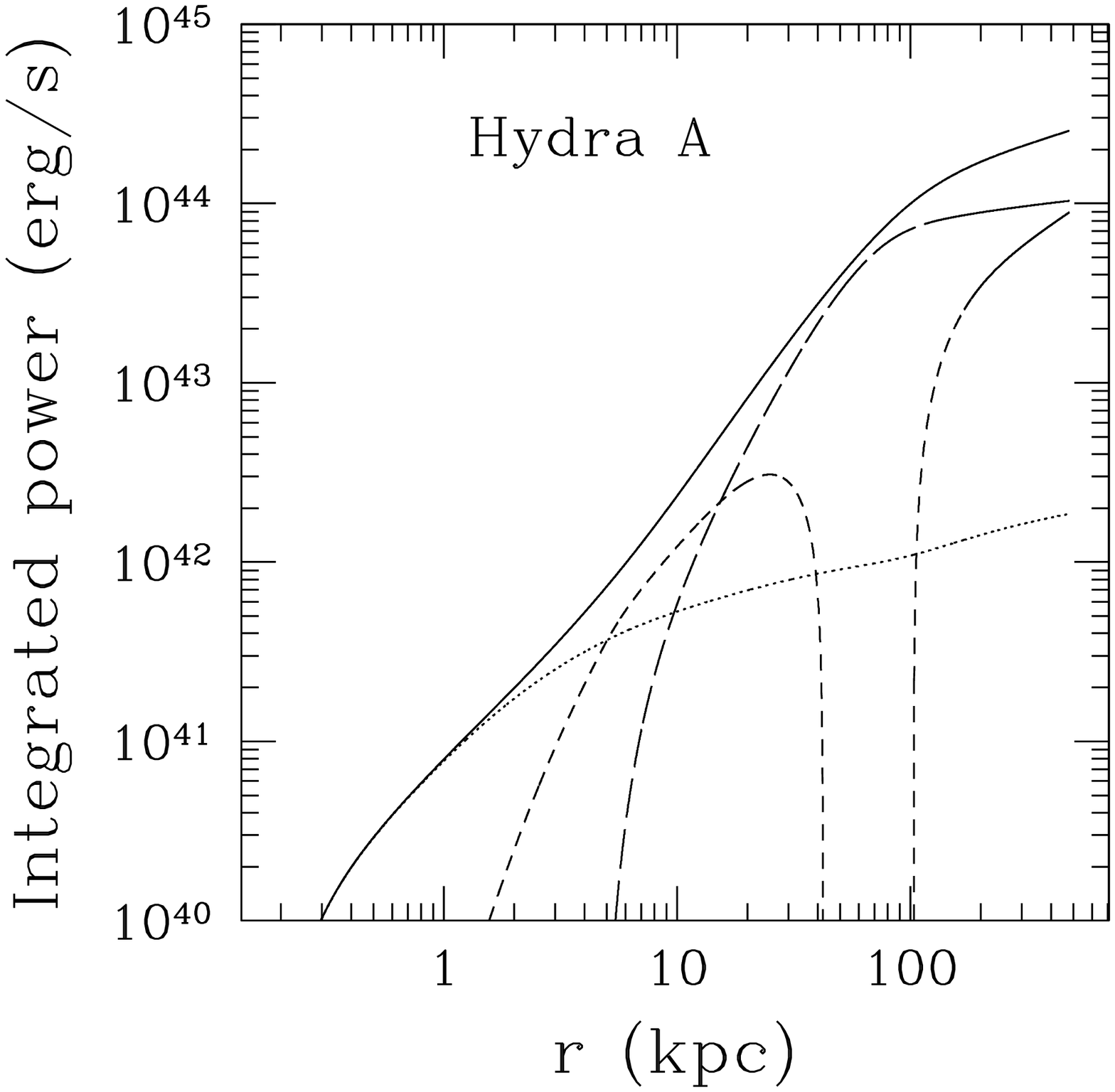}\hspace{0.05cm}\includegraphics[width=1.7in]{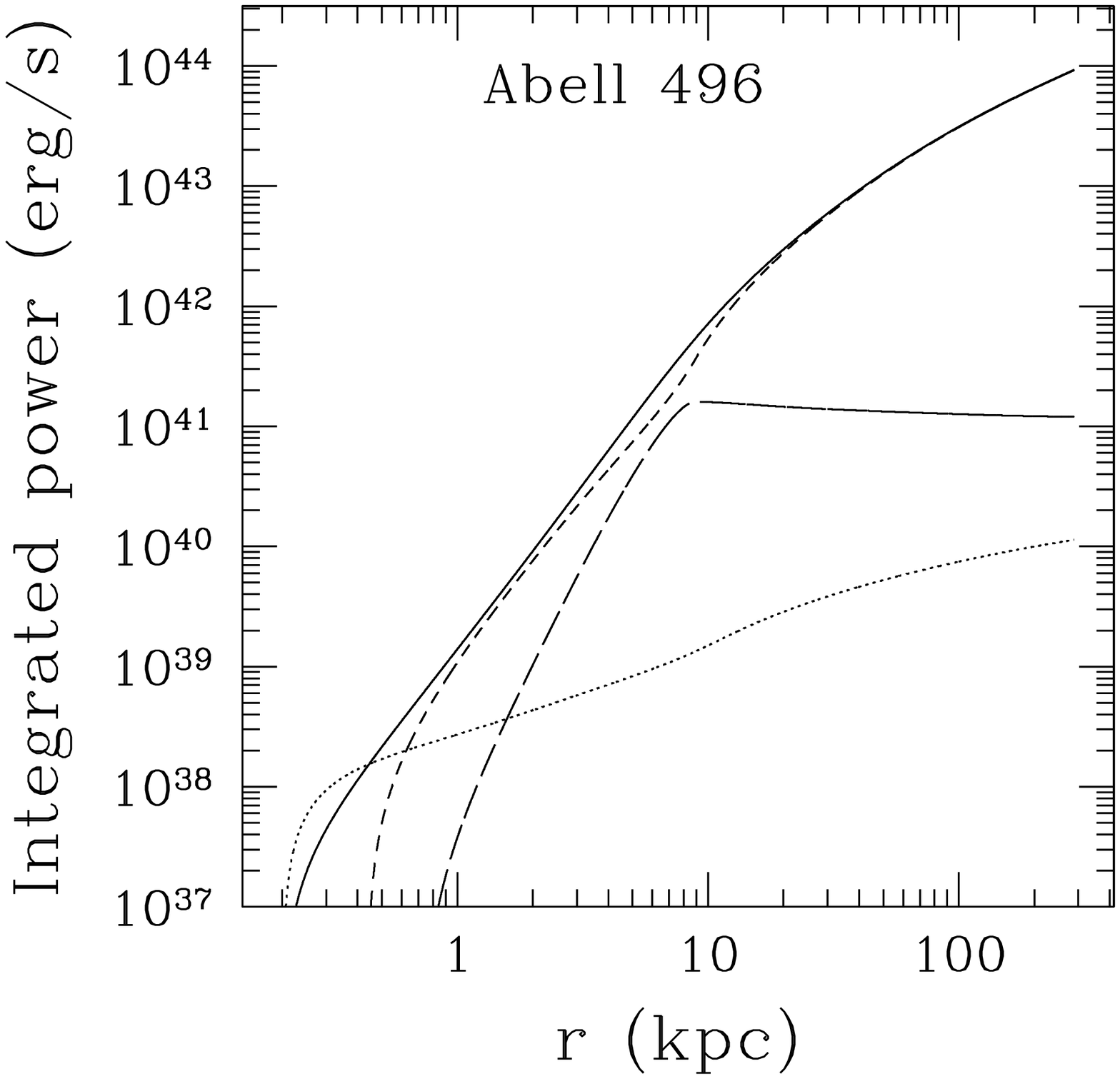}\hspace{0.05cm}\includegraphics[width=1.7in]{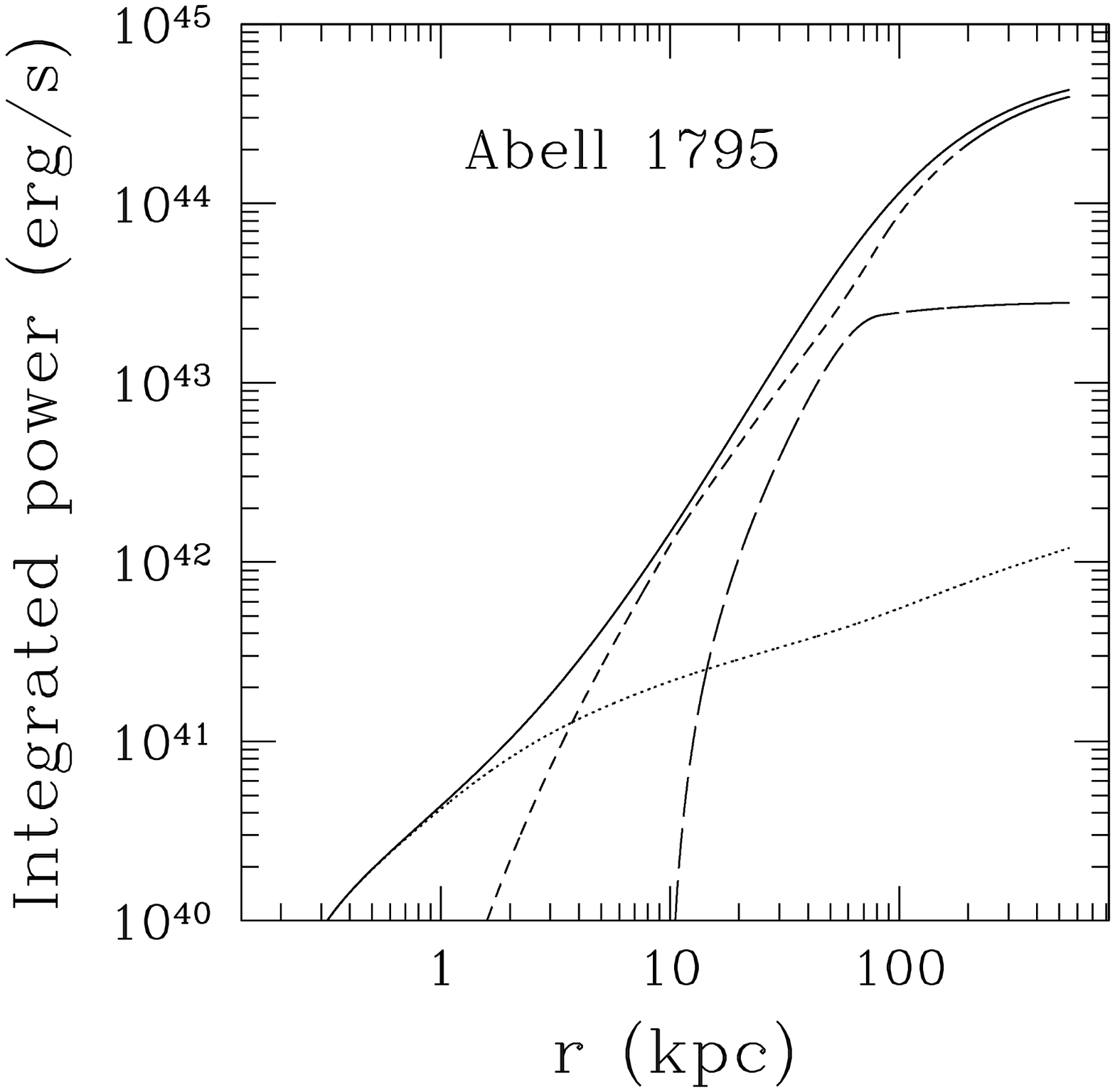}\hspace{0.05cm}\includegraphics[width=1.7in]{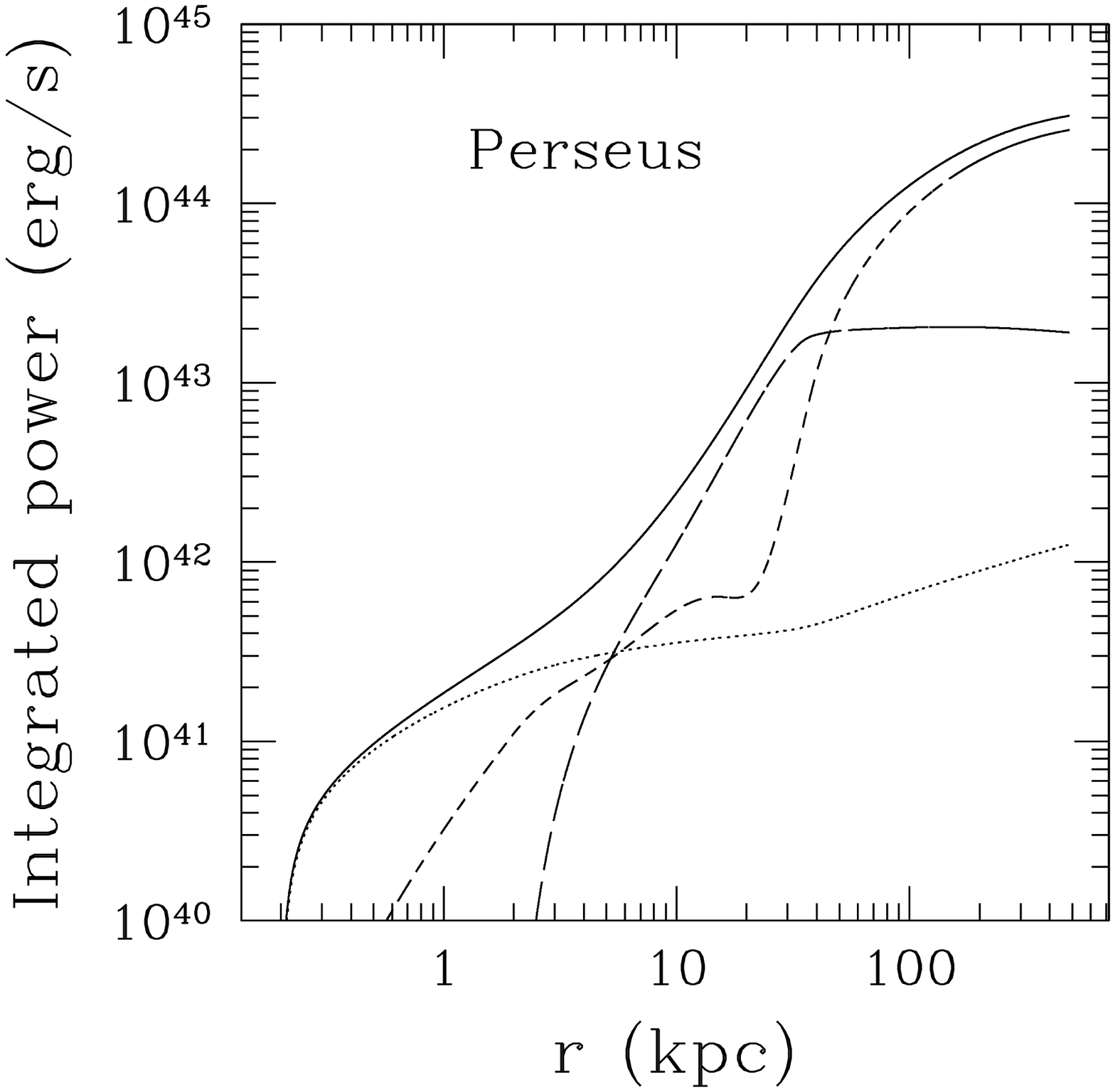}

\caption{\footnotesize The energy sources and sinks in
  the thermal plasma, integrated over volume from the inner
radius~$r_1=0.2$~kpc out to radius~$r$.  The solid line is the
  radiative losses (X-ray luminosity), the short dashed line is
  the heating from
  thermal conduction, the dotted line is power contributed by
the inflow associated with the mass accretion rate, and
  the long-dashed line is the  heating power
  due to the convective turbulence, which includes 
  both the turbulent diffusion of heat and
  $pdV$ work by cosmic-rays.
 \label{fig:f4} }
\end{figure}

In figure~\ref{fig:f4}, we plot the averages of $H_{\rm inflow}$
(dotted line), $H_{\rm conv}$ (long-dashed line), $H_{\rm tc}$
(short-dashed line), and $R$ (solid line), integrated over volume from
the inner radius of our model~($r_1=0.2$~kpc) out to
radius~$r$. We find that $\left \langle H_{\rm
  visc}\right\rangle \ll \langle H_{\rm conv}\rangle$ everywhere in
each cluster, and so we omit $H_{\rm visc}$ from the figures to keep the plots
easier to read.\footnote{We note that equation~(\ref{eq:avpe}) is not
  exactly satisfied by our model solutions.  In our model, we use the
  total-energy equation [equation~(\ref{eq:te})] instead of the plasma
  energy [equation~(\ref{eq:pe})].  Although equation~(\ref{eq:pe}) is
  exactly satisfied when equations~(\ref{eq:te}), (\ref{eq:momentum}), and
  (\ref{eq:cre}) are satisfied, our mixing-length
  approximation of the average of equation~(\ref{eq:pe}) 
is not exactly satisfied when our
  mixing-length approximations to the averages of 
equations~(\ref{eq:te}),
  (\ref{eq:momentum}), and (\ref{eq:cre}) are satisfied. This
  discrepancy is noticeable at the largest radii in Hydra A and
  Sersic 159-03.} 

We can divide the clusters into two groups, those with AGN-dominated
heating and those with conduction-dominated heating.  In Hydra~A and
Sersic~159-03, the heating within the central 100~kpc is dominated by 
convection driven by the central AGN.  In all of the other clusters,
the heating within the central 100~kpc is dominated by thermal
conduction.  The inability of conduction to balance cooling in Hydra~A
and Sersic~159-03 was previously noted by Zakamska \&
Narayan~(2003). These authors constructed density and temperature
profiles for clusters assuming that radiative cooling is balanced by
conductive heating, setting the thermal conductivity equal to a
constant~$f_c$ times the Spitzer thermal conductivity. For Hydra~A and
Sersic~159-03, they found that the values of~$f_c$ that best fit the
observations were 1.5 and 5.6, respectively, much larger than the
theoretically expected value of $f_c\simeq 0.1-0.2$.  In contrast, for 
Abell~1795 the best-fit value of~$f_c$ was~$0.2$.

What are the  characteristics of a cluster that determine
whether AGN feedback or thermal conduction is the dominant heat
source within the central 100~kpc? Given that $\kappa_S
 \propto T^{5/2}$, it seems clear that a lower average temperature~$T_{\rm
  avg}$ makes thermal conduction less able to balance radiative
cooling, leading in turn to a relatively greater role for AGN
feedback.  However, although $T_{\rm avg}$ is important for
determining the relative strength of conduction and AGN feedback, on
its own the value of~$T_{\rm avg}$ does not explain our results, since
Virgo and Abell~262 have temperatures comparable to that of
Sersic~159-03 and lower than that of Hydra~A.  An equally important
factor appears to be the baryon density in the core. In particular,
the clusters with AGN-dominated heating are significantly denser at a
given radius than clusters with conduction-dominated heating that have
similar average temperatures.  For example, for radii between 20 and
50~kpc, the electron density in Sersic~159-03 is 2-3 times greater
than in Virgo.  Similarly, at $r=100$~kpc the electron density in
Hydra~A is $\sim$60\% larger than in Abell~4059,\footnote{This ratio
  is based on the model density profile due to the offset in the
  radius of the observed densities, but a similar conclusion is
  reached by interpolating between data points.}  even though
Abell~4059 has a higher average temperature and larger virial
mass. (Typically, at a fixed radius the density is larger in hotter,
more massive clusters.)  It thus appears that the clusters in which
AGN feedback dominates most strongly over conduction are those in which
the clusters' ongoing formation channels unusually large quantities of
baryons towards the clusters' cores.

We conclude this section with a few additional comments relating to
figures~\ref{fig:f3} and~\ref{fig:f4}.  B\^irzan et al~(2004) found
that their observationally inferred values of~$L_{\rm mech}$ for
16~clusters were correlated with the X-ray luminosity inside the
cooling radius~$r_{\rm cool}$, denoted~$L_X$, supporting the idea that
AGN feedback is at least part of the solution to the cooling-flow
problem.  However, the level of the mechanical luminosity in their
study turns out to be a factor of 1 to 20 lower than the X-ray
luminosity, which raises the question of whether the mechanical
luminosity is sufficient to offset cooling in these clusters. Our
model solutions and figure~\ref{fig:f3} show that the mechanical
luminosity is indeed sufficient when thermal conduction is also
accounted for, and our discussion above regarding
AGN-dominated heating versus conduction-dominated heating offers an
explanation for the large variations in the ratio $L_X/L_{\rm
  mech}$. When heating is dominated by conduction, AGN feedback
heating is only a small fraction of the power radiated from within the
cooling radius, and $L_X/L_{\rm mech}$ is large. On the other hand,
when AGN-driven convection dominates the heating, the AGN heating
power is similar to the total power radiated from within~$r_{\rm
  cool}$, and $L_X\sim L_{\rm mech}$.

We note that figure~\ref{fig:f4}  shows that in Hydra~A,
Sersic~159-03, and Virgo,  conduction actually acts to cool the
plasma over a limited range of~$r$ due to the local maximum in the
temperature profile.  Also, in none of the clusters is the total
convective
heating rate equal to~$L_{\rm cr}$.  This is because the cosmic-ray
luminosity is the power deposited into the cosmic-ray fluid, and only
part of this is transferred to the thermal plasma through $pdV$
work. Additional plasma heating arises from the redistribution
(turbulent diffusion) of plasma thermal energy resulting from the
convective motions.  As can be seen from figure~\ref{fig:f4} and
table~\ref{tab:t1}, the total convective heating of the thermal plasma
is typically on the order of one-third of~$L_{\rm cr}$.

\section{Clusters with central cooling flows}
\label{sec:CF} 

An important point to emerge from figure~\ref{fig:f4} is the
appearance of central cooling flows in several clusters - Abell~262,
Sersic~159-03, Hydra~A, Abell~1795, and Perseus - with radii~$r_{\rm
  cf}$ that are typically a few~kpc. Heating is unable to balance
cooling at $r<r_{\rm cf}$ in these clusters for several reasons: the
AGN feedback heating is distributed over a large volume, thermal
conduction becomes less efficient at small~$r$ due to the lower
temperatures and the fact that $\kappa_T \propto T^{5/2}$, and the
radiative losses per unit volume peak sharply at small~$r$ due to
the large plasma densities. As a result, a cooling
flow develops in which the energy lost to cooling is replenished by
the inflow. Within this central region, we have the approximate
relation $t_{\rm cool} \sim t_{\rm inflow}$, where $t_{\rm inflow} =
r/|v_{\rm inflow}|$ is the inflow time and $t_{\rm cool} = 1.5 n k_B
T/R$ is the local cooling time.  Because $t_{\rm inflow} \sim t_{\rm
  cool}$, the mass accretion rate at $r=r_{\rm cf}$ is approximately
given by
\begin{equation}
\dot{M} \sim 4\pi r_{\rm cf}^2 \rho(r_{\rm cf}) 
\left[\frac{r_{\rm cf}}{t_{\rm cool}
(r_{\rm cf})}\right]
\sim \frac{M_{\rm cf}}{t_{\rm cool}(r_{\rm cf})},
\label{eq:mdotest} 
\end{equation} 
where $r_{\rm cf}$ is the radius of the central cooling
flow region, and $M_{\rm cf}$ is the mass of plasma contained
within the cooling flow region.  Because we
have no sources or sinks of plasma, $\dot{M}$ is independent
of~$r$ in our model.

How do these central cooling flows match onto adiabatic Bondi flow at
smaller radii?  In our model, as $r$ decreases from $r_{\rm cf}$
towards zero, $\rho$ rises and $T$ decreases until the Bondi accretion
rate at the fixed radius $r_1=0.2$~kpc matches the cooling-flow mass
accretion rate given by equation~(\ref{eq:mdotest}).  However, our
forcing the flow to become adiabatic at $r_1$ is artificial, and leads
to an unrealistic plasma profile near~$r_1$ with an abrupt transition in
the flow at~$r_1$.  A better approach was adopted by Quataert \&
Narayan (2000).  These authors investigated radial inflow with cooling
in the absence of thermal conduction and cosmic rays
using a numerical shooting method and solved all the way in to the
sonic point, $r = r_{\rm sonic}$, at which $v_r = -c_s$.  
They found 
a smooth transition from an outer cooling flow with
$t_{\rm inflow} \simeq t_{\rm cool}$ to an inner adiabatic Bondi
flow with $t_{\rm inflow} \ll t_{\rm cool}$, provided
that $r_{\rm sonic} \lesssim r_{\rm tr}$, where
\begin{equation}
r_{\rm tr}  \equiv \frac{GM_{\rm bh}}{\sigma^2} =
0.05 \mbox{ kpc} \left(\frac{M_{\rm bh}}{10^9 M_{\sun}}\right)
\left(\frac{\sigma}{300 \mbox{ km/s}}\right)^{-2}
\label{eq:rtr}
\end{equation}
is the radius within which gravity is dominated by the black hole and
$\sigma$ is the circular velocity of the BCG, which was taken to be
independent of~$r$. They also found that equation~(\ref{eq:mdotest})
was an accurate estimate of the numerically calculated mass accretion
rate in the absence of mass dropout.  For $r_{\rm sonic} < r_{\rm
  tr}$, Quataert \& Narayan's solution satisfies~$c_{\rm s} \sim
\sigma = \mbox{constant}$ at $r> r_{\rm tr}$.  In the absence of mass
dropout, the condition~$t_{\rm cool} = t_{\rm inflow}$ leads to the
relation~$\rho\propto r^{-3/2}$ within the cooling-flow part of their
solution. The Bondi accretion rate $\dot{M}_{\rm Bondi}$ [given by
  equation~(\ref{eq:mdot})] evaluated at a radius~$r$ within the
cooling-flow part of their solution thus increases towards smaller~$r$
like~$r^{-3/2}$.  At a sufficiently small value of~$r$, which we call
$r_{\rm ad}$, the Bondi accretion rate equals the rate $\dot{M}_{\rm
  cf}$ at which mass flows in through the cooling flow, and the flow
makes a transition to an adiabatic Bondi flow. At $r<r_{\rm ad}$, the
ratio~$t_{\rm cool}/t_{\rm inflow}$ increases towards smaller~$r$, and
so the neglect of cooling at $r<r_{\rm ad}$ is self-consistent.  We
note that the Bondi accretion formula can be applied in the model of
Quataert \& Narayan (2000) at the outer boundary of the adiabatic flow
region, $r_{\rm ad}$, even if $r_{\rm ad}$ lies outside the region in
which the black hole dominates the gravitational acceleration. This is
because the Bondi accretion rate depends only on the specific
entropy~$s$ of the plasma and~$M_{\rm bh}$, and $s$ is constant for
$r< r_{\rm ad}$.

It would be valuable to incorporate into our model an approach similar
to that of Quataert \& Narayan (2000), including cosmic rays, thermal
conduction, and the possibility of convection. Although such a
calculation is beyond the scope of this paper, we expect that in such
an analysis the mass accretion rate of the central accretion flow and
the plasma parameters at~$r_{\rm Bondi}$ are still controlled
by~$\dot{M}_{\rm cf}$, as in Quataert \& Narayan's (2000) work.
Because the central accretion flow is in some sense slaved to the
surrounding cooling flow, the AGN regulates the mass accretion rate
primarily by controlling the properties of the central cooling flow,
and in particular by controlling~$r_{\rm cf}$.  For example, if
$\dot{M}$ rises above the equilibrium value, the AGN-feedback heating
rises.  This then reduces~$r_{\rm cf}$, because one has to go to
smaller~$r$ in order for $\rho$ to rise enough that cooling exceeds
the convective heating rate. The reduction in~$r_{\rm cf}$ reduces
$\dot{M}$, as can be seen from equation~(\ref{eq:mdotest}), which then
causes $\dot{M}$ to drop back down to its equilibrium level.

As mentioned above, the existence of a smooth transition from a
cooling flow to an inner adiabatic Bondi flow requires that $r_{\rm
  sonic} <r_{\rm tr}$. Otherwise, as described by Quataert \& Narayan
(2000), the ratio $t_{\rm cool}/t_{\rm inflow}$ decreases inwards in
the supersonic region at $r_{\rm tr} < r < r_{\rm sonic}$, and the
plasma cools rapidly to very low temperature.  In this case, the
cooling plasma could still end up fueling the central black hole, but
it would do so through some process other than the one we have assumed
in our model, e.g., by forming stars whose winds then feed the black
hole or through infalling cold gas [see, e.g., Pizzolato \& Soker
  (2005) and Soker (2006)]. 

Under what conditions is $r_{\rm sonic} > r_{\rm tr}$? One factor that
can cause~$r_{\rm sonic}$ to exceed~$r_{\rm tr}$ is a small central
black hole mass, since a smaller~$M_{\rm bh}$ reduces~$r_{\rm tr}$.
In addition, as illustrated in Quataert \& Narayan's approximate
analytic results, if the black hole's contribution to gravity were
hypothetically ignored, $r_{\rm sonic}$ would increase with increasing
$\dot{M}$. Thus, a sufficiently large $\dot{M}$ can also cause $r_{\rm
  sonic}$ to exceed~$r_{\rm tr}$.  A large $\dot{M}$ results from
either a large $L_{\rm cr}$ or a small accretion efficiency~$\eta$.
As discussed in the previous section, $L_{\rm cr}$ is approximately
determined by the baryon density and temperature at the cooling radius
- a higher density and/or lower temperature at $r_{\rm cool}$ means
that thermal conduction can offset less of the radiative cooling
within the cooling radius, which in turn leads to a larger~$L_{\rm
  cr}$. Thus, to summarize, smaller values of $M_{\rm bh}$, $\eta$, or
$T(r_{\rm cool})$ and/or larger values of $\rho(r_{\rm cool})$ can
cause $r_{\rm sonic}$ to exceed~$r_{\rm tr}$, preventing a smooth
transition from a cooling flow to an inner Bondi flow, and causing the
cooling of intracluster plasma to low temperatures at $r>r_{\rm tr}$.

We note that if the accretion efficiency $\eta_{\rm cool}$ that arises
when plasma cools to low temperature outside~$r_{\rm tr}$ is much
smaller than the accretion efficiency~$\eta$ associated with Bondi
accretion, then a flow that cools rapidly outside $r_{\rm tr}$ will
need a much higher~$\dot{M}$ (and larger $r_{\rm cf}$) in order for
AGN feedback to provide the heating needed to offset cooling
within the cooling radius. A much larger $\dot{M}$, in conjunction
with plasma cooling to low temperatures outside $r_{\rm tr}$, would
imply a much larger star formation rate within the BCG. Thus, a
small~$M_{\rm bh}$ or large~$L_{\rm cr}$ could lead to the
condition~$r_{\rm sonic} > r_{\rm tr}$ and cause star formation at
rates that significantly exceed the Bondi accretion rates listed in
table~\ref{tab:t1}.

Returning to our model calculations, we list in table~\ref{tab:t1} the
values of $|\langle v_r\rangle|/c_s$ at $r=r_1$ for the clusters in our
sample. The value of $|\langle v_r\rangle|/c_s$ reaches its maximum at
$r=r_1$ in our solutions, and thus our solutions satisfy $|\langle
v_r\rangle|/c_s < 1$ at all radii. Our calculations are thus at
least marginally consistent with our assumptions of hydrostatic
equilibrium and Bondi accretion. We also note that the the sound
crossing time~$t_s$ is shorter than cooling time~$t_{\rm cool}$ at all
radii in our model solutions.  However, our model imposes an abrupt and
artificial transition in the flow at~$r=0.2$~kpc, which causes our
solution near $r=0.2$~kpc to be inaccurate. In addition, there is
significant uncertainty in the values of $M_{\rm bh}$ and $\eta$. It
is thus possible that some of the clusters reach a sonic transition
outside~$r_{\rm tr}$. Further investigation of this issue is needed.

\section{Summary}
\label{sec:disc} 

There is a growing consensus that AGN feedback holds the key to
solving the cooling-flow and overcooling problems for clusters of
galaxies.  However, the way in which an AGN's power is delivered to
the diffuse intracluster plasma is still not well understood. In this
paper, we suggest that an AGN's mechanical luminosity heats the
intracluster plasma by accelerating cosmic rays that cause the
intracluster medium to become convectively unstable. We explore this
idea by developing a steady-state, mixing-length-theory model. By
adjusting a single parameter in the model (the size of the cosmic-ray
acceleration region,~$r_{\rm source}$), we obtain a good match to the
observed density and temperature profiles in seven out of the eight
clusters in our sample. Our model underestimates the density in the
eighth cluster, Sersic~159-03, within the central $\sim 50$~kpc.  We
suggest that this discrepancy may result from the fact that the
parameters in our NFW mass model are determined neglecting the
cosmic-ray pressure. At the same time, Sersic 159-03 has the largest
soft x-ray excess of any cluster observed by XMM, and likely contains
a large population of non-thermal particles concentrated in the
cluster core. (Werner 2007) If the mass model were recalculated taking
the non-thermal pressure into account, the gravitational acceleration
would be larger, especially in the cluster core, which would increase
the plasma density at $r\lesssim 50$~kpc in our model calculations and
possibly bring the model into agreement with the observations.  We
also find that the cosmic-ray luminosities of the AGN in our sample
are strongly correlated with the observationally inferred mechanical
luminosities of these AGN.  Our results suggest that AGN-driven
convection is an important process in cluster cores.

In our model solutions, the radiative cooling rate is much more peaked
about $r=0$ than is the rate of convective heating.  As a result, a
compact central cooling flow arises in our model calculations
for several of the clusters in our sample. The radii,
$r_{\rm cf}$, of the cooling flows are typically a few~kpc.
The mass accretion rate onto the central AGN in these
clusters is roughly the plasma mass at $r< r_{\rm cf}$ divided by the
cooling time at~$r_{\rm cf}$. We suggest that the AGN regulates the
mass accretion rate in these clusters by controlling $r_{\rm cf}$: if
the AGN power rises above the equilibrium level, the size of the
central cooling flow decreases, the mass accretion rate drops, and the
AGN power then drops back down to the equilibrium level.

\acknowledgements We thank Eric Blackman, Nadia Zakamska, Jelle
Kaastra, and Eliot Quataert for helpful discussions.  
We acknowledge the usage of the HyperLeda database (http://leda.univ-lyon1.fr).
This work was partially supported by NASA's Astrophysical Theory Program under
grant NNG 05GH39G and by NSF under grant AST 05-49577.

\appendix

\section{The profiles of the cosmic-ray pressure and turbulent velocity}
\label{ap:prof}

\begin{figure}[t]
\hspace{-1cm}
\includegraphics[width=1.7in]{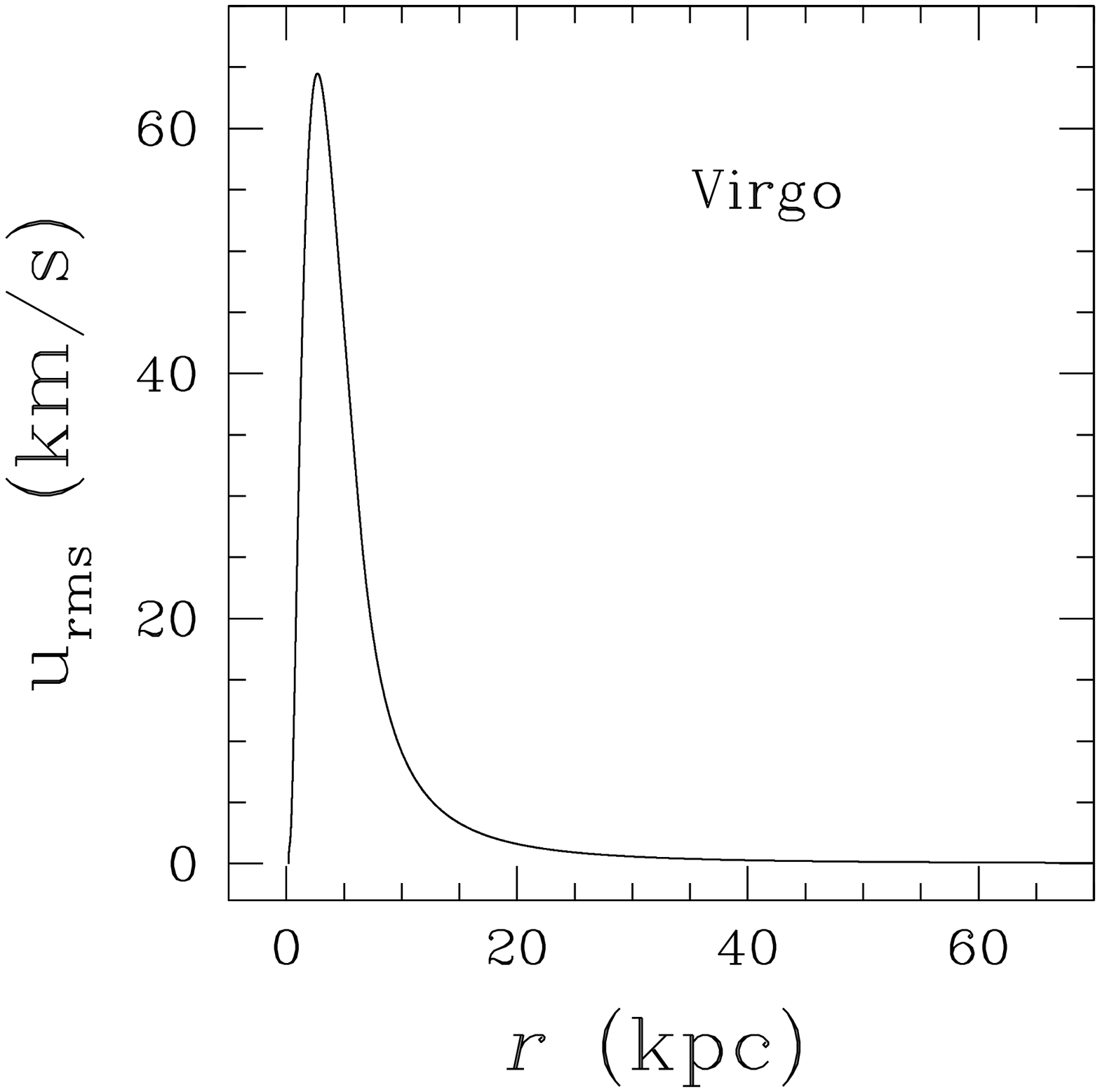}\hspace{0.05cm}\includegraphics[width=1.7in]{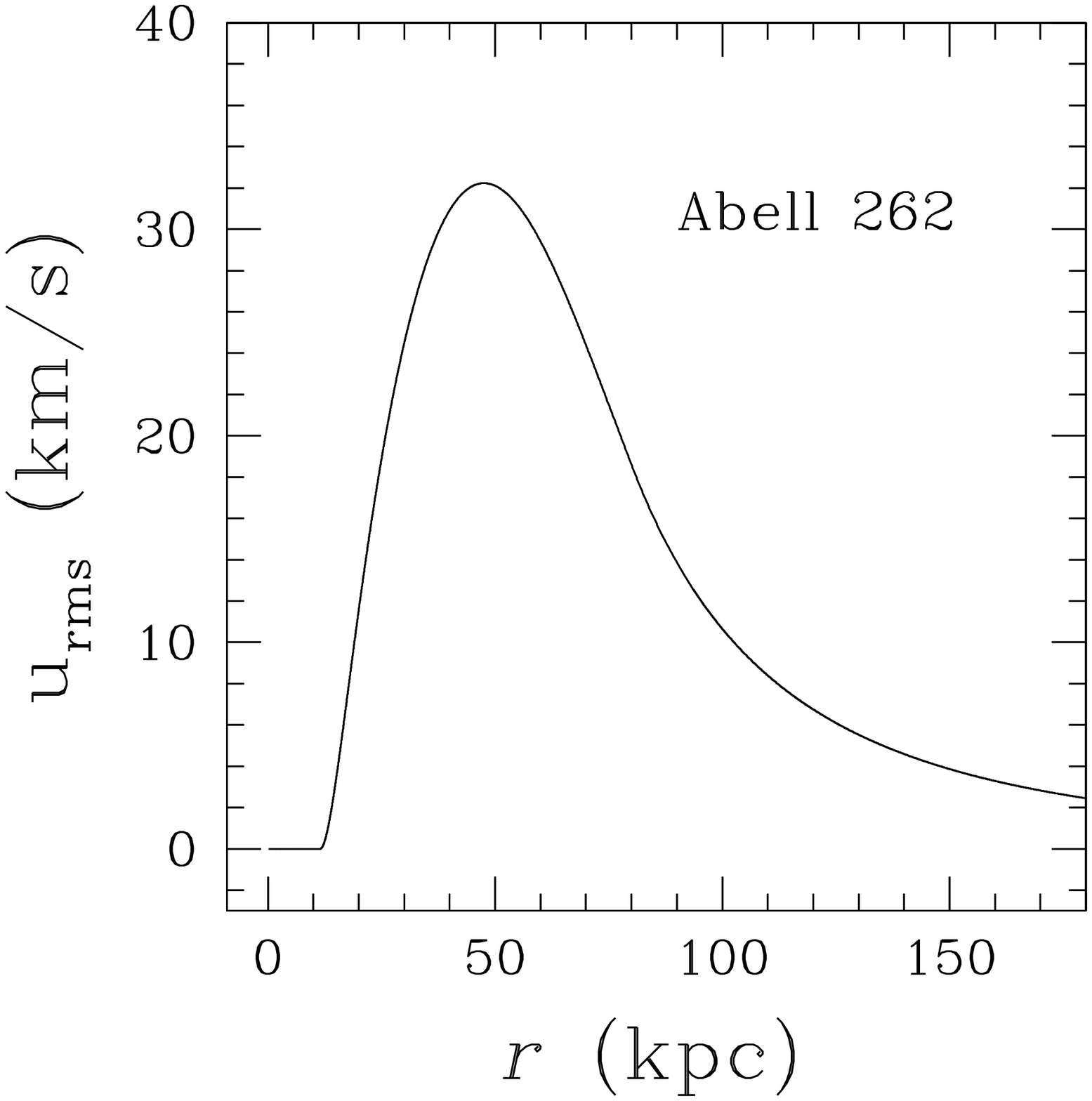}\hspace{0.05cm}\includegraphics[width=1.7in]{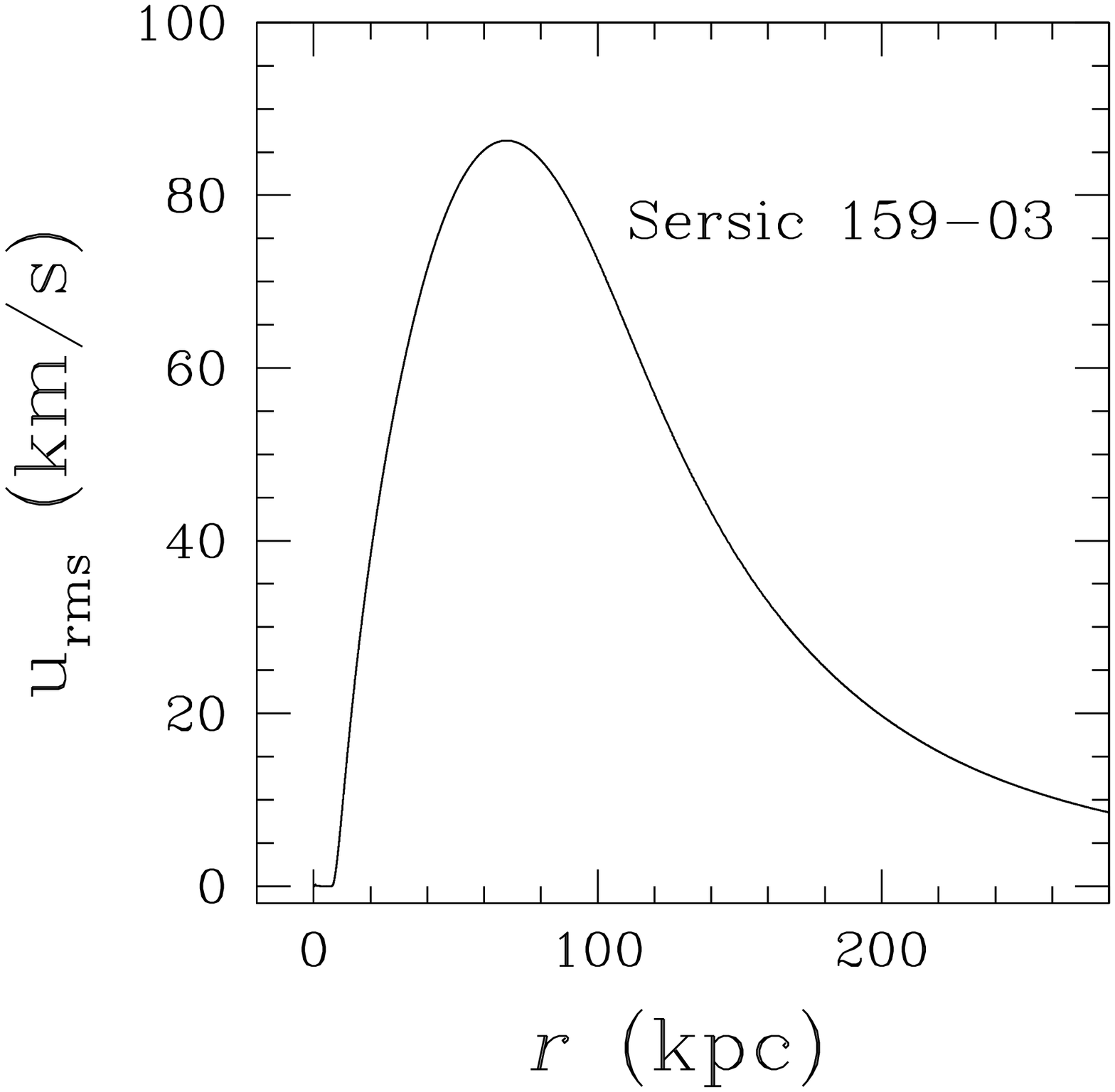}\hspace{0.05cm}\includegraphics[width=1.7in]{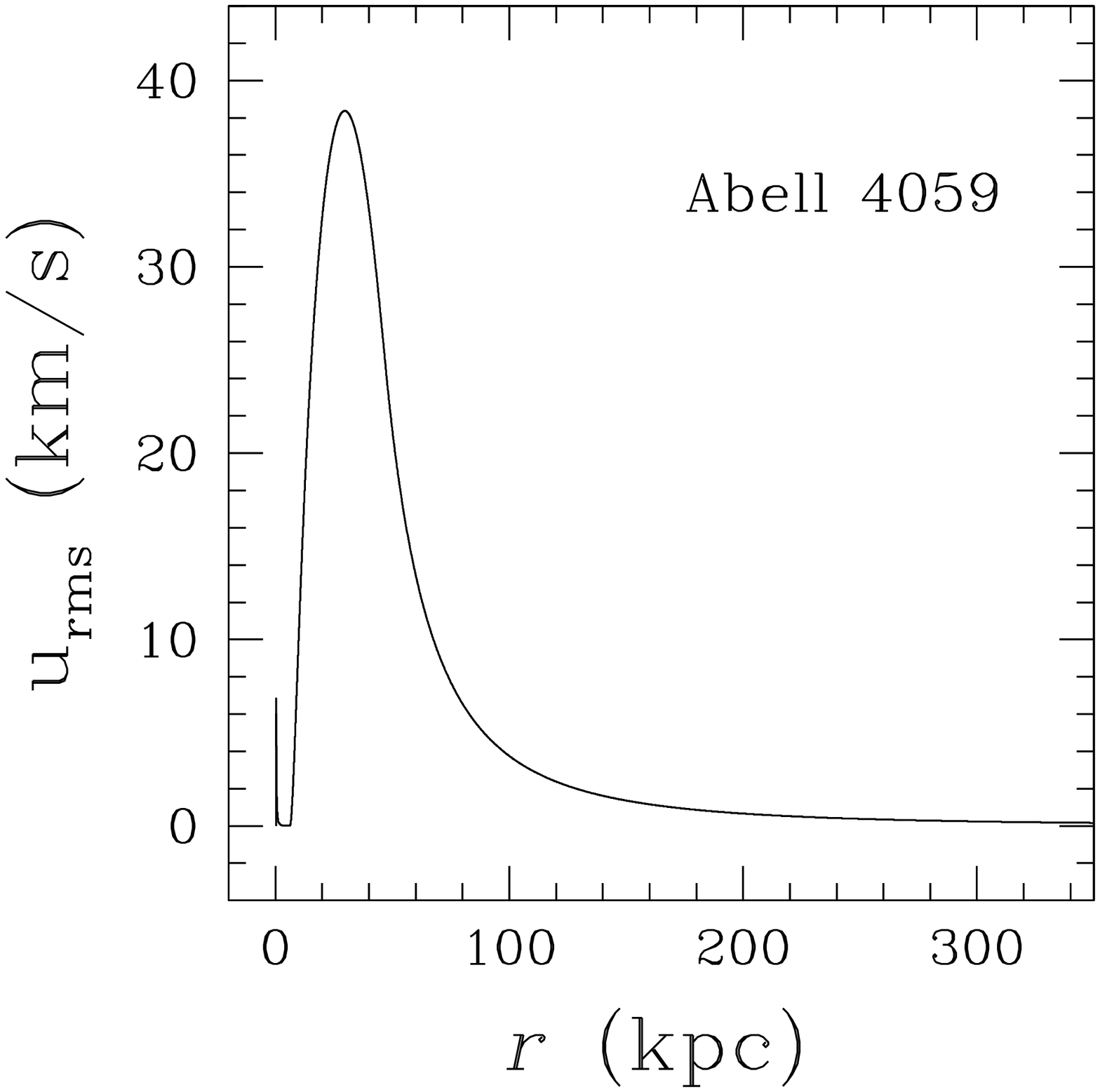}

\hspace{-1cm}
\includegraphics[width=1.7in]{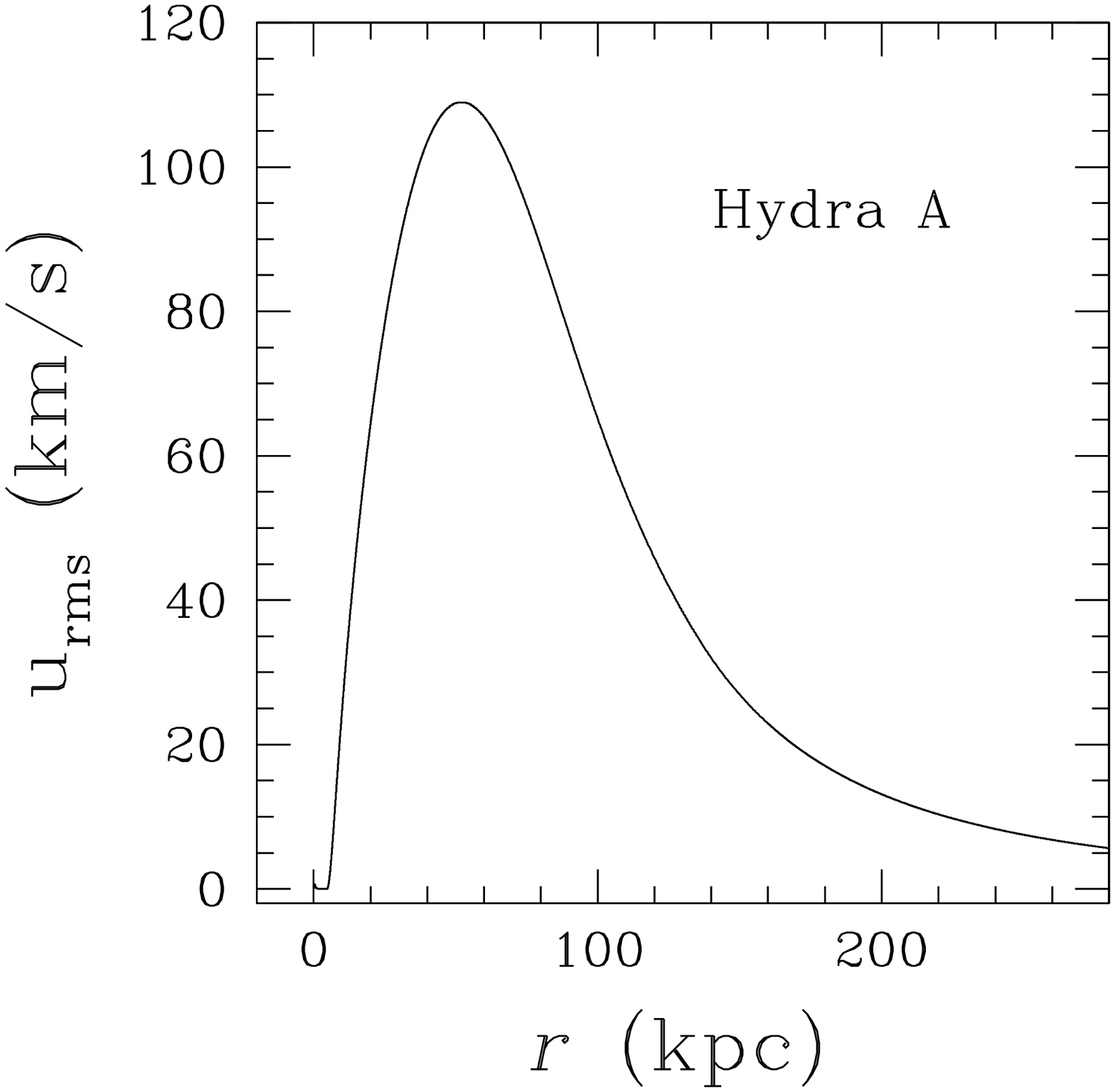}\hspace{0.05cm}\includegraphics[width=1.7in]{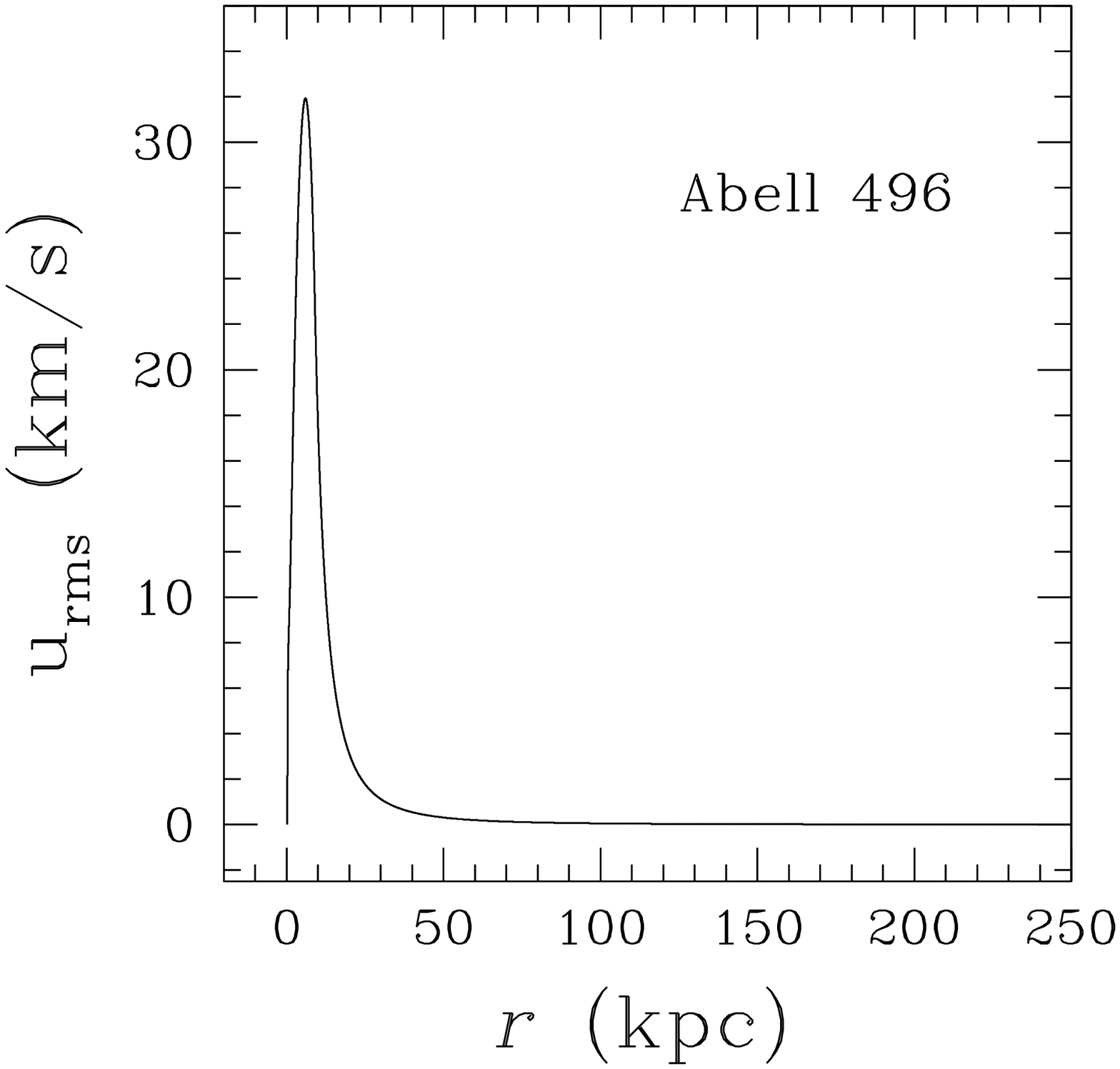}\hspace{0.05cm}\includegraphics[width=1.7in]{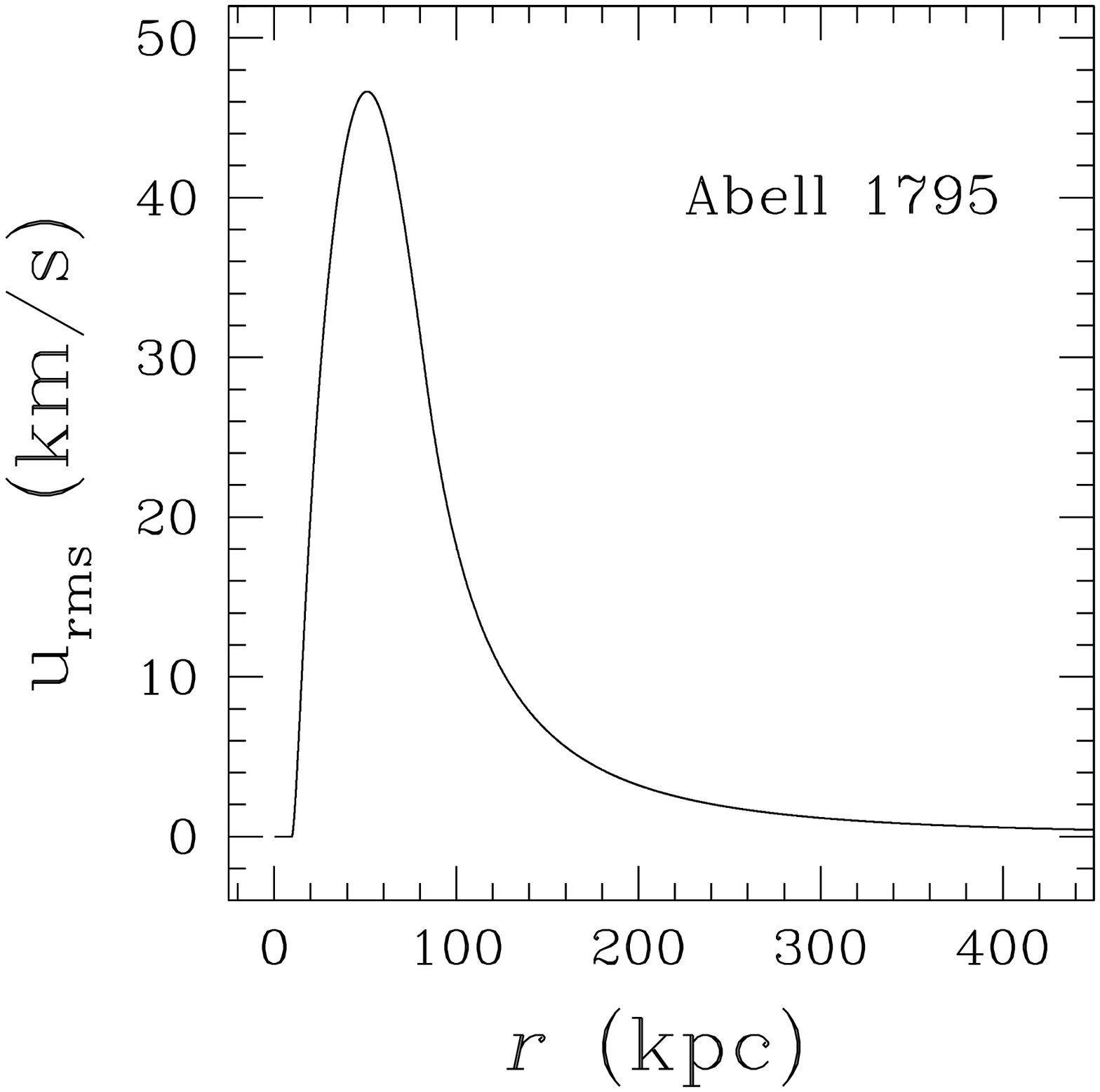}\hspace{0.05cm}\includegraphics[width=1.7in]{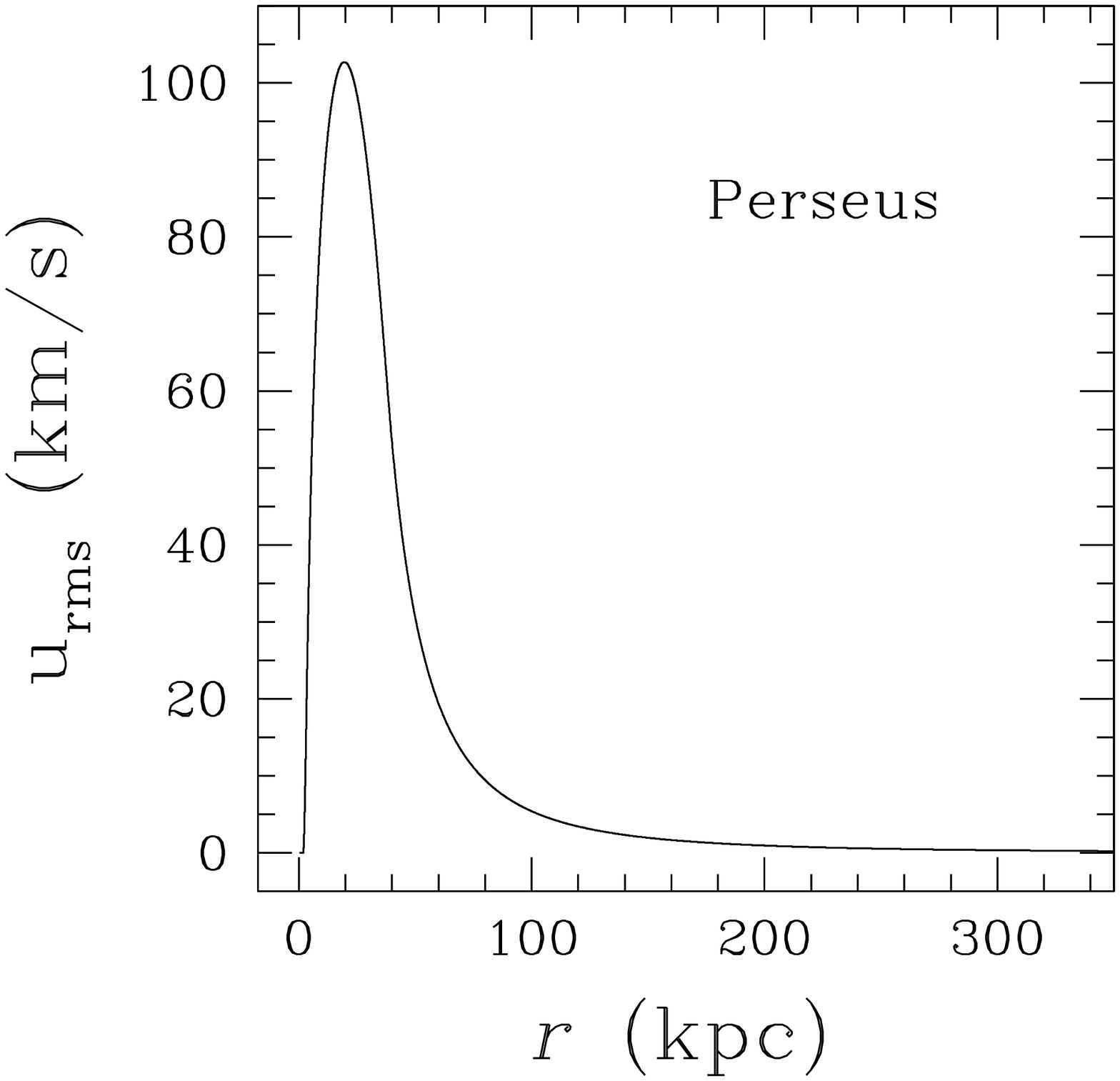}

\caption{\footnotesize The rms turbulent velocity as a function of radius in the
model solutions.
 \label{fig:f5} }
\end{figure}

\begin{figure}[t]
\hspace{-1cm}
\includegraphics[width=1.7in]{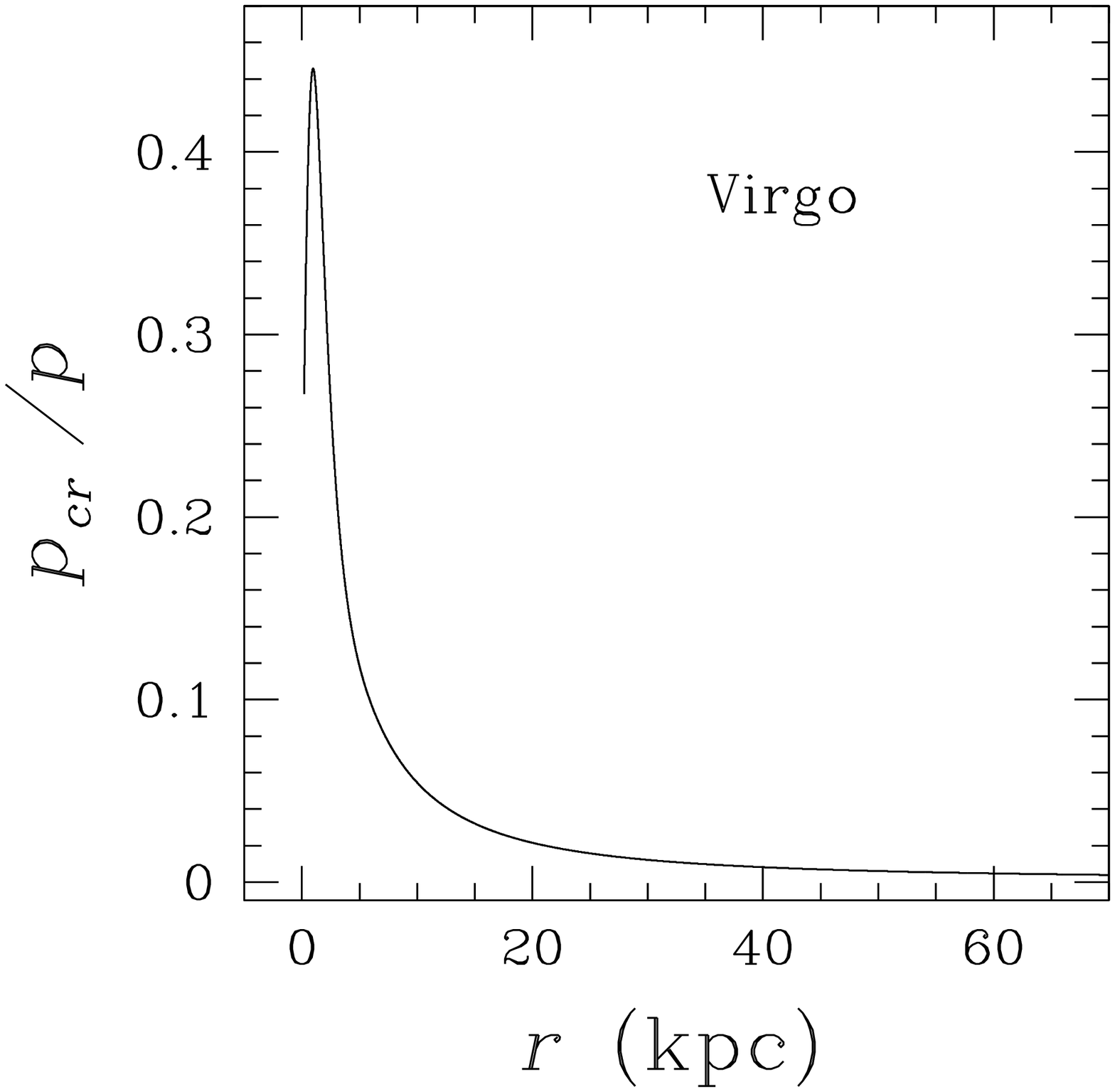}\hspace{0.05cm}\includegraphics[width=1.7in]{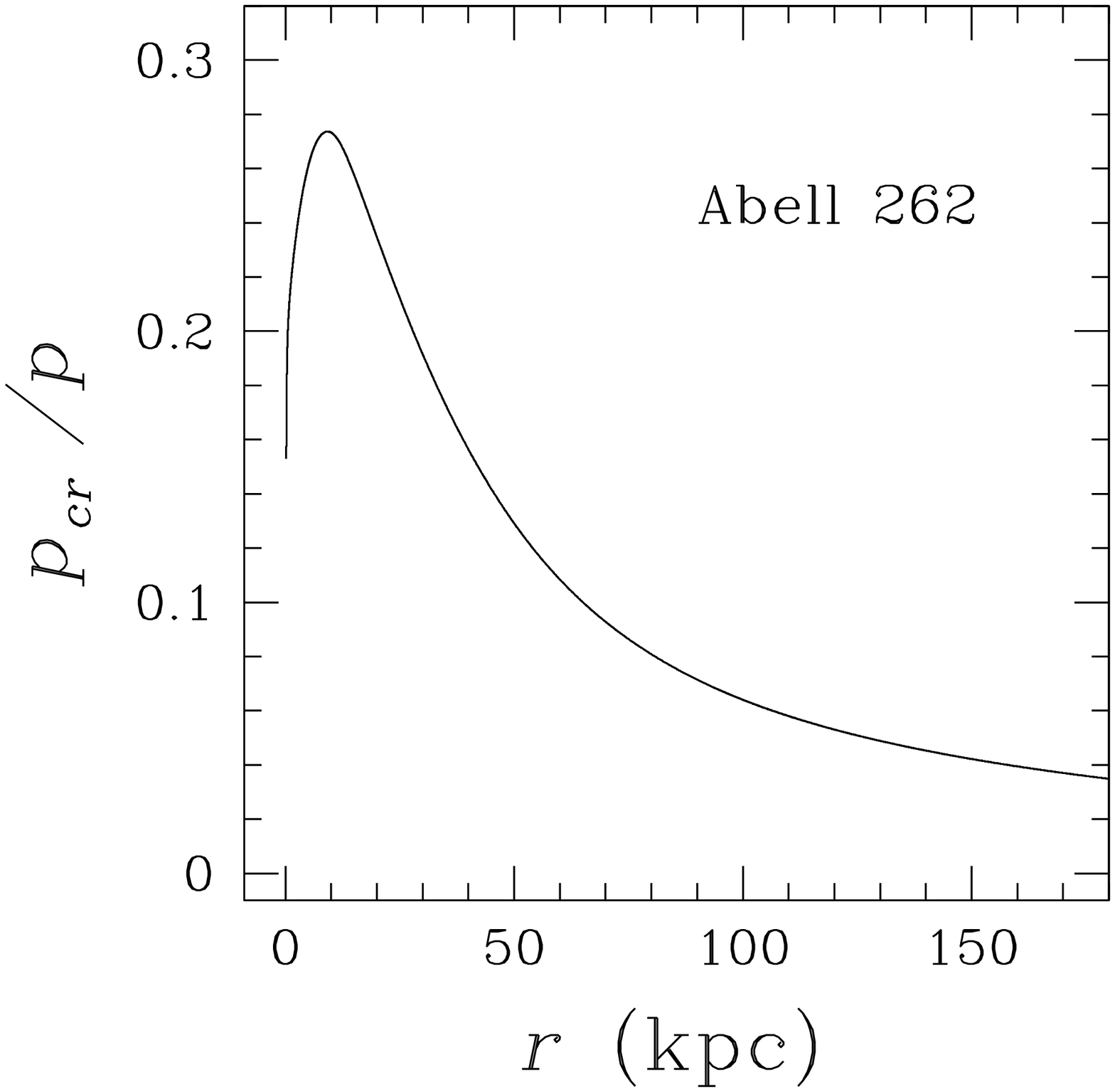}\hspace{0.05cm}\includegraphics[width=1.7in]{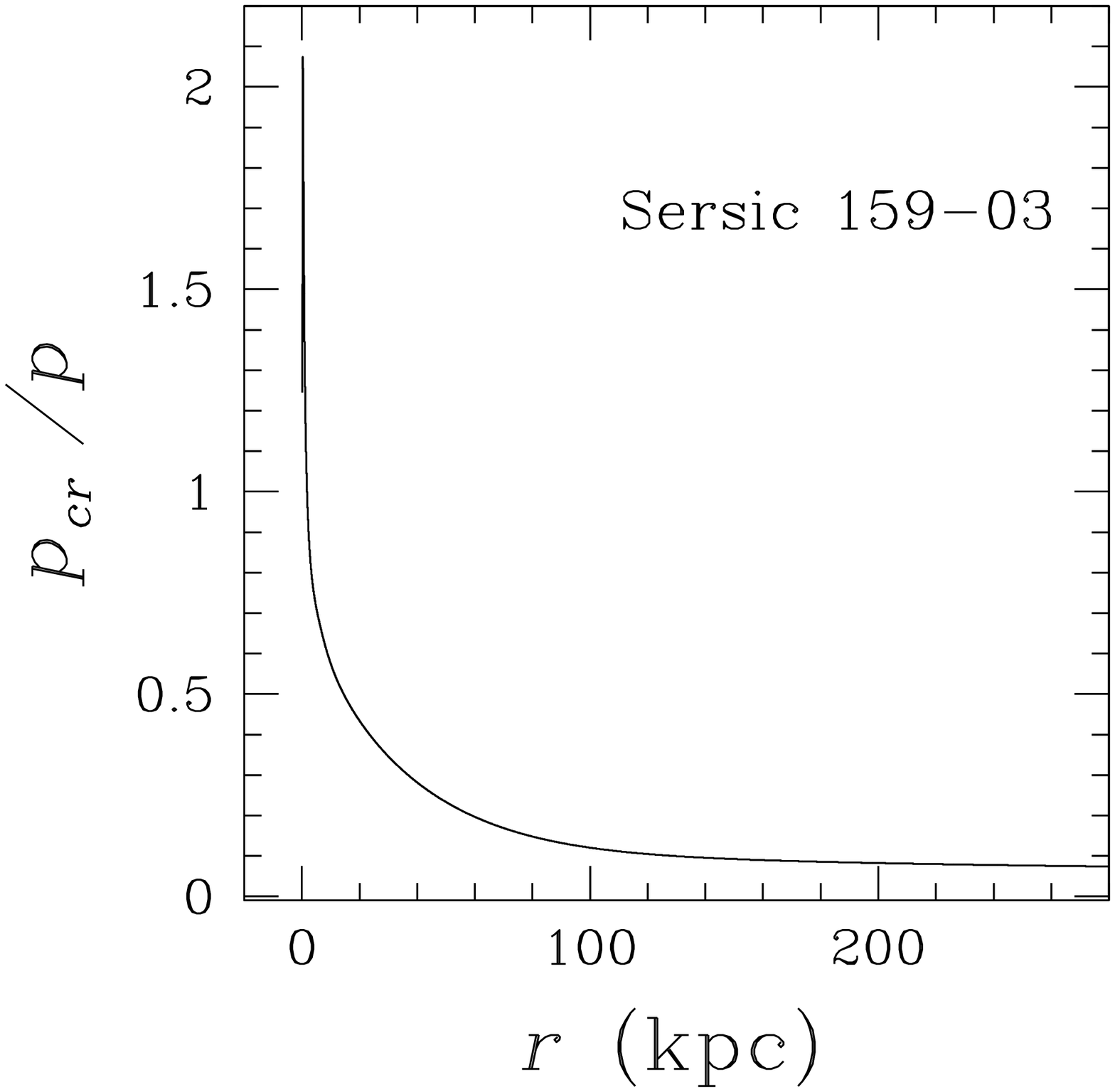}\hspace{0.05cm}\includegraphics[width=1.7in]{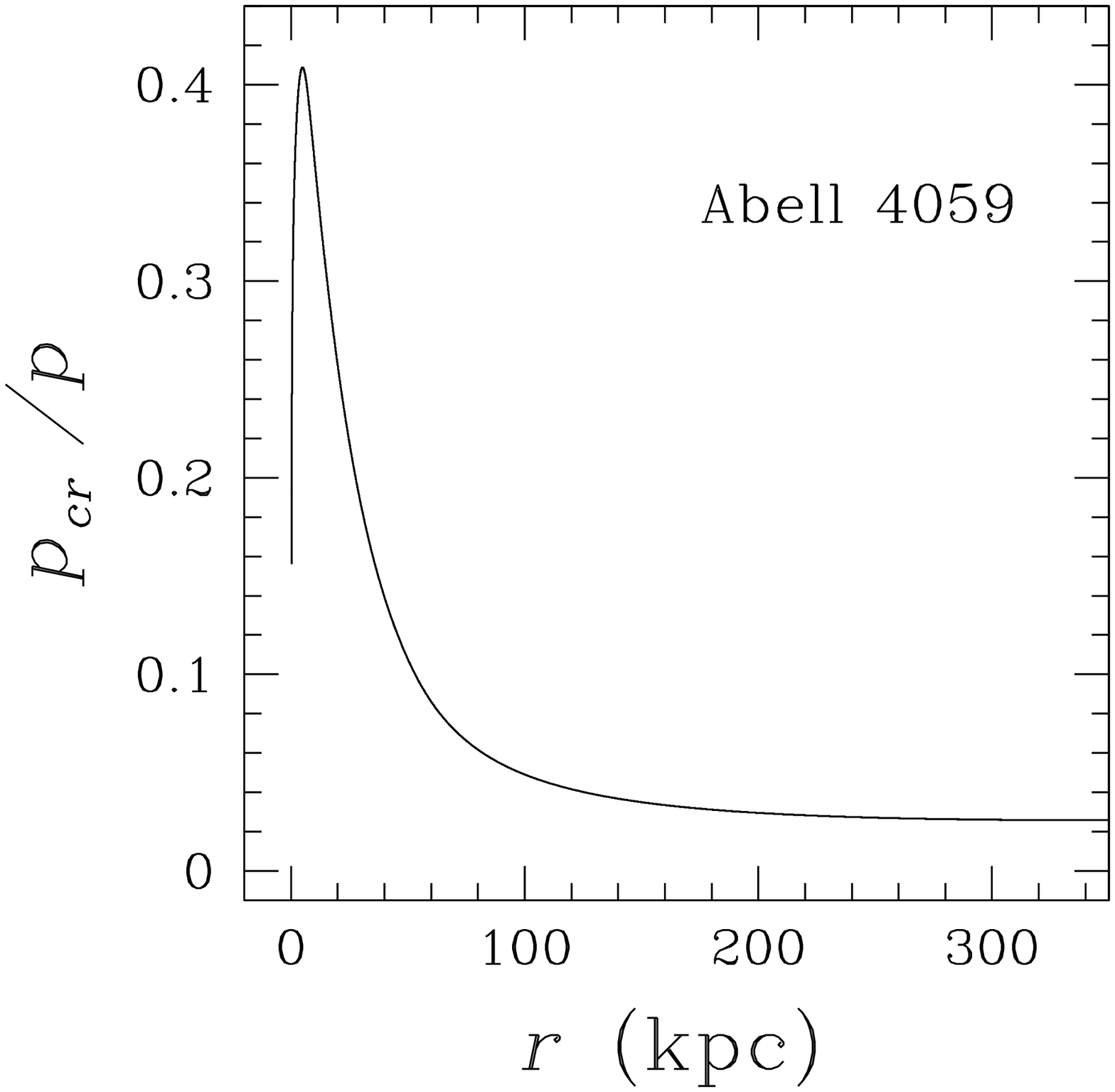}

\hspace{-1cm}
\includegraphics[width=1.7in]{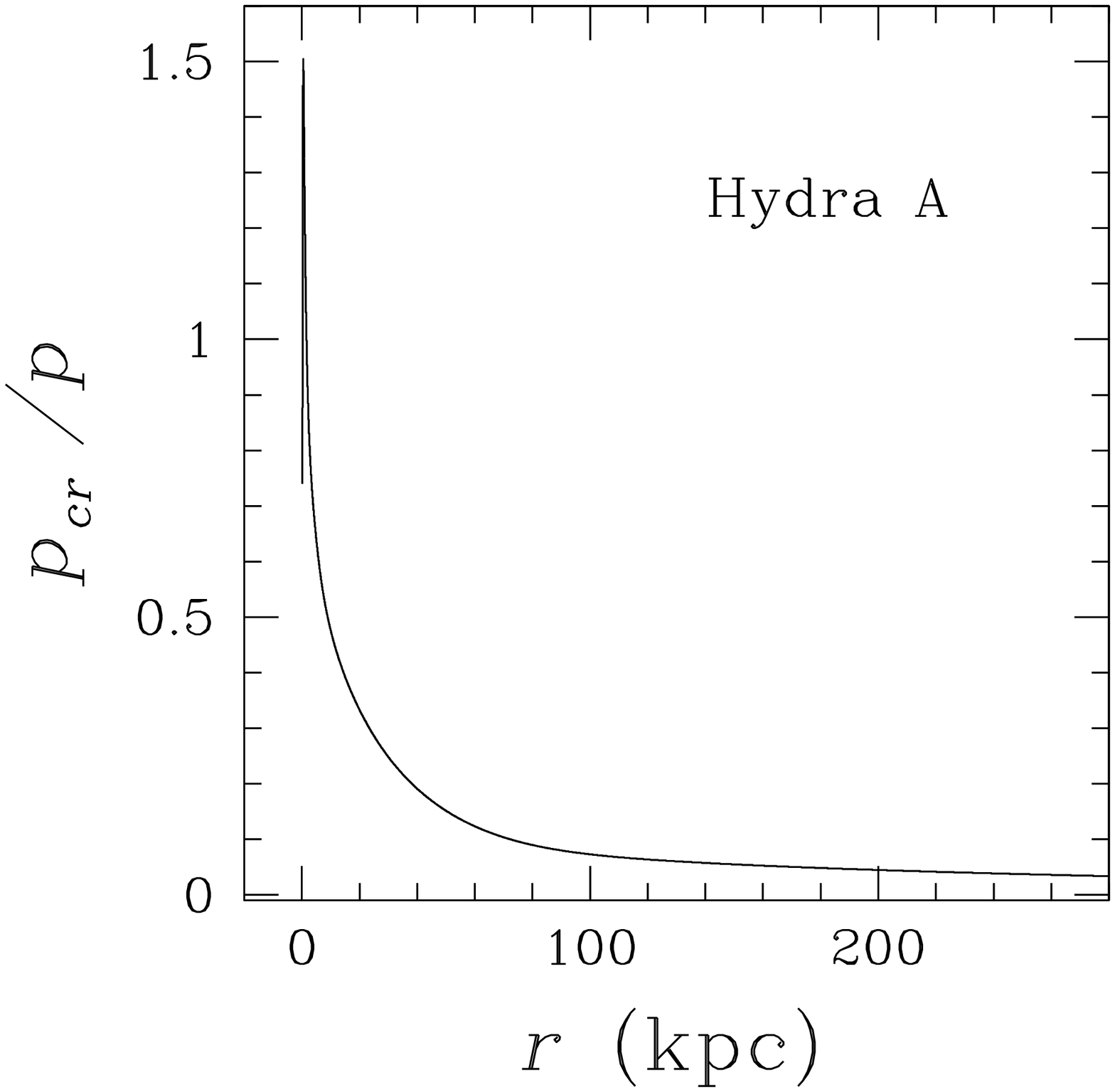}\hspace{0.05cm}\includegraphics[width=1.7in]{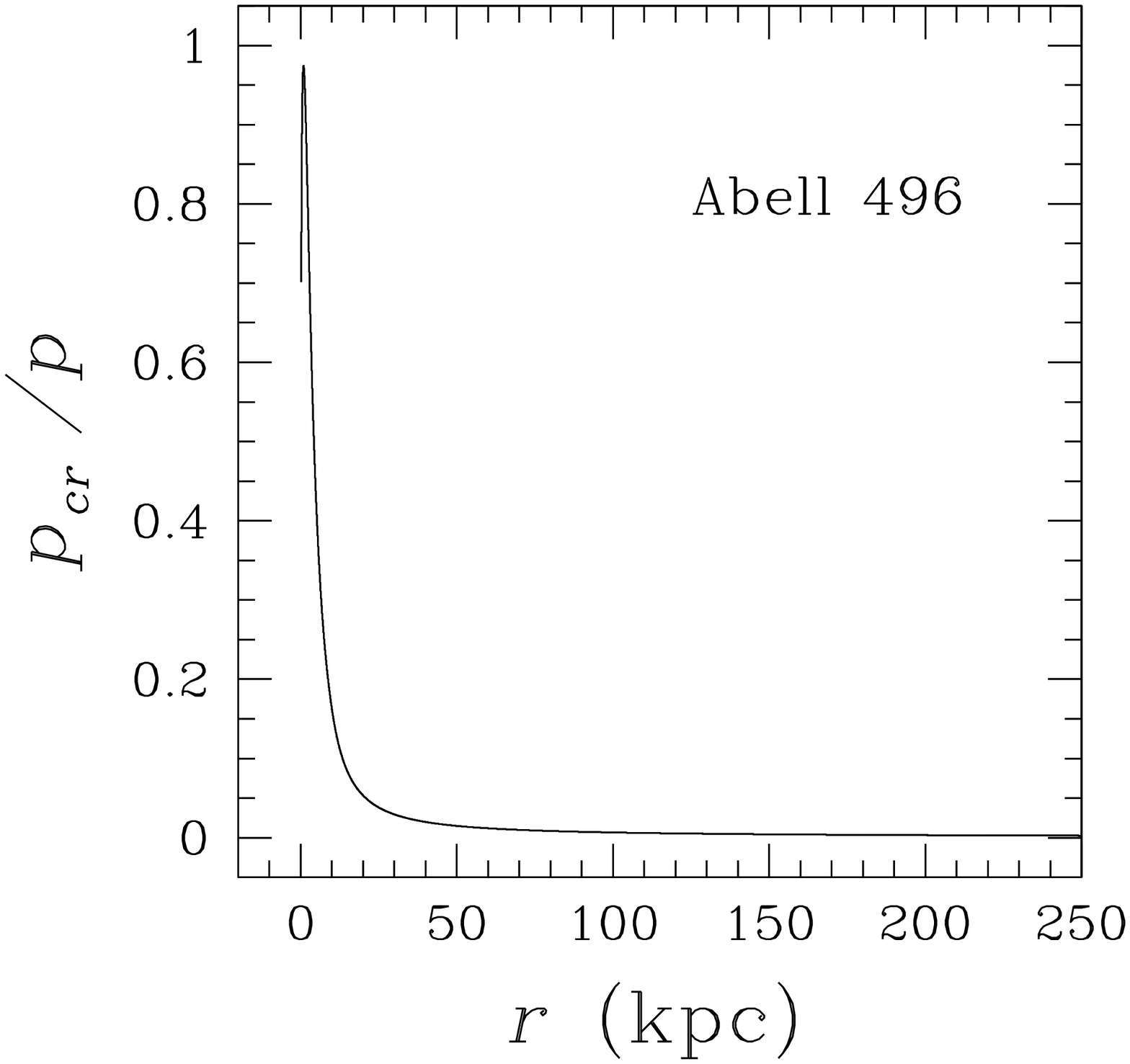}\hspace{0.05cm}\includegraphics[width=1.7in]{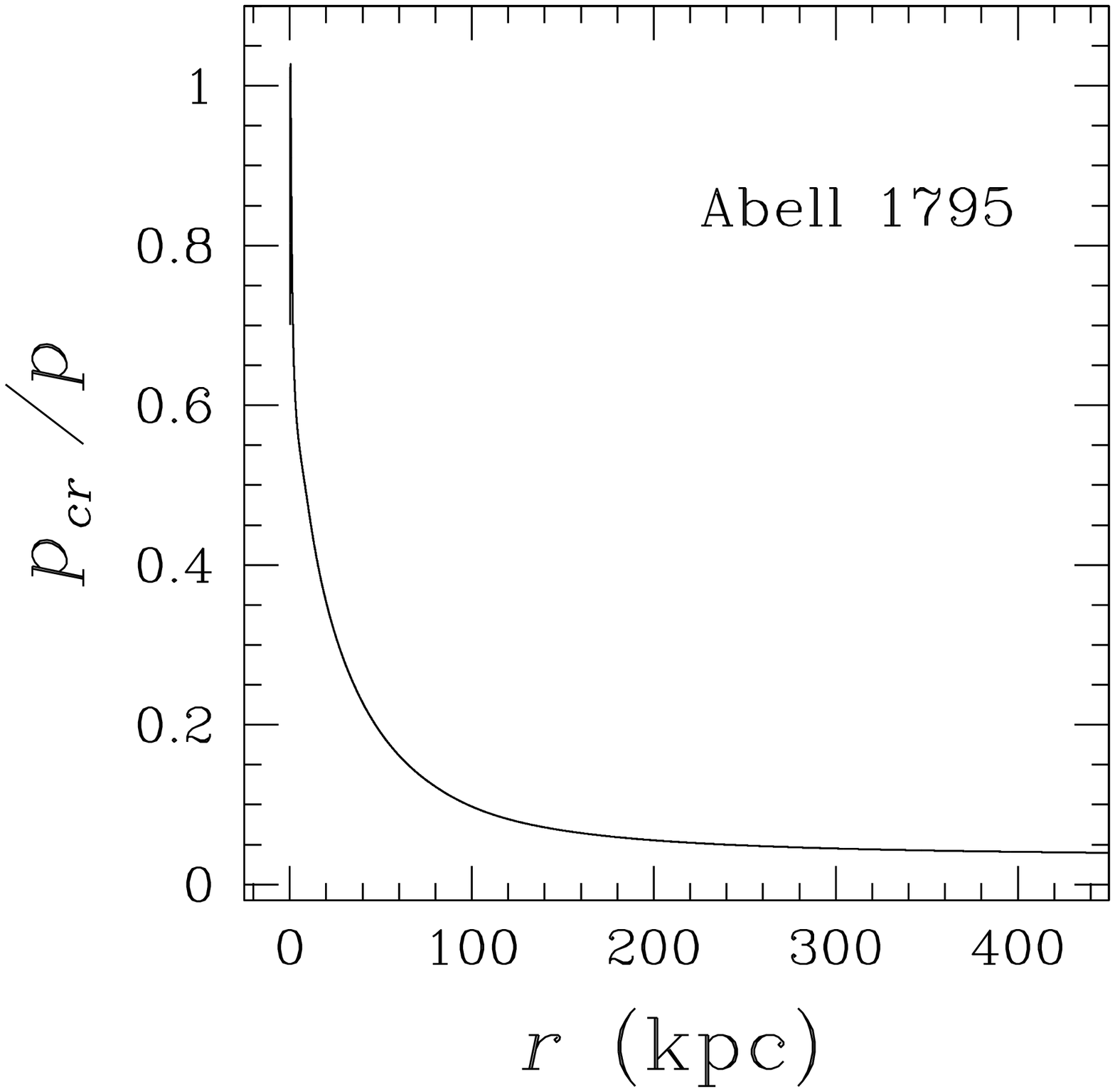}\hspace{0.05cm}\includegraphics[width=1.7in]{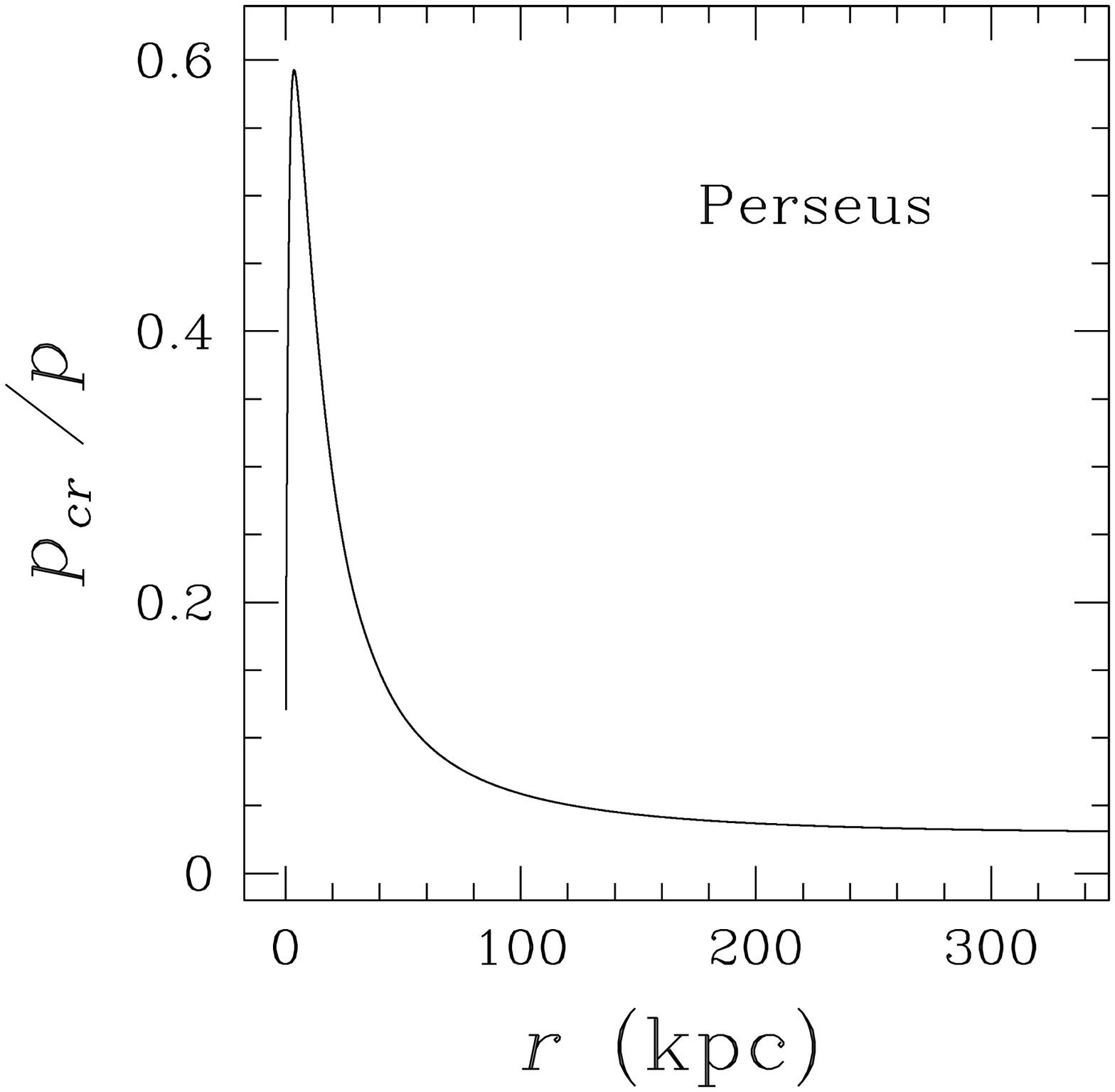}

\caption{\footnotesize The ratio of cosmic-ray pressure to thermal pressure in the
model solutions.
 \label{fig:f6} }
\end{figure}

The profiles of the rms turbulent velocity~$u_{\rm rms}$ [defined in
  equation~(\ref{eq:urms2})] and the cosmic-ray pressure (as a
fraction of the thermal pressure) are plotted in figures~\ref{fig:f5}
and~\ref{fig:f6}. Although it is difficult to see in the cases
of Sersic~159-03, Hydra~A, Abell~496, and Abell~1795, 
figure~\ref{fig:f6} shows that $(d/dr)(p_{\rm
  cr}/p)>0 $ at small~$r$. [This can also
be seen by comparing the figures with the
values of $p_{\rm cr}(r_1)/p(r_1)$ listed in table~\ref{tab:t1}.]
This is not because~$dp_{\rm cr}/dr >0$ (in
fact, $dp_{\rm cr}/dr<0$ at all~$r$ for each cluster), but instead
because the thermal pressure decreases with radius more rapidly than
the cosmic-ray pressure. We note that there was an error in one of the
plotting subroutines used for paper~II, which resulted in the
velocities plotted in figure~4 of paper~II being too large by a factor
of~3.

\section{The two-fluid mixing-length theory and numerical method}
\label{ap:mlt} 

In this appendix we describe the two-fluid mixing length
theory that we use to model the convective intracluster
medium.
We take all fluid quantities to be the sum of an
average value and a turbulent fluctuation, so that
\begin{equation}
\rho = \langle \rho \rangle + \delta \rho, 
\label{eq:avrho} 
\end{equation} 
\begin{equation}
\bm{v} = \langle \bm{v} \rangle + \delta \bm{v},
\label{eq:av2} 
\end{equation} 
etc. We take all average quantities to 
depend only on the radial
coordinate~$r$, and we set $\langle \bm{v}\rangle
= \langle v_r \rangle \hat{r}$.
We wish to solve for four quantities:
$\langle \rho \rangle $, $\langle
v_r \rangle$, $\langle T \rangle$, and $\langle p_{\rm cr}
\rangle$. To do so we take the averages of 
four equations: 
equations~(\ref{eq:cont}),
(\ref{eq:momentum}), and
(\ref{eq:cre}), as well as
the total-energy equation.
The latter  is obtained by taking the
dot product of equation~(\ref{eq:momentum}) with
$\bm{v}$ and adding the resulting equation to the sum
of equations~(\ref{eq:pe}) and (\ref{eq:cre}), which
yields:
\[
\frac{\partial}{\partial t}
\left(\frac{\rho v^2}{2} + \rho \Phi + \frac{p}{\gamma-1} 
+ \frac{p_{\rm cr}}{\gamma_{\rm cr} - 1} \right)
\]
\[
\,+\, \nabla \cdot
\left( \frac{\rho \mbox{\boldmath$v$} v^2}{2} + \rho \bm{v} \Phi
+\frac{\gamma \bm{v} p}{\gamma-1} + \frac{\gamma_{\rm cr} \bm{v}p_{\rm cr}}{
\gamma_{\rm cr} -1} + \bm{\Gamma}_{\rm visc} - 
{\bf \mathsf{\kappa}} \cdot \nabla T - 
\frac{{\bf \mathsf{D}} \cdot \nabla p_{\rm cr}}{\gamma_{\rm cr} - 1}\right)
\]
\begin{equation} 
= \,\rho \frac{\partial \Phi}{\partial t} - R + \dot{E}_{\rm source},
\label{eq:te} 
\end{equation} 
where $\Gamma_{\rm visc}$ is the viscous energy flux, and
where we have made use of the relation $H_{\rm visc} -
(\nabla \cdot \Pi_{\rm visc}) \cdot \bm{v} = - \nabla \cdot
\Gamma_{\rm visc}$.\footnote{The fact that the viscous terms
  can be written as a total divergence reflects
  the fact that viscosity is neither a source of 
energy nor a sink of energy, but instead merely converts
bulk-flow energy into thermal
  energy. Thus, when equation~(\ref{eq:te}) is integrated
  over volume, Gauss's law can be used to express the
  viscous terms as a surface integral, which vanishes if the
  boundary of the integration lies outside the plasma.}  We
assume that the viscous energy flux is much less than the
advective energy flux and thus drop $\Gamma_{\rm visc}$. We
also set $\partial \Phi/\partial t = 0$.

The average of equation~(\ref{eq:cont}) can be written
\begin{equation}
 \langle v_r \rangle  = \,-\, \frac{1}{\langle \rho \rangle}
\left(Q  + \frac{\dot{M}}{4\pi r^2}\right),
\label{eq:valvr}
\end{equation}
where
\begin{equation}
Q \equiv \langle \delta \rho \delta v_r \rangle,
\label{eq:defQ}
\end{equation}
and the mass accretion rate
$\dot{M} = -4 \pi r^2 \langle \rho v_r \rangle$ is
a constant.
The average of equation~(\ref{eq:momentum}) yields
\begin{equation}
\frac{d}{dr}\langle p_{\rm tot}\rangle = -\langle \rho\rangle \frac{d\Phi}{dr},
\label{eq:heq2} 
\end{equation} 
where
\begin{equation}
p_{\rm tot} = p + p_{\rm cr}
\label{eq:defptot2} 
\end{equation} 
is the total pressure. In writing equation~(\ref{eq:heq2}),
we have taken the convection to be subsonic, so that
the Reynolds stress can be neglected.  We have also
dropped the viscous stress, which is 
unimportant in the averaged equation.
The average of equation~(\ref{eq:cre}) can be written
\[
D_{\rm cr} \frac{d^2 \langle p_{\rm cr}
\rangle }{dr^2} = 
\frac{(1-\gamma) }{r^2}\frac{d}{dr}\left(r^2 F\right)
+ (1-\gamma_{\rm cr}) \left( W + \langle \dot{E}_{\rm source}
\rangle\right)
- \frac{d\langle p_{\rm cr}\rangle}{dr} \left[\frac{1}{r^2} \frac{d}{dr}
\left(r^2 D_{\rm cr}\right)\right]
\]
\begin{equation}
+\frac{ \gamma_{\rm cr}\langle p_{\rm cr}\rangle}{r^2}\frac{d}{dr}
\left(r^2\langle v_r\rangle
\right) + \langle v_r\rangle \frac{d\langle p_{\rm cr}\rangle}{dr},
\label{eq:avcre} 
\end{equation}
where 
\begin{equation}
F \equiv \frac{\langle \delta v_r \delta p\rangle}{\gamma - 1},
\label{eq:defF} 
\end{equation}
and
\begin{equation}
W \equiv \langle \delta p  \nabla
\cdot \delta \bm{v} \rangle.
\label{eq:defW} 
\end{equation} 
In writing equation~(\ref{eq:avcre}), we
have again made use of the fact that the convection is subsonic,
which implies that
 the total-pressure fluctuation is very small. As a result,
we can set $\delta p \simeq - \delta p_{\rm cr}$,  $
\langle \delta v_r \delta p_{\rm cr} \rangle
= - \langle \delta v_r \delta p\rangle$,
and  $\langle \delta p_{\rm cr}  \nabla
\cdot \delta \bm{v} \rangle
= - \langle \delta p  \nabla
\cdot \delta \bm{v} \rangle$.
The average of
equation~(\ref{eq:te}) yields
\[
\kappa_T \frac{d^2 \langle T\rangle}{dr^2 } 
= Q \frac{d\Phi}{dr} 
+ \frac{\gamma p}{(\gamma-1) r^2}\frac{d}{dr}
\left(r^2 \langle v_r \rangle \right)
- \frac{\langle v_r\rangle}{\gamma-1} \left(
\rho \frac{d\Phi}{dr} + \frac{dp_{\rm cr}}{dr}\right)
+ \frac{1}{r^2}\frac{d}{dr}\left(r^2 F\right)
\]
\begin{equation}
+ W + R - \frac{d\langle T\rangle}{dr} \left[
\frac{1}{r^2}\frac{d}{dr}\left(r^2 \kappa_T \right)\right],
\label{eq:avte} 
\end{equation}
where we have dropped the $\langle \rho \bm{v} v^2/2\rangle$ term since
it is much smaller than the other terms for subsonic convection.
In equation~(\ref{eq:avte}), $\kappa_T$ is the
average (isotropic) thermal conductivity given
by equations~(\ref{eq:kappas}) and~(\ref{eq:kappaT}), where
$T$ is set equal to~$\langle T\rangle$ in
equation~(\ref{eq:kappas}).

To solve equations~(\ref{eq:valvr}), (\ref{eq:heq2}),
(\ref{eq:avcre}), and (\ref{eq:avte}) for 
$\langle \rho \rangle$, $\langle v_r \rangle$, $
\langle T\rangle$, and~$\langle p_{\rm cr} \rangle$, we need
to express the quantities $Q$, $F$, and $W$ in terms of
these average fluid quantities, thereby closing the
equations. We accomplish this by using a two-fluid mixing
length theory.  Our approach is to first estimate $Q$,
$F$, and~$W$ using a local mixing length theory.  In the
local theory, the properties of the turbulence at some
radius are determined only by the average fluid properties
and gradients at that radius. We then use this local theory
as the basis for a non-local mixing length theory, as in
paper~II.  In the nonlocal theory, the
properties of the turbulence at some radius~$r$ are
determined by a weighted average of the turbulence
properties in the local mixing length theory over a range of
radii.  Our mixing-length theory differs from stellar
mixing-length theory (Cox \& Giuli 1968) in two important
ways: we include a cosmic-ray fluid, and we take into
account the fact that the diffusion of 
heat and cosmic rays occurs almost entirely along
magnetic field lines.

We derive quantities in the local mixing length theory as follows.  We
take the convective turbulence to have a correlation length~$l$, also
called the mixing length, where
\begin{equation}
l = \alpha r,
\label{eq:defl} 
\end{equation} 
$\alpha$ is a constant, and~$r$ is the distance from the center of the
cluster. Fluid parcels in convective regions are taken to rise or sink
a distance~$l$ before breaking up and mixing into the surrounding
plasma.  We take~$l$ to be much smaller than the pressure scale
height, so that~$\alpha$ is treated as~$\ll 1$.  (However, as in
standard mixing-length theory, after the mixing-length-theory
equations are derived, we relax the requirement that $\alpha \ll 1$,
and set~$\alpha = 0.4 $ when applying the model to clusters in
section~\ref{sec:comp}.)  We treat the fluctuations as small
quantities, and take $\left\langle \sqrt{(\delta \rho)^2}\right
\rangle/\langle \rho\rangle$, $\left\langle \sqrt{(\delta T)^2}\right
\rangle/\langle T\rangle$, $\left\langle \sqrt{(\delta p_{\rm
    cr})^2}\right \rangle/\langle p_{\rm cr}\rangle$, and
$\left\langle \sqrt{|\delta \bm{v}|^2}\right \rangle/c_{\rm s}$ to
be~$\sim {\cal O}(\alpha)$ (meaning of order~$\alpha$), where $c_{\rm
  s}$ is the sound speed.  We then expand the equations in powers
of~$\alpha$, and keep only the lowest-order non-vanishing terms in
this expansion.  The quantities $Q$, $F$, and~$W$ involve products of
fluctuating quantities, and are thus~$\sim {\cal O}(\alpha^2)$.  We
take $R$, $\langle \dot{E}_{\rm source}\rangle$, $\kappa_T$,
$\kappa_\parallel$, $D_{\rm cr}$, $D_\parallel$, and $H_{\rm visc}$ to
be~$\sim {\cal O}( \alpha^2)$, so that, e.g., radiative cooling,
conduction, and turbulent heating are of the same order in~$\alpha$ in
equations~(\ref{eq:avcre}) and~(\ref{eq:avte}).  (Since the
fluctuations are small, we can write, e.g., that $ \langle p \rangle
\simeq k_B \langle \rho\rangle \langle T \rangle/ \mu m_H$, where
$\mu$ is the mean molecular weight.)  There are two contributions to
the average velocity~$\langle \bm{v} \rangle$.  One is driven by the
turbulent fluctuations, and, as will be seen below, is of
order~$\alpha^2$. The second arises from the net inflow of mass
towards the center of the cluster. In our model, the mass accretion
rate is set by the Bondi accretion rate calculated from the plasma
parameters at the inner radius~$r_1 = 0.2$~kpc, as described in
section~\ref{sec:model}.  We treat this second contribution to
$\langle \bm{v}\rangle$ as also of order~$\alpha^2$.

We now proceed to estimate typical values for~$\delta T$, $\delta
\rho$, and~$\delta p_{\rm cr}$ to first order in~$\alpha$ - i.e.,
ignoring terms of order~$\alpha^2$.  Because we take the fluid
displacement and the correlation length to be comparable, we need to
use a Lagrangian approach to formally integrate the equations.  We
take the initial position of a fluid element at time~$t=0$ to be
denoted~$\bm{r}_0$, and its position at time~$t$ to be
\begin{equation}
\bm{r}(\bm{r}_0,t) = \bm{r}_0 + \bm{\xi}(\bm{r}_0,t),
\label{eq:defxi} 
\end{equation} 
where~$\bm{\xi}$ is the displacement of the fluid
element.  We use the shorthand notation that $\rho(t)
$, $T(t)$, $p(t)$, $p_{\rm cr}(t)$, and $\bm{\xi}(t)$
are the density, temperature, pressure, cosmic-ray
pressure, and displacement at time~$t$ of the fluid
element that started at position~$\bm{r}_0$ at $t=0$.
The velocity at time~$t$ of the fluid element that
starts at~$\bm{r}_0$ at~$t=0$ is given by
\begin{equation}
\bm{v} = \left.\frac{\partial \bm{r}}{\partial t}\right|_{r_0}
= \left.\frac{\partial \bm{\xi}}{\partial t}\right|_{r_0}.
\label{eq:lagvel} 
\end{equation}
We then have that
\begin{equation}
\left.\frac{\partial }{\partial t}\right|_{r_0}
= \left.\frac{\partial }{\partial t}\right|_{r}
+ \bm{v} \cdot \nabla,
\label{eq:lagder} 
\end{equation}
where the spatial derivatives on the right-hand side are
with respect to~$\bm{r}$, not~$\bm{r}_0$.
The Jacobian
matrix~$\underline{\underline{J}}$
 for the transformation from~$\bm{r}_0$
to~$\bm{r}$ is given by the equation
\begin{equation}
J_{ij} = \frac{\partial r_i}{\partial r_{0j}}.
\label{eq:defJ} 
\end{equation}
The determinant of this matrix,
denoted~$J$, satisfies the equation
\begin{equation}
\left.\frac{\partial}{\partial t}\ln J \right|_{r_0}
= \nabla \cdot \bm{v},
\label{eq:djdt} 
\end{equation}
where the spatial derivatives on the right-hand side are again with
respect to~$\bm{r}$, not~$\bm{r}_0$.\footnote{Equation~\ref{eq:djdt}
  can be shown as follows. We define $M_{ij}$ to be the determinant of
  the~$2\times2$ matrix obtained by deleting the~$i^{\rm th}$ row
  and~$j^{\rm th}$ column of the matrix~$\underline{\underline{J}}$.
  We can then write that 
$\partial J/\partial J_{ij} = (-1)^{i+j}M_{ij}$. The inverse of
$\underline{ \underline{ J}}$, which we denote $\underline{
\underline{ B}}$, satisfies $B_{ij} J_{jk} = \delta_{ik}$, and is given by
 $B_{ij} = (-1)^{i+j}  M_{ji}/J$.  We can thus write $\left.\partial
  J/\partial t\right|_{r_0} = \left. (\partial J/\partial
  J_{ij})(\partial J_{ij}/\partial t) \right|_{r_0}  = J B_{ji}
  \left.(\partial J_{ij}/\partial t)\right|_{r_0} $.  From the chain
  rule, $(\partial {r_{0i}}/\partial r_j) (\partial r_j/\partial
  r_{0k}) = \delta_{ik}$. Thus, $B_{ij} = \partial r_{0i}/\partial
  r_j$, and $(1/J) \left.\partial J/\partial t \right|_{r_0} =
  (\partial r_{0j}/\partial r_i)[\left.  (\partial/\partial
  t)\right|_{r_0} (\partial r_i/\partial r_{0j})] = (\partial
  r_{0j}/\partial r_i) (\partial v_i/\partial r_{0j}) = \partial
  v_i/\partial r_i.$} Using equations~(\ref{eq:lagder})
and~(\ref{eq:djdt}), we rewrite equation~(\ref{eq:cont}) as
\begin{equation}
\left.\frac{\partial \ln \rho}{\partial t}\right|_{r_0}
 = - \left.\frac{\partial \ln J}{\partial t}\right|_{r_0}.
\label{eq:lagcont} 
\end{equation}
We integrate equation~(\ref{eq:lagcont}) in time from $t=0$
to $t=  \Delta t $ holding $\bm{r}_0$ fixed (as in the time
integrals below), where
\begin{equation}
\Delta t = \frac{l}{u_{\rm L}}
\label{eq:defdt} 
\end{equation} 
is the ``mixing time,''
and $u_L$ is the rms radial velocity
in the local mixing length theory.
We set $\bm{\xi}(0) = 0$ and 
$\rho(0) = \langle
\rho \rangle$, where $\langle \rho \rangle$ is shorthand
notation for the average density at~$r_0$. We then
obtain
\begin{equation}
\frac{\rho (\Delta t)}{\langle \rho\rangle} =
\frac{1 }{J(\Delta t)},
\label{eq:lagrho0} 
\end{equation}
where $J(\Delta t)$ is the Jacobian at~$t=\Delta t$
evaluated for the initial position~$\bm{r}_0$. If we
start at~$t=0$
with a fluid element of infinitesimal volume~$d^3 r_0$ 
centered at~$\bm{r}_0$, then at time~$\Delta t$ its volume
is~$d^3 r = J(\Delta t) d^3 r_0$. Thus,
equation~(\ref{eq:lagrho0}) is a statement
of mass conservation.
The Lagrangian density perturbation at
time~$\Delta t$ is
\begin{equation}
\Delta \rho_{\rm Lag} \equiv \rho(\Delta t) - \rho(0) 
 = \rho(\Delta t) - \langle \rho \rangle .
\label{eq:defDrho}
\end{equation}
Combining equations~(\ref{eq:lagrho0}) and~(\ref{eq:defDrho}), we
write
\begin{equation}
\frac{\Delta \rho_{\rm Lag}}{\langle \rho\rangle}
= \frac{1}{J(\Delta t)} - 1.
\label{eq:valDrho}
\end{equation}

To solve for the pressure fluctuation, we write
equation~(\ref{eq:pe})  in the form
\begin{equation}
\left.\frac{\partial p}{\partial t}\right|_{r_0}
 = - \gamma p \left.\frac{\partial \ln J}{\partial t}\right|_{r_0}
+ (\gamma - 1)  \delta H_{\rm tc},
\label{eq:dplag1} 
\end{equation} 
where 
\begin{equation}
H_{\rm tc} = \nabla \cdot (\kappa_\parallel \hat{b}\hat{b} \cdot \nabla T)
\label{eq:defhtc} 
\end{equation} 
is the rate of heating due to thermal conduction,
and $\delta H_{\rm tc}$ is the deviation
of~$H_{\rm tc}$ from its average
value. In writing equation~(\ref{eq:dplag1}), we have dropped terms
of order~$\alpha^2$.
[$\delta H_{\rm tc} \sim {\cal O}(\alpha)$
since $\delta T \sim {\cal O}(\alpha)$,
$\kappa_\parallel \sim {\cal O}(\alpha^2)$, and
$\nabla^2 \delta T \sim \delta T/l^2 \sim
{\cal O}(\alpha^{-1})$.]
We integrate equation~(\ref{eq:dplag1})   from
$t=0$ to $t= \Delta t$, setting 
$\bm{\xi}(0) = 0$ and
$p(0) = \langle p\rangle$, where $\langle p \rangle $ is
the average density at~$r_0$.
To first order in~$\alpha$, we can 
replace $\gamma p \left. \partial \ln J/\partial t\right|_{r_0}$
with $\gamma \langle p\rangle
 \left. \partial \ln J/\partial t\right|_{r_0}$.
We thus obtain
\begin{equation}
\Delta p_{\rm lag} = - \gamma \langle p\rangle \ln J(\Delta t)
 + (\gamma-1)\int_0^{\Delta t}
\delta H_{\rm tc}(t) dt,
\label{eq:Dp1}
\end{equation}
where
\begin{equation}
\Delta p_{\rm Lag} \equiv p(\Delta t)  - p(0) 
= p(\Delta t) - \langle p\rangle,
\label{eq:defDp}
\end{equation}
is the Lagrangian pressure perturbation at time~$\Delta t$.
Because the density fluctuations are small, the value of
$J(\Delta t)$ is very close to 1. 
Writing $J(\Delta t) = 1 + x$, the quantity~$x$ is
of order~$\alpha$. Thus, $ \ln J(\Delta t)  = x + {\cal O}(\alpha^2)
= 1 - [J(\Delta t)]^{-1} +{\cal O}(\alpha^2) $
and
\begin{equation}
 \ln J(\Delta t) = - \frac{\Delta \rho_{\rm Lag}}
{\langle \rho\rangle}
 + {\cal O}(\alpha^2).
\label{eq:logJ}
\end{equation}
To first order in~$\alpha$, 
we can thus rewrite equation~(\ref{eq:Dp1}) as
\begin{equation}
\Delta p_{\rm lag} 
=
\frac{\gamma \langle p\rangle \Delta \rho_{\rm Lag}}{\langle
\rho \rangle}
 + (\gamma-1)\int_0^{\Delta t}
\delta H_{\rm tc}(t) dt.
\label{eq:Dp2}
\end{equation}

\begin{figure}[t]
\includegraphics[width=3in]{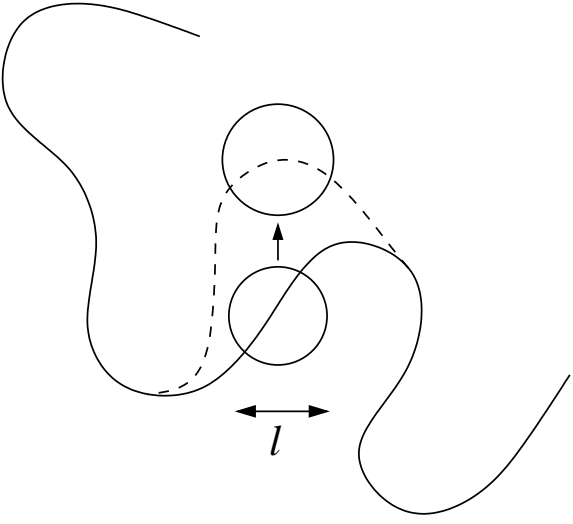}
\caption{\footnotesize Schematic diagram of a
  rising fluid parcel.  The solid line is a
  magnetic field line passing through the parcel's
  initial location. The dashed line is an
  idealization of how the field line changes as a
  result of the fluid parcel's displacement.
\label{fig:f7}}
\end{figure}

We have been analyzing a fluid element that starts off
at~$t=0$ as an ``average'' fluid element, and we thus
take $\delta T=0$ and $\delta H_{\rm tc} =0$ at~$t=0$.
When this fluid element is displaced radially outwards
a distance~$l$, it remains magnetically connected to
the same set of fluid elements to which it was
initially connected (at least until it is mixed into
the surrounding fluid, at which point it is assumed
that the magnetic field in the fluid parcel is
randomized).  This is depicted schematically in
figure~\ref{fig:f7}.  If its temperature remains
unchanged as it moves [i.e., if $\delta T_{\rm Lag}
\equiv T(t) - T(0) =0$], then the effects of thermal
conduction are unchanged and~$\delta H_{\rm tc}$ will
remain zero.  However, if its temperature decreases,
then more heat will be conducted into the fluid
element, and $\delta H_{\rm tc} $ will increase.
Thermal conduction will thus act to restore the
temperature to its initial value (i.e. to keep~$\delta
T_{\rm Lag}=0$), with $\delta H_{\rm tc} \propto
-\delta T_{\rm Lag}$ for small values of~$\delta T_{\rm
  Lag}$.  To estimate~$\delta H_{\rm tc}$, we note that
the mixing length~$l$ is comparable to the correlation
lengths of both the convective turbulence and the
temperature fluctuations.  Thus, $|\nabla \delta T_{\rm
  Lag}| \sim \delta T_{\rm Lag}/l$, and we have the
order-of-magnitude relation
\begin{equation}
\delta H_{\rm tc} \sim - \frac{ \kappa_\parallel
 \delta T_{\rm Lag}}{l_T^2}.
\label{eq:defHtc} 
\end{equation} 
We estimate that
\begin{equation}
(\gamma -1) \int_0^{\Delta t}  \delta H_{\rm tc} \,dt
= - \frac{0.3 (\gamma -1) \kappa_\parallel
\Delta T_{\rm Lag} \Delta t}{l^2},
\label{eq:dhtc2} 
\end{equation} 
where $\Delta T_{\rm Lag}
= \delta T_{\rm Lag}(\Delta t)$ is the Lagrangian temperature perturbation
at time~$ \Delta t$, and the numerical factor of~0.3 is chosen
somewhat arbitrarily to reflect (1) our expectation that the length
scale of the temperature fluctuations is somewhat larger than~$l$,
which is just the radial component of the displacement vector, not the
full modulus of~$\bm{\xi}$, and (2) the fact that $\delta H_{\rm tc}$
 increases from zero to its maximum value as $t$ ranges
from~0 to~$\Delta t$, so the typical value of $\delta H_{\rm tc}$ is
less than its value at~$t=\Delta t$.  Finally, we obtain the relation
\begin{equation}
\Delta p_{\rm lag} = \frac{ \gamma \langle p\rangle \Delta \rho_{\rm
Lag}}{\langle \rho \rangle}
 - \, \frac{0.3 (\gamma -1) \kappa_\parallel
\Delta T_{\rm Lag} \Delta t}{l^2}.
\label{eq:dp3.5} 
\end{equation} 
To first order in~$\alpha$,
\begin{equation}
\frac{\Delta T_{\rm Lag}}{\langle T \rangle} 
= \frac{\Delta p_{\rm Lag}}{\langle p \rangle }
 - \frac{\Delta \rho_{\rm Lag}}{\langle \rho \rangle},
\label{eq:lin} 
\end{equation} 
where $\langle T \rangle$ is the average temperature
at~$r_0$.
Equations (\ref{eq:dp3.5}) and~(\ref{eq:lin}) combine to give
\begin{equation}
\Delta p_{\rm Lag} = \Delta \rho_{\rm Lag} \frac{\langle
p \rangle}{\langle \rho \rangle}
\left(\frac{\gamma + a_1 }{1 + a_1}\right),
\label{eq:dp4} 
\end{equation} 
where
\begin{equation} 
a_1 = \frac{0.3(\gamma-1)\kappa_\parallel \langle
T\rangle}{lu_{\rm L} \langle p\rangle }
\label{eq:defa1} 
\end{equation} 
is roughly the ratio of the mixing time
$\Delta t$ to the time for
heat to diffuse a distance~$l$ along the magnetic
field. When~$a_1\ll 1$ the thermal plasma expands
adiabatically, and when~$a_1 \gg 1$ the thermal
plasma expands isothermally.

To solve for the cosmic-ray pressure fluctuation, we write
equation~(\ref{eq:cre})  in the form
\begin{equation}
\left.\frac{\partial p_{\rm cr}}{\partial t}\right|_{r_0}
 = - \gamma_{\rm cr} p_{\rm cr}
 \left.\frac{\partial \ln J}{\partial t}\right|_{r_0}
+  \delta H_{\rm diff} + (\gamma_{\rm cr} - 1) \delta \dot{E}_{\rm source},
\label{eq:dpcrlag1} 
\end{equation} 
where 
\begin{equation}
H_{\rm diff} = \nabla \cdot (D_\parallel \hat{b}\hat{b} \cdot \nabla p_{\rm cr}),
\label{eq:defhdiff} 
\end{equation} 
$\delta H_{\rm diff}$ is the deviation of~$H_{\rm diff}$ from its
average value, $\dot{E}_{\rm source}$ is the cosmic-ray energy per
unit volume generated by the central radio source, and
$\delta\dot{E}_{\rm source}$ is the deviation of $\dot{E}_{\rm
  source}$ from its average value. 
As discussed in section~\ref{sec:cra}, $\delta \dot{E}_{\rm source}$
can be significantly larger than $\langle \dot{E}_{\rm source}\rangle$
if the cosmic rays are accelerated in a small fraction of the volume.
We thus treat $\delta \dot{E}_{\rm source}$ as $\cal{O}(\alpha)$
and $\langle \dot{E}_{\rm source}\rangle$ as ${\cal O}(\alpha^2)$.
We integrate
equation~(\ref{eq:dpcrlag1}) from $t=0$ to $t= \Delta t$, setting
$\bm{\xi}(0) = 0$ and $p_{\rm cr}(0) = \langle p_{\rm
  cr}\rangle$, where $ \langle p_{\rm cr}\rangle$ is the average
cosmic-ray pressure at~$r_0$. To first order in~$\alpha$, we can
replace $\gamma_{\rm cr} p_{\rm cr} \left. \partial \ln J/\partial
t\right|_{r_0}$ with $\gamma_{\rm cr} \langle p_{\rm
  cr}\rangle \left. \partial \ln J/\partial
t\right|_{r_0}$.  Using equation~(\ref{eq:logJ}), we obtain
\begin{equation}
\Delta p_{\rm cr, Lag} = \frac{\gamma_{\rm cr} \langle p_{\rm cr}
\rangle
\Delta \rho_{\rm Lag}}{\langle \rho\rangle}
 + \int_0^{\Delta t}
\left[\delta H_{\rm diff} + (\gamma_{\rm cr} - 1) \delta \dot{E}_{\rm source}\right] dt,
\label{eq:Dpcr1}
\end{equation}
where
\begin{equation}
\Delta p_{\rm cr, Lag} = p_{\rm cr}(\Delta t) 
- \langle p_{\rm cr}\rangle,
\label{eq:defDpcr}
\end{equation}
is the Lagrangian cosmic-ray
pressure perturbation at time~$\Delta t$.

We treat parallel cosmic-ray diffusion 
in the same way as parallel
thermal conduction and make the estimate 
\begin{equation}
\int_0^{\Delta t}  \delta H_{\rm diff} \,dt
= - \frac{0.3 D_\parallel
\Delta p_{\rm cr,Lag} \Delta t}{l^2}.
\label{eq:dhdiff2} 
\end{equation} 
We assume that $\delta \dot{E}_{\rm source}$
is typically positive in outwardly moving fluid
elements and negative in inwardly moving fluid elements.
Here, we are focusing on a  a fluid element that 
is moving outwards (the discussion can be repeated with
little alteration for inwardly moving fluid parcels),
and thus we treat $\delta \dot{E}_{\rm source}$
as positive. We then make the estimate that 
\begin{equation}
\int_0^{\Delta t}  \delta\dot{E}_{\rm source} \,dt
= \delta \dot{E}_{\rm rms} \Delta t,
\label{eq:ds1} 
\end{equation} 
where $\delta \dot{E}_{\rm rms}$ is the rms value
of $\delta \dot{E}_{\rm source}$, which is
determined from
equation~(\ref{eq:eta2}). Substituting
 (\ref{eq:dhdiff2}),
and~(\ref{eq:ds1}) into
equation~(\ref{eq:Dpcr1}), we obtain
\begin{equation}
\Delta p _{\rm cr,Lag} = \frac{\gamma_{\rm cr}
\langle p_{\rm cr}\rangle \Delta \rho_{\rm Lag}}{
(1 + a_2)\langle \rho\rangle} 
+ (\gamma_{\rm cr}-1)\delta \dot{E}_{\rm rms} \tau,
\label{eq:Dpcr2} 
\end{equation} 
where
\begin{equation}
\tau = \left(\frac{u_L}{l } + \frac{0.3 D_\parallel }{l^2}
\right)^{-1}
\label{eq:deftau} 
\end{equation} 
is the effective time during which cosmic rays
can accumulate in the fluid parcel as a result of
the cosmic-ray source term. ($\tau$ is roughly
the shorter of the mixing time~$l/u_L$
and the diffusion time $l^2/D_\parallel$.)
The quantity $a_2$ is given by
\begin{equation}
a_2 = \frac{0.3 D_\parallel}{lu_{\rm L}}
\label{eq:defa2} 
\end{equation} 
and is approximately the ratio of~$\Delta t$ to the time for cosmic
rays to diffuse a distance~$l$ along the magnetic field.  When $a_2
\ll 1$, the cosmic rays expand adiabatically if $\delta \dot{E}_{\rm
  rms} = 0$.  When $a_2 \gg 1$ the cosmic ray pressure in the fluid
element remains constant as the element is displaced if $\delta
\dot{E}_{\rm rms} = 0$, as in the linear Parker instability in the
large-$D_\parallel$ limit (Parker 1966, Shu 1974, Ryu et~al~2003).

Since it is assumed that the convection is
subsonic, the total pressure in the fluid element
remains approximately the same as the average
total pressure in the fluid element's
surroundings. We take 
\begin{equation}
\hat{r} \cdot \bm{\xi}(\Delta t) = l,
\label{eq:rxi}
\end{equation}
and thus to first order in~$\alpha$
\begin{equation} 
\Delta p_{\rm Lag} + \Delta p_{\rm cr,Lag}
 = l\frac{d}{dr}\langle p_{\rm tot}\rangle.
\label{eq:valptot} 
\end{equation} 
Adding
equations~(\ref{eq:dp4}) and (\ref{eq:Dpcr2}) and
using equation~(\ref{eq:valptot}), we obtain
\begin{equation}
l \frac{d}{dr}\langle p_{\rm tot}\rangle=
c_{\rm eff}^2 \Delta \rho_{\rm Lag} + (\gamma_{\rm cr}
- 1) \delta \dot{E}_{\rm rms}\tau,
\label{eq:drho1} 
\end{equation} 
where
\begin{equation}
c_{\rm eff} = \left[\left(\frac{\gamma + a_1}
{1 + a_1}\right)
\frac{\langle p\rangle}{\langle \rho\rangle}
 + \frac{\gamma_{\rm cr}
\langle p_{\rm cr}\rangle }{(1 + a_2)\langle \rho\rangle}
\right]^{1/2}
\label{eq:defceff} 
\end{equation} 
is an effective sound speed for the medium.

The fluctuating quantities appearing in
equations~(\ref{eq:defQ}), (\ref{eq:defF}),
and~(\ref{eq:defW}) are Eulerian fluctuations,
in that they involve
the difference between some quantity and the
average of that quantity at the same location.  To
first order in~$\alpha$, we can write the Eulerian
density fluctuation of our outwardly displaced fluid
element at time~$\Delta t$, denoted $\Delta \rho$, as
\begin{equation}
\Delta \rho = \Delta \rho_{\rm Lag} - l \frac{d\langle \rho \rangle}
{dr}.
\label{eq:drhoE}
\end{equation}
Combining equations (\ref{eq:drho1}) and~(\ref{eq:drhoE}), we find that 
\begin{equation}
\Delta \rho = l \left(\frac{1}{c_{\rm eff}^2}
\frac{d\langle p_{\rm tot}\rangle}{dr} - \frac{d
\langle \rho\rangle }{dr}\right)
- \frac{(\gamma_{\rm cr} - 1) \delta \dot{E}_{\rm rms}
\tau}{c_{\rm eff}^2}.
\label{eq:drho}
\end{equation} 
The fluid is convectively stable if an outwardly displaced
parcel is heavier than its surroundings (i.e.,
if $\Delta \rho >0$) for any value of $u_L$.
We note that as $u_L$ increases, $a_1$ and $a_2$
decrease, $c_{\rm eff}^2$ increases, and $\tau$
decreases. Since $d\langle p_{\rm tot}\rangle /dr<0$, 
it follows that
\begin{equation}
\frac{d}{du_L}\Delta \rho > 0.
\end{equation} 
Thus, if $\Delta \rho >0$ as $u_L\rightarrow 0$,
then $\Delta \rho >0$ for any $u_L$ ($u_L$ is by
definition non-negative), and the medium is
convectively stable.  On the other hand, if
$\Delta \rho <0$ as $u_L \rightarrow 0$, then the
medium is convectively unstable. The necessary and
sufficient condition for convective stability is
thus that $\Delta \rho$ be positive in the limit
$u_L\rightarrow 0$. As $u_L\rightarrow 0$, we
have that $a_1\rightarrow \infty$, $a_2\rightarrow
\infty$, $\tau \rightarrow l^2/(0.3 D_\parallel)$,
and $c_{\rm eff}^2 \rightarrow \langle p
\rangle /\langle \rho\rangle$.
For constant mean molecular weight~$\mu$, this then
leads to the stability criterion
\begin{equation}
l\left(n k_B \frac{dT}{dr} + \frac{dp_{\rm cr}}{dr}\right)
- \frac{(\gamma_{\rm cr} - 1) \delta \dot{E}_{\rm rms} l^2}{0.3 D_\parallel} > 0,
\label{eq:stabcrit0} 
\end{equation} 
where $n = \rho/(\mu m_H)$ is the number density
of thermal particles. If one sets $\delta \dot{E}_{\rm rms}$
to zero, then equation~(\ref{eq:stabcrit0}) reduces
to the stability criterion derived by
Chandran (2005) and Chandran \& Dennis (2006).
Here, we have kept the fluctuations in $\dot{E}_{\rm source}$,
which act to destabilize
the medium to convection, since localized
excesses in the cosmic-ray pressure lead to
pockets of buoyant, lower-density fluid.

If the convective stability criterion is
satisfied at some radius, 
we set the local convective velocity~$u_{\rm
  L}$ to zero at that
radius.  Otherwise the fluid
is convectively unstable, and we estimate
the value of~$u_{\rm
  L}$ by solving the 
polynomial equation
\begin{equation}
\frac{\langle \rho \rangle
u_{\rm L}^2}{2} = \left| \frac{l\Delta \rho}{16} \frac{d\Phi}{dr}\right|.
\label{eq:eqnu} 
\end{equation} 
Equation~(\ref{eq:eqnu}) states that the mean radial kinetic
energy of the fluid element is the mixing length times the
buoyancy force on the fully displaced parcel times the
numerical factor of~$1/16$ that is commonly used in
one-fluid mixing length theory (Cox \& Giuli 1968).
Once~$u_{\rm L}$ is found, 
we determine $\Delta \rho_{\rm Lag}$ and $\Delta
p_{\rm Lag}$ using
equations~(\ref{eq:drho1}) and (\ref{eq:dp4}), respectively.
The Eulerian pressure perturbation at time~$\Delta t$,
denoted~$\Delta p$, is
then given by  the equation
\begin{equation}
\Delta p = \Delta p_{\rm Lag} - l \frac{d\langle p \rangle}
{dr}.
\end{equation}
We then estimate the quantity~$F $ in equation~(\ref{eq:defF})
to be
\begin{equation}
F_{\rm L} = \frac{c_{\rm avg} u_{\rm L} \Delta p}{\gamma-1}.
\end{equation} 
Here, as below, the L subscript is used to denote
the estimate obtained using local mixing length theory.
We set 
\begin{equation}
c_{\rm avg} = 1/2
\end{equation} 
to match
standard treatments of local one-fluid mixing length
theory (Cox \& Giuli 1968).
We estimate the quantity~$Q$ in equation~(\ref{eq:defQ}) to be
\begin{equation}
Q_L = c_{\rm avg} u_L \Delta \rho,
\end{equation}
with $\Delta \rho$ determined from
equation~(\ref{eq:drho}).
We estimate the quantity~$W$ in equation~(\ref{eq:defW})
by noting that
$\displaystyle
\int_0^{\Delta t} \nabla \cdot \bm{v}dt = - \ln \left[
\frac{\rho(\Delta t)}{\langle \rho\rangle}\right] = - \ln\left[
1 + \frac{\Delta \rho_{\rm Lag}}{\langle \rho\rangle}\right]
\simeq - \frac{\Delta \rho_{\rm Lag}}{\langle \rho \rangle}.$
Thus, the typical value of~$\nabla \cdot \delta
\bm{v}$ is $\displaystyle
\frac{1}{\Delta t} \left( - \frac{\Delta \rho_{\rm Lag}}{\langle
\rho\rangle}\right) = - \frac{u \Delta \rho_{\rm Lag}}{l \langle \rho \rangle}$.
We thus set
\begin{equation}
W_{\rm L} = \,-\,
\frac{c_{\rm avg} u_L \Delta 
p \Delta \rho_{\rm Lag} }{l\langle \rho\rangle}.
\end{equation}

Having estimated~$Q$, $F$, and~$W$ using local mixing
length theory, we now use these estimates as the basis
for a nonlocal theory.
In the study of Ulrich~(1976),
the nonlocal heat flux is given by a weighted
spatial average of the heat flux obtained from local
mixing length theory. We adopt
the same approach and set
\begin{equation}
F_{\rm NL}(z) = \int_{-\infty}^{\infty} dz_1 F_{\rm L}(z_1) \psi_F(z-z_1),
\label{eq:fnl1} 
\end{equation} 
\begin{equation}
W_{\rm NL}(z) = \int_{-\infty}^{\infty} dz_1 W_{\rm L}(z_1) \psi_W(z-z_1),
\label{eq:wnl1} 
\end{equation} 
and 
\begin{equation}
Q_{\rm NL}(z) = \int_{-\infty}^{\infty} dz_1 Q_{\rm L}(z_1) \psi_Q(z-z_1),
\label{eq:unl1} 
\end{equation} 
where 
\begin{equation}
z = \ln \left(\frac{r}{r_{\rm ref}}\right),
\label{eq:defz} 
\end{equation} 
$r_{\rm ref}$ is an unimportant constant,
and the NL subscripts denote values in our nonlocal
theory.
Different forms for the kernel function~$\psi_F$
were considered by Ulrich~(1976). Here, we adopt the
following values:
\begin{equation}
\psi_Q(x) = \psi_F(x) = \left\{ 
\begin{array}{ll}
\alpha^{-1} e^{-x/\alpha} & \mbox{ \hspace{0.3cm} if $x>0$} \\
0 & \mbox{ \hspace{0.3cm} if $x\leq 0$}
\end{array}
\right.  ,
\label{eq:defpsifu} 
\end{equation} 
and 
\begin{equation}
\psi_W(x)  = \left\{ 
\begin{array}{ll}
\alpha_W^{-1} e^{-x/\alpha_W} & \mbox{ \hspace{0.3cm} if $x>0$} \\
0 & \mbox{ \hspace{0.3cm} if $x\leq 0$}
\end{array}
\right.  .
\label{eq:defpsiW} 
\end{equation} 
Equations (\ref{eq:fnl1}) through
(\ref{eq:defpsiW}) are equivalent to the
differential equations
\begin{equation}
\alpha r\,\frac{dF_{\rm NL}}{dr} + F_{\rm NL} = F_{\rm L},
\label{eq:fnl2} 
\end{equation} 
\begin{equation}
\alpha r\,\frac{Q_{\rm NL}}{dr} + Q_{\rm NL} = Q_{\rm L},
\label{eq:Qnl2} 
\end{equation} 
and
\begin{equation}
\alpha_W r\,\frac{W_{\rm  NL}}{dr} + W_{\rm  NL} = W_{\rm L}.
\label{eq:wnl2} 
\end{equation} 
For $r>r_{\rm conv}$, where $r_{\rm conv}$ is the
largest radius at which the fluid is locally
unstable to convection, $F_L=0$ and
$F_{\rm NL} \propto r^{-1/\alpha}$. When $F$
and~$W$ are set equal to~$F_{\rm NL}$
and~$W_{\rm NL}$ in equations~(\ref{eq:avcre})
and~(\ref{eq:avte}), the
terms containing~$F_{\rm NL}$ are~$\propto r^{-1
  -1/\alpha}$ for~$r>r_{\rm conv}$.  To obtain the
same scaling for the terms containing~$W_{\rm 
  NL}$, the value
of~$\alpha_W$ is determined from the equation
\begin{equation}
\alpha_W^{-1} = \alpha^{-1} + 1.
\label{eq:alphaw} 
\end{equation} 
Equations~(\ref{eq:defpsifu}) and (\ref{eq:defpsiW})
represent a one-sided average, in the sense that
the nonlocal quantities $F_{\rm NL}$, $Q_{\rm NL}$,
and $W_{\rm NL}$ depend only on
the values of~$F_L$, $Q_L$, and $W_L$ at smaller radii.  A more
sophisticated nonlocal theory could be developed along
different lines (see e.g.  Travis \& Matsushima 1973,
Ulrich~1976, Xiong~1991, Grossman, Narayan, \&
Arnett~1993), but is beyond the scope of this paper.

The final equations for our mixing-length model are then
equations~(\ref{eq:valvr}), (\ref{eq:heq2}), (\ref{eq:avcre}),
(\ref{eq:avte}), (\ref{eq:fnl2}), (\ref{eq:Qnl2}),
and~(\ref{eq:wnl2}), which form a system of two second-order ordinary
differential equations (ODEs), four first-order ODEs, and one
algebraic equation [equation~(\ref{eq:valvr})] for the seven variables
$\langle \rho \rangle$, $\langle v_r \rangle$ $\langle T\rangle$,
$\langle p_{\rm cr}\rangle$, $F_{\rm NL}$, $Q_{\rm NL}$, and $W_{\rm
  NL}$.  Eight boundary conditions are required to specify a
solution. Two boundary conditions are obtained by 
requiring that the model density and temperature match
the observed values $\rho_{\rm outer}$ and $T_{\rm outer}$
at the outer radius~$r_{\rm outer}$.
For seven of the eight clusters in our sample
(all except Virgo), we choose $r_{\rm outer}$ to be the center of the
first radial bin outside the cooling radius~$r_{\rm cool}$.
Values of $r_{\rm cool}$ for each cluster are given
in table~\ref{tab:t0}. For Virgo, we take $r_{\rm outer}$
to be the outermost data point, which lies inside of $r_{\rm cool}$.
Since we do not solve all the
way in to the sonic point, we are forced to pick inner boundary
conditions (at~$r_1 = 0.2$~kpc) in a somewhat arbitrary way. We take
$d\langle T\rangle/dr$, $d\langle p_{\rm cr} \rangle/dr$, $F_{\rm
  NL}$, $W_{\rm NL}$, and $Q_{\rm NL}$ to vanish at~$r=r_1$.  These
``no-flux'' boundary conditions set the diffusive and turbulent energy
fluxes to zero at the inner boundary. Although this choice is
undoubtedly inaccurate, we expect that it has only a small effect on
our solution for the structure of the intracluster medium at~$r\gg
r_1$. The eighth boundary condition is obtained by assuming
that~$\langle p _{\rm cr}\rangle \rightarrow 0$ as~$r \rightarrow
\infty$. This condition is translated into a condition on~$\langle
p_{\rm cr}\rangle$ at~$r_{\rm outer}$ as follows. The value of $r_{\rm
  outer}$ is chosen to be significantly greater than~$r_{\rm source}$
and much greater than~$D_0/v_d$, so that for $r> r_{\rm outer}$,
$\dot{E}_{\rm source}$ is negligible and $D_{\rm cr} \simeq v_d
r$.\footnote{The case $v_d=0$ requires a different approach and is not
  treated in this paper.}  In addition, $r_{\rm outer}$ is taken to
lie outside $r_{\rm conv}$, the largest radius at which the
intracluster medium is locally convectively unstable, so that $F_{\rm
  NL} = F_{\rm outer} (r/r_{\rm outer})^{-1/\alpha}$ and $W_{\rm NL} =
W_{\rm outer} (r/r_{\rm outer})^{-1 - 1/\alpha}$ for $r>r_{\rm
  outer}$, where $F_{\rm outer}$ and $W_{\rm outer}$ are the values of
$F_{\rm NL}$ and $W_{\rm NL}$ at $r=r_{\rm outer}$. We also take
$\langle v_r \rangle$ to be negligible for~$r>r_{\rm outer}$. This
latter assumption is reasonable, since $Q_{\rm NL} = Q_{\rm outer}
(r/r_{\rm outer})^{-1/\alpha}$ for~$r>r_{\rm conv}$, where~$Q_{\rm
  outer}$ is the value of~$Q_{\rm NL}$ at~$r_{\rm outer}$. The
resulting value of~$\langle v_r\rangle$ is significantly less
than~$v_d$ for $r>r_{\rm conv}$ in the numerical solutions we present
in section~\ref{sec:comp}, and thus~$\langle v_r\rangle$ plays only a
small role in equation~(\ref{eq:avcre}) at~$r>r_{\rm conv}$.  Solving
equation~(\ref{eq:avcre}) and requiring that $\langle p _{\rm
  cr}\rangle\rightarrow 0$ as $r\rightarrow \infty$, we find that for
$r\geq r_{\rm outer}$
\begin{equation}
\frac{d\langle p_{\rm cr}\rangle}{dr}  = \frac{\chi}{v_d r^{1 + 1/\alpha}} - \frac{2 \langle p_{\rm cr}\rangle}{r},
\label{eq:bc8} 
\end{equation} 
where
\begin{equation}
\chi = (2\alpha -1) (\gamma-1) F_{\rm outer} r_{\rm outer}^{1/\alpha}
+ \alpha(\gamma_{\rm cr} - 1) W_{\rm outer} r_{\rm outer}^{1 +1 /\alpha}.
\label{eq:defchi} 
\end{equation} 
Equation~(\ref{eq:bc8}) applied at $r=r_{\rm outer}$
provides the eighth boundary condition.  We then solve
our system of equations using a shooting method. We guess the
values of~$\langle \rho\rangle$,~$\langle T\rangle$,
and~$\langle p_{\rm cr} \rangle$ at~$r=r_1$ and then
integrate the equations from~$r=r_1$ to~$r=r_{\rm
  outer}$. We then update our three guesses using Newton's
method until the three boundary conditions at~$r_{\rm
  outer}$ are met.

To compare to future observations and to analyze the turbulent
diffusion of metals in the ICM [see, e.g., Rebusco et al (2005)], it
is of interest to calculate the rms turbulent velocity. We define a
nonlocal turbulent velocity, $u_{\rm NL}$, through the equation
\begin{equation}
\alpha r\,\frac{du_{\rm NL}}{dr} + u_{\rm NL} = u_{\rm L}.
\label{eq:urms1} 
\end{equation} 
Since $u_L$ (and thus $u_{\rm NL}$) is an estimate of the radial component of
the velocity of a convective fluid element,
the full rms turbulent velocity is roughly 
\begin{equation}
u_{\rm rms} = \sqrt{3} \, u_{\rm NL},
\label{eq:urms2} 
\end{equation} 
which is the quantity plotted in figure~\ref{fig:f5}.

\references

Allen, S. W., Dunn, R. J. H., Fabian, A. C., Taylor, G. B., Reynolds, C. S.
2006, MNRAS, 372, 21

Balbus, S. 2000, ApJ, 534, 420

Balbus, S. 2001, ApJ, 562, 909

Begelman, M. C. 2001, in ASP Conf. Proc., 240, {\em Gas and Galaxy Evolution},
ed. J. E. Hibbard, M. P. Rupen, \& J. H. van Gorkom
(San Fransisco: ASP), 363

Begelman, M. C. 2002, in ASP Conf. Proc., 250, {\em Particles and Fields
in Radio Galaxies}, ed. R. A. Laing, \& K. M. Blundell 
(San Fransisco: ASP), 443

Bieber, J. Matthaeus, W., Smith, C., Wanner, W., Kallenrode, M.,
\& Wibberenz, G. 1994, ApJ, 420, 294

Binney, J., \& Tabor, G. 1995, MNRAS, 276, 663

Birzan, L., Rafferty, D., McNamara, B., Wise, M., \& Nulsen, P. 2004,
ApJ, 607, 800

Blanton, E., Sarazin, C., \& McNamara, B. 2003, ApJ, 585, 227

B\"{o}hringer, H. et~al 2001, A\&A, 365, L181

B\"{o}hringer, H., \& Morfill, G. 1988, ApJ, 330, 609

B\"{o}hringer, H., Matsushita, K., Churazov, E., \& Finoguenov, A. 2004a,
in {\em The Riddle of Cooling Flows and Clusters
of Galaxies}, ed. Reiprich, T., Kempner, J., \& Soker, N., E3,
\newline{\bf http://www.astro.virginia.edu/coolflow/proc.php}

Bondi, H. 1952, MNRAS, 112, 159

Braginskii, S. I. 1965, in {\em Reviews of Plasma Physics}, vol. 1,
ed. M. A. Leontovich (New York: Consultants Bureau), 205

Cattaneo, A., \& Teyssier, R.\ 2007, \mnras, 192

Chandran, B. 2000a, Phys. Rev. Lett., 85, 4656

Chandran, B. 2004, ApJ, 616, 169 (Paper~I)

Chandran, B. 2005, ApJ, 632, 809 (Paper~II)

Chandran, B., \& Cowley, S. 1998, Phys. Rev. Lett., 80, 3077

Chandran, B., \& Dennis, T. 2006, ApJ, 642, 140

Chandran, B., Maron, J. 2004, ApJ, 602, 170

Churazov, E., Br\"{u}ggen, M., Kaiser, C., B\"{o}hringer, H., \& Forman, W. 2001,
ApJ, 554, 261

Churazov, E., Forman, W., Jones, C., Sunyaev, R., \& B\"{o}hringer, H. 2004,
MNRAS, 347, 29

Churazov, E., Sunyaev, R., Forman, W., \& B\"{o}hringer, H. 2002, MNRAS, 332, 729

Ciotti, L.,  \& Ostriker, J. 1997, ApJ, 487, L105

Ciotti, L., \& Ostriker, J. 2001, ApJ, 551, 131

Ciotti, L., \& Ostriker, J., \& Pellegrini, S. 2004, 
in {\em Plasmas in the Laboratory and in the Universe: New Insights and New Challenges},

Cox, D. P. 2000, Allen's Astrophysical Quantities (New York: AIP)

David, L. P., Nulsen, P. E. J., McNamara, B. R., Forman, W., Jones, C., Ponman, T., Robertson, B., Wise, M. 2001, ApJ, 557, 546

Dennis, T., \& Chandran, B. 2005, 622, 205

Di Matteo, T., Allen, S.~W., Fabian, A.~C., Wilson, A.~S., \& Young, A.~J.\ 2003, \apj, 582, 133 

Drury, L., \& Volk, H. 1981, ApJ, 248, 344

Eilek, J. 2004, in {\em The Riddle of Cooling Flows and Clusters
of Galaxies}, ed. Reiprich, T., Kempner, J., \& Soker, N., E13,
{\bf http://www.astro.virginia.edu/coolflow/proc.php}

Fabian, A. C. 1994, Ann. Rev. Astr. Astrophys., 32, 277

Fabian, A. C., Sanders, J., Allen, S., Crawford, C., Iwasawa, K., Johnstone, M.,
Schmidt, R., \& Taylor, G. 2003, MNRAS, 344, L43

Fabian, A. C. 1994, Ann. Rev. Astr. Astrophys., 32, 277

Goldreich, P. \& Sridhar, S. 1995, ApJ, 438, 763

Graham, A., Merritt, D., Moore, B., Diemand, J., \& Terzi\'c, B. 2006, ApJ, 132, 2711

Graham, A., Lauer, T.~R., Colless, M., \& Postman, M.\ 1996, ApJ, 465, 534

Grossman, S., Narayan, R., \& Arnett, D. 1993, ApJ, 407, 284

Hernquist, L. 1990, ApJ, 356, 359

Hoeft, M., \& Br\"{u}ggen, M. 2004, ApJ, 617, 896

Kaastra, J. S., Tamura, T., Peterson, J., Bleeker, J., Ferrigno, C.,
Kahn, S., Paerels, F., Piffaretti, R., Branduardi-Raymont, G., \& B\"ohringer, H.
2004, A\&A, 413, 415

Kronberg, P. 1994, Rep. Prog. Phys., 57, 325

Jones, T., \& Kang, H. 1990, ApJ, 363, 499

Kim, W., \& Narayan, R. 2003, ApJ, 596, L139

Lauer, T. R., Faber, S. M., Richstone, D., Gebhardt, K., Tremaine, S., Postman, M., Dressler, A., Aller, M. C., Filippenko, A. V., Green, R., Ho, L. C., Kormendy, J., Magorrian, J., Pinkney, J. 2007, ApJ, 662, 808

Lazarian, A. 2006, ApJL, 645, 25

Lewis, G. F., Babul, A., Katz, N., Quinn, T., Hernquist, L., \& Weinberg, D. 2000,
ApJ, 536, 623

Loewenstein, M., \& Fabian, A. 1990, MNRAS, 242 120

Loewenstein, M., Zweibel, E., \& Begelman, M. 1991, ApJ, 377, 392

Maron, J., Chandran, B., \& Blackman, E. 2004, Phys. Rev. Lett., 92, id. 045001

McLaughlin, D. 1999, ApJ, 512, L9

Molendi, S., \& Pizzolatao, F. 2001, ApJ, 560, 194

Nagai, D., \& Kravtsov, A. 2004, in IAU Colloq. 195, Outskirts of
Galaxy Clusters: Intense Life in the Suburbs, ed. A. Diaferio (Cambridge:
Cambridge Univ. Press), 296

Narayan, R., \& Medvedev, M. 2001, ApJ, 562, 129

Navarro, J., Frenk, C., \& White, S. 1997, ApJ, 490, 493

Nulsen, P. 2004,
in {\em The Riddle of Cooling Flows and Clusters
of Galaxies}, ed. Reiprich, T., Kempner, J., \& Soker, N., E30,
{\bf http://www.astro.virginia.edu/coolflow/proc.php}

Owen, F., \& Ledlow, M. 1997, ApJS, 108, 410

Parrish, I., \& Stone, J., 2005, ApJ, 633, 334

Parrish, I., \& Stone, J. 2006, astro-ph/0612195

Paturel, G.; Petit, C.; Prugniel, Ph.; Theureau, G.; Rousseau, J.; Brouty, M.; Dubois, P.; Cambrésy, L. 2003, A\&A, 412, 45-55

Peterson, J. R., et al 2001, A\&A, 365, L104

Peterson, J. R., Kahn, S., Paerels, F., Kaastra, J., Tamura, T., Bleeker, J.,
Ferrigo, C., \& Jernigan, J. 2003, ApJ, 590, 207

Piffaretti, R.,  Jetzer, Ph., Kaastra, J., Tamura, T. 2005, A\&A, 433, 101

Piffaretti, R.,  \&  Kaastra, J. 2006, A\&A, 453, 423

Pizzolato, F., \& Soker, N. 2005, ApJ, 632, 821

Quataert, E. 1998, ApJ, 500, 978

Quataert, E., \& Narayan, R. 2000, ApJ, 528, 236

Rasera, Y., \& Chandran, B. 2007, ApJ, submitted

Rebusco, P., Churazov, E., Böhringer, H., Forman, W. 2005, MNRAS,
359, 1041

Rechester, R., \& Rosenbluth, M. 1978, Phys. Rev. Lett., 40, 38

Reynolds, C. S. 2002, in ASP Conf. Proc., 250, {\em Particles and Fields
in Radio Galaxies}, ed. R. A. Laing, \& K. M. Blundell 
(San Fransisco: ASP), 449

Reynolds, C. S., McKernan, B., Fabian, A., Stone, J., \& Vernaleo, J. 2005,
MNRAS

Rosner, R., \& Tucker, W. 1989, ApJ, 338 761

Ruszkowski, M., \& Begelman, M. 2002, 581, 223

Ruszkowski, M., Bruggen, M., \& Begelman, M. 2004a, ApJ, 611, 158

Ruszkowski, M., Bruggen, M., \& Begelman, M. 2004b, ApJ, 615, 675

Sazonov, S.~Y., 
Ostriker, J.~P., Ciotti, L., \& Sunyaev, R.~A.\ 2005, MNRAS, 358, 168 

Schekochihin, A. A., Cowley, S. C., Kulsrud, R. M., Hammett, G. W.,
\& Sharma, P. 2006, ApJ, 629, 139

Schekochihin, A. A. \& Cowley, S. C. 2006, Phys. Plasmas,
13, 056501 

Schombert, J.~M.\ 1987, APJS, 64, 643

Soker, N. 2006, New Astronomy, 12, 38

Spitzer, L, \& Harm, R. 1953, Phys. Rev., 89, 977

Springel, V., Di Matteo, T., \& Hernquist, L.\ 2005, \mnras, 361, 776

Suginohara, T.,  \& Ostriker, J. 1998, ApJ, 507, 16

Tabor, G., \& Binney, J. 1993, MNRAS, 263, 323

Tan, J., \& Blackman, E. 2005, MNRAS, 362, 983

Taylor, G., Fabian, A., Allen, S. 2002, MNRAS, 334, 769  

Taylor, G., Govoni, F., Allen, S., Fabian, A. 2001, MNRAS, 326, 2

Tamura, T. et~al 2001, A\&A, 365, L87

Tornatore, L., Borgani, S., Springel, V., Matteucci, F.,
Menci, N., \& Murante, G. 2003, MNRAS, 342, 1025

Tozzi, P., \& Norman, C. 2001, ApJ, 546, 63

Travis, L., \& Matsushima, S. 1973, ApJ, 180, 975

Ulrich, R. 1976, ApJ, 207, 564

Vogt, C., \& Ensslin, T. 2003, A\&A, 412, 373

Vogt, C., \& Ensslin, T. 2005, A\&A, 434, 67

Voigt, L., \& Fabian, A. 2004, MNRAS, 347, 1130

Werner, N., Kaastra, J. S., Takei, Y., Lieu, R., Vink, J.,
\& Tamura, T. 2007, A\&A, 468, 849

Worthey, G. 1994, ApJ, 95, 107

Xiong, D. R. 1991, Proc. Astr. Soc. Australia, 9, 26

Yan, H., \& Lazarian, A. 2004, ApJ, 614, 757

Zakamska, N., \& Narayan, R. 2003, ApJ, 582, 162

\end{document}